# Nonparametric Depth and Quantile Regression for Functional Data

JOYDEEP CHOWDHURY[1,*] PROBAL CHAUDHURI[1,**]

[1]*Statistics and Mathematics Unit, Indian Statistical Institute, Kolkata, India.*
*E-mail:* [*]joydeepchowdhury01@gmail.com; [**]probal@isical.ac.in

We investigate nonparametric regression methods based on spatial depth and quantiles when the response and the covariate are both functions. As in classical quantile regression for finite dimensional data, regression techniques developed here provide insight into the influence of the functional covariate on different parts, like the center as well as the tails, of the conditional distribution of the functional response. Depth and quantile based nonparametric regression methods are useful to detect heteroscedasticity in functional regression. We derive the asymptotic behavior of the nonparametric depth and quantile regression estimates, which depend on the small ball probabilities in the covariate space. Our nonparametric regression procedures are used to analyze a dataset about the influence of per capita GDP on saving rates for 125 countries, and another dataset on the effects of per capita net disposable income on the sale of cigarettes in some states in the US.

## 1. Introduction

Nonparametric regression with functional covariate and real valued response has been extensively studied in the recent literature (see [33], [20], [19], [36], [6], [7], etc.). Since the publication of the seminal paper by Koenker and Bassett [26], quantile regression has emerged as a powerful statistical tool for investigating the nature of dependence of a response on a covariate. Main advantage of quantile regression is that it provides information about the influence of the covariate on all parts of the conditional distribution of the response unlike the usual mean regression, which focuses only on the center of the conditional distribution. Linear quantile regression, where the response is scalar and the covariate is a function, is studied in [5] and [25]. For similar situation, a semiparametric approach in quantile regression is considered in [14]. Nonparametric quantile regression with real valued response and functional covariate is studied in [20] and [21]. The notion of spatial quantiles developed and studied in [2], [12] and [28] extends the concept of univariate quantiles to multivariate data. It was shown in [28] that spatial quantiles completely characterize a multivariate distribution function like the univariate quantiles characterize a univariate distribution. Spatial quantile regression was considered in [9] and [15] for problems where both the response and the covariate are finite dimensional.

Recently, nonparametric spatial quantile regression for finite dimensional response and functional covariate was investigated in [10] and [11].

To probe into different parts of the conditional distribution, which are off-center, we develop nonparametric regression methods based on the conditional spatial depth and quantiles when both the response and the covariate are functions. Spatial depth for multivariate data was developed in [42] and [37], based on the ideas of spatial quantiles in [12] and [28]. The concept of spatial depth is extended to infinite dimensional data in [8]. In [38], multivariate spatial depth was employed in a functional data context by first discretizing the sample curves. In [35], the topological aspect of a formal definition of depth for functional data was investigated. Some aspects of the integrated depth for functional data were studied in [34]. In this paper, we employ conditional spatial depth and quantiles for functional data to investigate the spread of the conditional distribution of the response and detect the presence of heteroscedasticity.

We model functional data as random elements in infinite dimensional spaces. The response is assumed to be an element in a separable Hilbert space while the covariate is assumed to be an element in a complete separable metric space. The conditional spatial quantiles and conditional maximal depth sets are defined in such a setup in section 2. We construct measures of conditional spread based on the quantiles and the depth sets. In section 3, kernel based nonparametric estimates for the conditional spatial quantiles, the maximal depth sets and the measures of conditional spread are constructed, which can be used to investigate possible presence of heteroscedasticity in the data. We investigate the asymptotic properties of the estimates in section 4. The estimates of the conditional spatial quantiles and the maximal depth sets are demonstrated using simulated and real data in section 5. Some computational details for the sample conditional quantiles are described in Appendix A. The proofs of the theorems are presented in Appendix B.

## 2. Conditional spatial depth and quantiles

Given a data cloud, *statistical depth* of a point is a measure of the relative position of the point with respect to that cloud. If a point is deep inside the data cloud, the direction vectors from the observations to that point tend to nullify one another. But, if a point is on the periphery of the data cloud, most of the observations are concentrated around a few directions from the point, and consequently the direction vectors cannot nullify one another. So, the length of the average of all direction vectors originating from the observations towards a particular point in the data cloud will be small if the point is deep inside the data cloud, while this length will be close to 1 for points at the outer regions of the data cloud. We demonstrate this in a bivariate dataset. We generate 20 observations from a bivariate normal distribution with mean **0** and the identity matrix as dispersion. In Figure 1, this data is presented in two sub-plots. In the left sub-plot, the vectors from a few data points towards a given point, marked as point A, near the central region of the data cloud are shown. It is clear that the direction vectors tend to cancel out one another for point A so that the resultant length of the average direction vector will be close to zero. In the right sub-plot, a few vectors are plotted for another point, marked as point B, in the periphery of the data cloud. Here, the vectors have very similar directions, and instead of canceling one another, they add up so that the average direction vector has a length close to 1. This idea was used in [42] and [37] to define the *spatial depth* of a point relative to a data cloud in a multivariate setup. We now extend the same idea to define the conditional depth for the response lying in an infinite dimensional space.

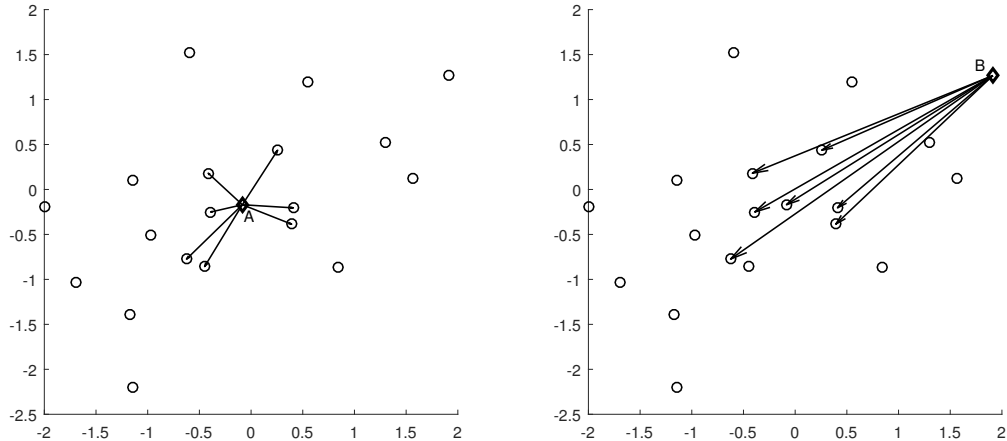

**Figure 1**. Spatial depth obtained using direction vectors

Let the response $\mathbf{Y}$ and the covariate $\mathbf{X}$ be random elements in a separable Hilbert space $\mathcal{H}$ and a complete separable metric space $(\mathcal{M}, d)$, respectively ($d$ is the metric on the space $\mathcal{M}$). We adopt the convention of defining $\|\mathbf{v}\|^{-1}\mathbf{v} = \mathbf{0}$, when $\mathbf{v} = \mathbf{0} \in \mathcal{H}$. Let $\mathbf{x} \in \mathcal{M}$. The conditional spatial distribution $S(\mathbf{y} \mid \mathbf{x})$ of $\mathbf{Y}$ given $\mathbf{X} = \mathbf{x}$ at $\mathbf{y} \in \mathcal{H}$ is defined as $S(\mathbf{y} \mid \mathbf{x}) = E[\|\mathbf{y} - \mathbf{Y}\|^{-1}(\mathbf{y} - \mathbf{Y}) \mid \mathbf{X} = \mathbf{x}]$. The conditional spatial depth $SD(\mathbf{y} \mid \mathbf{x})$ of a point $\mathbf{y} \in \mathcal{H}$ given $\mathbf{X} = \mathbf{x}$ is defined as $SD(\mathbf{y} \mid \mathbf{x}) = 1 - \|S(\mathbf{y} \mid \mathbf{x})\|$ (cf. [8]). Defining the conditional depth in the above way ensures that a point near the center of the conditional distribution of the response has higher depth than a point at the peripheral regions of the conditional distribution, and also that the conditional depth lies between 0 and 1. When the Hilbert space $\mathcal{H}$ is the Euclidean space $\mathbb{R}^p$, our definitions of conditional spatial distribution and depth coincide with the multivariate conditional spatial distribution and depth.

We next define conditional quantiles in Hilbert spaces. The development of multivariate quantiles as minimizers of convex functions in [12] motivates our definition of conditional quantiles in Hilbert spaces. In fact, the definition of multivariate quantiles can be naturally extended to define quantiles in infinite dimensional Hilbert spaces. Let $\boldsymbol{\tau} \in \mathcal{H}$ with $\|\boldsymbol{\tau}\| < 1$. Define $g(\cdot \mid \mathbf{x}) : \mathcal{H} \to \mathbb{R}$ by

$$g(\mathbf{Q} \mid \mathbf{x}) = E[\|\mathbf{Q} - \mathbf{Y}\| - \|\mathbf{Y}\| \mid \mathbf{X} = \mathbf{x}] - \langle \boldsymbol{\tau}, \mathbf{Q} \rangle.$$

The conditional $\boldsymbol{\tau}$-quantile $\mathbf{Q}(\boldsymbol{\tau} \mid \mathbf{x})$ of $\mathbf{Y}$ given $\mathbf{X} = \mathbf{x}$ is defined as a minimizer of $g(\cdot\,|\,\mathbf{x})$ (cf. [8]). It is easy to see that $g(\mathbf{Q}\,|\,\mathbf{x})$ is finite for every $\mathbf{Q}$ and $\mathbf{x}$, even when $E[\|\mathbf{Y}\|\,|\,\mathbf{X} = \mathbf{x}] = \infty$, and if $E[\|\mathbf{Y}\|\,|\,\mathbf{X} = \mathbf{x}] < \infty$, then a minimizer of $g(\mathbf{Q}\,|\,\mathbf{x})$ is same as a minimizer of $E[\|\mathbf{Q} - \mathbf{Y}\|\,|\,\mathbf{X} = \mathbf{x}] - \langle \boldsymbol{\tau}, \mathbf{Q} \rangle$. Using the convexity of $g(\mathbf{Q}\,|\,\mathbf{x})$ and the fact that $g(\mathbf{Q}\,|\,\mathbf{x}) \to \infty$ as $\|\mathbf{Q}\| \to \infty$, one can show that $\mathbf{Q}(\boldsymbol{\tau}\,|\,\mathbf{x})$ exists in $\mathcal{H}$. If the support of the conditional distribution of $\mathbf{Y}$ given $\mathbf{X} = \mathbf{x}$ is not contained in a straight line in $\mathcal{H}$, we get that $g(\mathbf{Q}\,|\,\mathbf{x})$ is strictly convex, and hence $\mathbf{Q}(\boldsymbol{\tau}\,|\,\mathbf{x})$ becomes the unique minimizer of $g(\mathbf{Q}\,|\,\mathbf{x})$. If in addition, the conditional distribution of $\mathbf{Y}$ given $\mathbf{X} = \mathbf{x}$ is non-atomic, it follows that $\mathbf{Q}(\boldsymbol{\tau}\,|\,\mathbf{x})$ is the unique solution of $S(\mathbf{y}\,|\,\mathbf{x}) = \boldsymbol{\tau}$ for $\mathbf{y} \in \mathcal{H}$. The point $\mathbf{Q}(\mathbf{0}\,|\,\mathbf{x})$ is defined as the conditional spatial median. Recall Figure 1 again. Note that $S(\mathbf{y}\,|\,\mathbf{x})$ is the average direction vector from the response $\mathbf{Y}$ to the point $\mathbf{y}$ given $\mathbf{X} = \mathbf{x}$. So, when $\boldsymbol{\tau}$ is near $\mathbf{0}$, we see that $\mathbf{Q}(\boldsymbol{\tau}\,|\,\mathbf{x})$ lies near the conditional spatial median $\mathbf{Q}(\mathbf{0}\,|\,\mathbf{x})$, which is the deepest point in the data, and we have $SD(\mathbf{Q}(\boldsymbol{\tau}\,|\,\mathbf{x})\,|\,\mathbf{x})$ close to 1 for such $\mathbf{Q}(\boldsymbol{\tau}\,|\,\mathbf{x})$. When $\|\boldsymbol{\tau}\|$ is near 1, $\mathbf{Q}(\boldsymbol{\tau}\,|\,\mathbf{x})$ is an extreme quantile, and we have $SD(\mathbf{Q}(\boldsymbol{\tau}\,|\,\mathbf{x})\,|\,\mathbf{x})$ close to 0. This explains the connection between the spatial depth and the spatial quantiles.

When the response $\mathbf{Y}$ is a real random variable, for $\boldsymbol{\tau} = 0$, $\mathbf{Q}(\boldsymbol{\tau}\,|\,\mathbf{x})$ becomes the usual conditional median of $\mathbf{Y}$ given $\mathbf{X} = \mathbf{x}$, and for $\boldsymbol{\tau} = 2\alpha - 1$, $\mathbf{Q}(\boldsymbol{\tau}\,|\,\mathbf{x})$ is the conditional $\alpha$-quantile of $\mathbf{Y}$ given $\mathbf{X} = \mathbf{x}$ for any $0 < \alpha < 1$.

We shall now briefly describe some equivariance properties of the conditional quantiles. Let $\boldsymbol{\Omega}$ be an invertible norm preserving linear operator on $\mathcal{H}$, i.e., $\|\boldsymbol{\Omega}(\mathbf{u})\| = \|\mathbf{u}\|$ for all $\mathbf{u} \in \mathcal{H}$. Then, for any $\mathbf{w} \in \mathcal{H}$, it is straight-forward to verify that $\mathbf{w} + \boldsymbol{\Omega}(\mathbf{Q}(\boldsymbol{\tau}\,|\,\mathbf{x}))$ is the conditional $\boldsymbol{\Omega}(\boldsymbol{\tau})$-quantile of the transformed response $\mathbf{w} + \boldsymbol{\Omega}(\mathbf{Y})$ given $\mathbf{X} = \mathbf{x}$ (cf. Fact 2.2.1 and related discussion in [12]). As a special case, we get that $\mathbf{Q}(\boldsymbol{\tau}\,|\,\mathbf{x})$ is equivariant under any location transformation. Further, one can also easily verify that $\mathbf{Q}(\boldsymbol{\tau}\,|\,\mathbf{x})$ is equivariant under any scale transformation, i.e., $c\mathbf{Q}(\boldsymbol{\tau}\,|\,\mathbf{x})$ is the conditional quantile of the scale-transformed response $c\mathbf{Y}$ given $\mathbf{X} = \mathbf{x}$ for any positive number $c$.

## 2.1. Conditional maximal depth sets

When the response $Y$ is a real random variable, the conditional spatial depth $SD(y\,|\,\mathbf{x})$ simplifies to $SD(y\,|\,\mathbf{x}) = 1 - |2F(y\,|\,\mathbf{x}) - 1|$, where $F(\cdot\,|\,\mathbf{x})$ is the conditional distribution function of $Y$ given $\mathbf{X} = \mathbf{x}$. So, the conditional spatial median of $Y$ given $\mathbf{X} = \mathbf{x}$ is same as the usual conditional median of $Y$ given $\mathbf{X} = \mathbf{x}$. The conditional inter-quartile interval $I(\mathbf{x})$ of $Y$ given $\mathbf{X} = \mathbf{x}$ is $[Q(-0.5\,|\,\mathbf{x}), Q(0.5\,|\,\mathbf{x})]$, where $Q(-0.5\,|\,\mathbf{x})$ and $Q(0.5\,|\,\mathbf{x})$ are the conditional first and third quartiles respectively. The conditional inter-quartile range of $Y$ given $\mathbf{X} = \mathbf{x}$ is $Q(0.5\,|\,\mathbf{x}) - Q(-0.5\,|\,\mathbf{x})$. Denote $\mathcal{I}$ to be the collection of all intervals $I$ such that $P[Y \in I\,|\,\mathbf{X} = \mathbf{x}] \geq 0.5$ and $SD(y_1\,|\,\mathbf{x}) \geq SD(y_2\,|\,\mathbf{x})$ for every $y_1 \in I$ and $y_2 \in I^c$. Then $I(\mathbf{x}) = \bigcap_{I \in \mathcal{I}} I$. We can generalize this property of the conditional inter-quartile interval when the response space $\mathcal{H}$ is a separable Hilbert space.

Given $\alpha > 0$, we define the spatial depth based conditional $\alpha$-trimmed set $B(\alpha\,|\,\mathbf{x})$ for $\mathbf{Y}$ given $\mathbf{X} = \mathbf{x}$ as $B(\alpha\,|\,\mathbf{x}) = \{\mathbf{y} \in \mathcal{H}\,|\,SD(\mathbf{y}\,|\,\mathbf{x}) \geq \alpha\}$ (cf. [43]). Clearly, $\alpha_1 \geq \alpha_2$ implies that $B(\alpha_1\,|\,\mathbf{x}) \subseteq B(\alpha_2\,|\,\mathbf{x})$. For $0 < p < 1$, let

$$\mathcal{A}_p = \{\alpha > 0\,|\,P[\mathbf{Y} \in B(\alpha\,|\,\mathbf{x})\,|\,\mathbf{X} = \mathbf{x}] \geq p\}.$$

The set $\mathcal{A}_p$ is bounded above as $SD(\mathbf{y} \,|\, \mathbf{x}) \leq 1$. Denote $\alpha_p = \sup \mathcal{A}_p$. We define the conditional $100p\%$ maximal depth set of $\mathbf{Y}$ given $\mathbf{X} = \mathbf{x}$ as $B(\alpha_p \,|\, \mathbf{x})$. So, the set $B(\alpha_p \,|\, \mathbf{x})$ contains $100p\%$ of the conditional probability mass with its elements having the highest conditional spatial depth. Clearly, the conditional spatial median of $\mathbf{Y}$ given $\mathbf{X} = \mathbf{x}$ belongs to the conditional $100p\%$ maximal depth set. Recall that $\mathbf{Q}(\boldsymbol{\tau} \,|\, \mathbf{x})$ is the unique solution of $S(\mathbf{y} \,|\, \mathbf{x}) = \boldsymbol{\tau}$ when the conditional distribution of $\mathbf{Y}$ given $\mathbf{X} = \mathbf{x}$ is non-atomic, and its support is not contained in a straight line in $\mathcal{H}$. So, we get

$$B(\alpha \,|\, \mathbf{x}) = \{\mathbf{y} \in \mathcal{H} \,|\, \|S(\mathbf{y} \,|\, \mathbf{x})\| \leq 1 - \alpha\} = \{\mathbf{Q}(\boldsymbol{\tau} \,|\, \mathbf{x}) \in \mathcal{H} \,|\, \|\boldsymbol{\tau}\| \leq 1 - \alpha\}.$$

Hence, the spatial depth based conditional $100p\%$ maximal depth set of $\mathbf{Y}$ given $\mathbf{X} = \mathbf{x}$ is $\{\mathbf{Q}(\boldsymbol{\tau} \,|\, \mathbf{x}) \in \mathcal{H} \,|\, \|\boldsymbol{\tau}\| \leq 1 - \alpha_p\}$. Note that the conditional $50\%$ maximal depth set reduces to the conditional inter-quartile interval when $\mathcal{H} = \mathbb{R}$.

The functional box plot defined in [41] is a concept closely related to our maximal depth set, which we define in a conditional set up while the concept of functional box plot was introduced in an unconditional set up. The functional box plot was defined using modified band depths of the sample functional observations. Unlike our maximal depth set, no population version of the functional box plot was provided in [41]. If one constructs the functional box plot for the population using the idea of pointwise range in [41], it can be shown using the isolated outliers concept in Section 2.1 in [24] that the functional box plot may sometimes become unbounded and hence a non-informative set.

## 2.2. Measures of conditional spread

We define two measures of conditional spread, one based on conditional maximal depth sets and another using conditional spatial quantiles. The measure $D_1(p \,|\, \mathbf{x})$ of conditional spread is defined as the diameter of the conditional $100p\%$ maximal depth set of $\mathbf{Y}$ given $\mathbf{X} = \mathbf{x}$, i.e., $D_1(p \,|\, \mathbf{x}) = \sup\{\|\mathbf{y_1} - \mathbf{y_2}\| \,|\, \mathbf{y_1}, \mathbf{y_2} \in B(\alpha_p \,|\, \mathbf{x})\}$. Equivalently,

$$D_1(p \,|\, \mathbf{x}) = \sup\{\|\mathbf{Q}(\boldsymbol{\tau}_1 \,|\, \mathbf{x}) - \mathbf{Q}(\boldsymbol{\tau}_2 \,|\, \mathbf{x})\| \,|\, \|\boldsymbol{\tau}_1\|, \|\boldsymbol{\tau}_2\| \leq 1 - \alpha_p\}.$$

Note that $D_1(0.5 \,|\, \mathbf{x})$ generalizes the concept of the conditional interquartile range that we have for a real valued response. We also define a measure of directional spread, based on conditional spatial quantiles. The measure $D_2(\boldsymbol{\tau} \,|\, \mathbf{x})$ of conditional spread is defined as $D_2(\boldsymbol{\tau} \,|\, \mathbf{x}) = \|\mathbf{Q}(\boldsymbol{\tau} \,|\, \mathbf{x}) - \mathbf{Q}(-\boldsymbol{\tau} \,|\, \mathbf{x})\|$, where $\|\boldsymbol{\tau}\| < 1$. Note that $D_2(\boldsymbol{\tau} \,|\, \mathbf{x})$ depends only on the conditional quantiles in the direction of $\boldsymbol{\tau}$, while $D_1(p \,|\, \mathbf{x})$ is a 'global' measure of the conditional spread in the sense that its definition involves conditional quantiles in all directions. The measure $D_2(\boldsymbol{\tau} \,|\, \mathbf{x})$ also reduces to the conditional inter-quartile range when the response is real valued and $\boldsymbol{\tau} = 0.5$ because $\mathbf{Q}(-\boldsymbol{\tau} \,|\, \mathbf{x})$ and $\mathbf{Q}(\boldsymbol{\tau} \,|\, \mathbf{x})$ become the conditional first and third quartiles respectively.

Both $D_1(p \,|\, \mathbf{x})$ and $D_2(\boldsymbol{\tau} \,|\, \mathbf{x})$ can be used to investigate the presence of heteroscedasticity in regression problems involving functional data, and this will be demonstrated in section 5 using real and simulated datasets. Investigation of heteroscedasticity using linear regression quantiles for a real valued response and finite dimensional covariate was done as early as in 1982 by Koenker and Bassett (see [27]). Subsequently, nonparametric quantile regression was used in [22] and [13] to study heteroscedasticity in problems involving real valued response and finite dimensional covariate.

## 3. Nonparametric estimates

We describe here the construction of kernel estimates of $S(\mathbf{y} \,|\, \mathbf{x})$, $SD(\mathbf{y} \,|\, \mathbf{x})$ and $\mathbf{Q}(\boldsymbol{\tau} \,|\, \mathbf{x})$. Given the sample $(\mathbf{X}_1, \mathbf{Y}_1), \cdots, (\mathbf{X}_n, \mathbf{Y}_n)$, a Kernel function $K(\cdot)$ supported on $[0,1]$ with a bandwidth $h_n$, and for $\mathbf{y} \in \mathcal{H}$, the kernel estimator $\widehat{S}(\mathbf{y} \,|\, \mathbf{x})$ of $S(\mathbf{y} \,|\, \mathbf{x})$ is defined as

$$\widehat{S}(\mathbf{y} \,|\, \mathbf{x}) = \frac{\sum_{i=1}^{n} (\|\mathbf{y} - \mathbf{Y}_i\|^{-1}(\mathbf{y} - \mathbf{Y}_i)) K(h_n^{-1} d(\mathbf{x}, \mathbf{X}_i))}{\sum_{i=1}^{n} K(h_n^{-1} d(\mathbf{x}, \mathbf{X}_i))}.$$

The kernel estimate $\widehat{SD}(\mathbf{y} \,|\, \mathbf{x})$ of $SD(\mathbf{y} \,|\, \mathbf{x})$ is defined as $\widehat{SD}(\mathbf{y} \,|\, \mathbf{x}) = 1 - \|\widehat{S}(\mathbf{y} \,|\, \mathbf{x})\|$.

### 3.1. Estimation of conditional quantiles

When the response space $\mathcal{H}$ is finite dimensional, the conditional sample $\boldsymbol{\tau}$-quantile $\widehat{\mathbf{Q}}_n(\boldsymbol{\tau} \,|\, \mathbf{x})$ can be defined as a minimizer of the function

$$\widehat{g}_n(\mathbf{Q} \,|\, \mathbf{x}) = \frac{\sum_{i=1}^{n} \|\mathbf{Q} - \mathbf{Y}_i\| K(h_n^{-1} d(\mathbf{x}, \mathbf{X}_i))}{\sum_{i=1}^{n} K(h_n^{-1} d(\mathbf{x}, \mathbf{X}_i))} - \langle \boldsymbol{\tau}, \mathbf{Q} \rangle$$

in $\mathcal{H}$. $\widehat{\mathbf{Q}}_n(\boldsymbol{\tau} \,|\, \mathbf{x})$ can be computed using iterative methods (see [12]). When $\mathcal{H}$ is an infinite dimensional separable Hilbert space, let $\{\mathbf{e}_n\}$ be an orthonormal basis of $\mathcal{H}$. For $\mathbf{v} \in \mathcal{H}$, let $\{v_k\}$ satisfy $\mathbf{v} = \sum_{k=1}^{\infty} v_k \mathbf{e}_k$. Let $\{d_n\}$ be a sequence of positive integers increasing to infinity, and let $\mathcal{Z}_n = \operatorname{span}\{\mathbf{e}_1, \mathbf{e}_2, \cdots, \mathbf{e}_{d_n}\}$. Define $\mathbf{v}^{(n)} = \sum_{k=1}^{d_n} v_k \mathbf{e}_k$ for $\mathbf{v} \in \mathcal{H}$. We define the function $\widehat{g}_n(\cdot \,|\, \mathbf{x})$ on $\mathcal{H}$ as

$$\widehat{g}_n(\mathbf{Q} \,|\, \mathbf{x}) = \frac{\sum_{i=1}^{n} \|\mathbf{Q} - \mathbf{Y}_i^{(n)}\| K(h_n^{-1} d(\mathbf{x}, \mathbf{X}_i))}{\sum_{i=1}^{n} K(h_n^{-1} d(\mathbf{x}, \mathbf{X}_i))} - \langle \boldsymbol{\tau}^{(n)}, \mathbf{Q} \rangle. \tag{3.1}$$

The conditional sample $\boldsymbol{\tau}$-quantile $\widehat{\mathbf{Q}}_n(\boldsymbol{\tau} \,|\, \mathbf{x})$ is defined as a minimizer of $\widehat{g}_n(\mathbf{Q} \,|\, \mathbf{x})$ in $\mathcal{Z}_n$. This method of computing conditional spatial quantiles in an infinite dimensional space is similar to the procedure described in [8] for computing unconditional quantiles.

The function $\widehat{g}_n(\mathbf{Q} \,|\, \mathbf{x})$ is not Fréchet differentiable at $\mathbf{Q} = \mathbf{Y}_i^{(n)}$ for any $i$. So, we cannot compute $\widehat{\mathbf{Q}}_n(\boldsymbol{\tau} \,|\, \mathbf{x})$ by directly solving the equation $\widehat{g}_n^{(1)}(\mathbf{Q} \,|\, \mathbf{x}) = \mathbf{0}$ in $\mathcal{Z}_n$ using a straightforward Newton-Raphson type iterative method, where $\widehat{g}_n^{(1)}(\mathbf{Q} \,|\, \mathbf{x})$ is the Fréchet derivative of $\widehat{g}_n(\mathbf{Q} \,|\, \mathbf{x})$ w.r.t. $\mathbf{Q}$. Instead, we first check if any of the $\mathbf{Y}_i^{(n)}$s minimize $\widehat{g}_n(\mathbf{Q} \,|\, \mathbf{x})$ in $\mathcal{Z}_n$. We apply the Newton-Raphson method if $\widehat{g}_n(\mathbf{Q} \,|\, \mathbf{x})$ is not minimized at

any of the $\mathbf{Y}_i^{(n)}$s. Details of this computational procedure are provided in Appendix A. Further details of the estimation procedure and the choices of the bandwidth $h_n$, the basis $\{\mathbf{e}_n\}$ and $\{d_n\}$ are discussed in section 5.

Like the population conditional quantile $\mathbf{Q}(\boldsymbol{\tau} \,|\, \mathbf{x})$, the conditional sample quantile $\widehat{\mathbf{Q}}_n(\boldsymbol{\tau} \,|\, \mathbf{x})$ is also equivariant under all invertible and distance preserving affine transformations. Further, the sample conditional quantiles are scale equivariant in the sense that for any positive constant $c$, $c\widehat{\mathbf{Q}}_n(\boldsymbol{\tau} \,|\, \mathbf{x})$ is the conditional sample quantile for the scale-transformed responses $c\mathbf{Y}_i$, $i = 1, \cdots, n$.

### 3.2. Estimation of maximal depth sets and conditional spread

We estimate the conditional $100p\%$ maximal depth set of $\mathbf{Y}$ given $\mathbf{X} = \mathbf{x}$ as follows. We order the sample of responses $\mathbf{Y}_1, \cdots, \mathbf{Y}_n$ by their conditional sample spatial depth $\widehat{SD}(\mathbf{Y}_i \,|\, \mathbf{x})$, and denote the ordered responses as $\mathbf{Y}_{[1]}, \cdots, \mathbf{Y}_{[n]}$, where $\widehat{SD}(\mathbf{Y}_{[i]} \,|\, \mathbf{x}) \geq \widehat{SD}(\mathbf{Y}_{[i+1]} \,|\, \mathbf{x})$ for $i = 1, \cdots, n-1$. Given $p \in (0,1)$, let $i_p$ be the smallest integer such that

$$\frac{\sum_{i=1}^{i_p} K(h_n^{-1} d(\mathbf{x}, \mathbf{X}_{[i]}))}{\sum_{i=1}^{n} K(h_n^{-1} d(\mathbf{x}, \mathbf{X}_{[i]}))} \geq p.$$

The conditional sample $100p\%$ maximal depth set of $\mathbf{Y}$ given $\mathbf{X} = \mathbf{x}$ is the set $\{\mathbf{Y}_{[1]}, \cdots, \mathbf{Y}_{[i_p]}\}$, which contains $100p\%$ of the sample observations having the highest conditional sample spatial depth. Define

$$\widehat{D}_1(p \,|\, \mathbf{x}) = \max\{\|\mathbf{Y}_{[i]} - \mathbf{Y}_{[j]}\| \,|\, i, j = 1, \cdots, i_p\},$$

which is an estimate of the conditional spread measure $D_1(p \,|\, \mathbf{x})$. The other measure of conditional spread, $D_2(\boldsymbol{\tau} \,|\, \mathbf{x})$, is estimated by

$$\widehat{D}_2(\boldsymbol{\tau} \,|\, \mathbf{x}) = \|\widehat{\mathbf{Q}}_n(\boldsymbol{\tau} \,|\, \mathbf{x}) - \widehat{\mathbf{Q}}_n(-\boldsymbol{\tau} \,|\, \mathbf{x})\|.$$

## 4. Asymptotic Properties of estimates

We now proceed to derive the asymptotic properties of the estimates of conditional spatial distribution, depth and quantiles. Recall that the response space is a separable Hilbert space $\mathcal{H}$, and the kernel function is denoted by $K(\cdot)$. Define the small ball probability function $\phi(\cdot \,|\, \mathbf{x})$ as $\phi(h \,|\, \mathbf{x}) = P[d(\mathbf{x}, \mathbf{X}) \leq h]$. Denote the conditional probability measure of $\mathbf{Y}$ given $\mathbf{X} = \mathbf{z}$ as $\mu(\cdot \,|\, \mathbf{z})$. We make the following assumptions on $\phi(\cdot \,|\, \mathbf{x})$, $K(\cdot)$ and $\mu(\cdot \,|\, \mathbf{z})$.

C(i) $\phi(h \,|\, \mathbf{x}) > 0$ for all $h > 0$. Also, there exists a function $\rho(\cdot \,|\, \mathbf{x}) : [0,1] \to [0,1]$ such that, for every $0 \leq s \leq 1$,

$$\frac{\phi(hs \,|\, \mathbf{x})}{\phi(h \,|\, \mathbf{x})} \to \rho(s \,|\, \mathbf{x}) \text{ as } h \to 0^+.$$

C(ii) The kernel function $K(\cdot)$ is bounded and supported on $[0,1]$ with $K(1) > 0$, and it has a continuous bounded derivative on $(0,1)$ such that $K'(u) \leq 0$ for all $0 < u < 1$.

C(iii) $\mu(\cdot \,|\, \mathbf{z}) \to \mu(\cdot \,|\, \mathbf{x})$ *weakly* as $d(\mathbf{x}, \mathbf{z}) \to 0$.

Assumptions C(i), C(ii) and C(iii) will be considered to be true throughout this section, Appendix B and section 2 in the supplement [17], and we may not always explicitly mention them. It is easy to see that the function $\rho(s \,|\, \mathbf{x})$ in assumption C(i) is non-decreasing in $s$, and $\rho(1 \,|\, \mathbf{x}) = 1$. Below we describe some situations when the above assumptions are satisfied.

- Assumption C(i) holds if the covariate $\mathbf{X}$ is finite dimensional with a positive probability density at $\mathbf{x}$ or a fractal-type process (see [20, p. 207]), as then $\rho(s \,|\, \mathbf{x}) = s^d$ for some $d > 0$. On the other hand, if $\mathbf{X}$ is an infinite dimensional process like a continuous Gaussian Markov process on an interval equipped with the $L_p$-norm with $1 \leq p \leq \infty$, and $\mathbf{x}$ belongs to the associated reproducing kernel Hilbert space, then $\phi(h \,|\, \mathbf{x}) \sim c_1(\mathbf{x}) \exp[-c_2 h^{-2}]$ as $h \to 0$, where $c_1(\mathbf{x}) > 0$, and $c_2 > 0$ does not depend on $\mathbf{x}$ (see Theorem 3.1 in [31] and Theorem 1.1 in [30]). Also, from Theorem 2.1 in [23] and Theorem 1.1 in [30] it follows that that for any continuous Gaussian Markov process on an interval equipped with the $L_2$-norm and for any $\mathbf{x}$, $\phi(h \,|\, \mathbf{x}) \sim c_1(\mathbf{x}) \exp[-c_2 h^{-2}]$ as $h \to 0$. In both the above cases, one can show that $\rho(s \,|\, \mathbf{x}) = I(s = 1)$. In particular, if the covariate $\mathbf{X}$ is a Brownian motion or a Brownian bridge or an Ornstein–Uhlenbeck process equipped with the $L_2$-norm, we have $\rho(s \,|\, \mathbf{x}) = I(s = 1)$. See Proposition 1 in [19] for other examples of $\rho(s \,|\, \mathbf{x})$.
- Kernel functions satisfying condition C(ii) have been considered earlier in [19]. These are type I kernel functions (see Definition 4.1 in [20, p. 42]), which are popular in nonparametric regression involving functional data (see, e.g., [18], [33], [3], [32]), satisfying the additional condition that their derivatives are nonnegative. The simplest choice of $K(\cdot)$ satisfying C(ii) is the indicator kernel $K(u) = I(0 \leq u \leq 1)$. Other examples of such kernels are the truncated Gaussian kernel $K(u) = \exp[-u^2/2]I(0 \leq u \leq 1)$, the truncated triangular kernel $K(u) = [1 - u/2]I(0 \leq u \leq 1)$, etc.
- Assumption C(iii) states the continuity of the conditional probability measure of $\mathbf{Y}$ given $\mathbf{X} = \mathbf{z}$ at $\mathbf{X} = \mathbf{x}$, and this holds in many standard models. For example, consider the location-scale model: $\mathbf{Y} = \mathbf{m}(\mathbf{X}) + f(\mathbf{X})\mathbf{G}$, where $\mathbf{X}$ and $\mathbf{G}$ are independent random elements in $(\mathcal{M}, d)$ and $\mathcal{H}$ respectively, and the functions $\mathbf{m}(\cdot) : \mathcal{M} \to \mathcal{H}$ and $f(\cdot) : \mathcal{M} \to \mathbb{R}$ are both continuous at $\mathbf{x}$. It is easy to verify that assumption C(iii) holds in this model. Evidently, this model covers every situation where the response is a Gaussian process equipped with the $L_2$-norm given the covariate, and its conditional mean and conditional covariance operator are continuous functions of the covariate.

From assumption C(ii), we get $0 < l \leq K(u) \leq L < \infty$ for all $u \in [0,1]$, where $L = K(0)$ and $l = K(1)$. Consequently,

$$l^j \phi(h \,|\, \mathbf{x}) \leq E[K^j(h^{-1} d(\mathbf{x}, \mathbf{X}))] \leq L^j \phi(h \,|\, \mathbf{x})$$

for any $h > 0$ and any positive integer $j$. Denote

$$F_{(j)}(h\,|\,\mathbf{x}) = \frac{E[K^j(h^{-1}d(\mathbf{x},\mathbf{X}))]}{\phi(h\,|\,\mathbf{x})}.$$

Using assumptions C(i), C(ii) and arguments similar to those in the proof of Lemma 2 in [19], we get

$$F_{(1)}(h\,|\,\mathbf{x}) \to K(1) - \int_0^1 \rho(s\,|\,\mathbf{x})K'(s)ds = E_{(1)}(\mathbf{x}) \text{ (say)}$$

and

$$F_{(2)}(h\,|\,\mathbf{x}) \to K^2(1) - 2\int_0^1 \rho(s\,|\,\mathbf{x})K(s)K'(s)ds = E_{(2)}(\mathbf{x}) \text{ (say)}$$

as $h \to 0$. Clearly, $0 < l^j \leq E_{(j)}(\mathbf{x}) \leq L^j$ for $j = 1, 2$. Denote $E_n = E[K(h_n^{-1}d(\mathbf{x},\mathbf{X}))]$. Define the bilinear operator $\gamma(\mathbf{y}\,|\,\mathbf{z})(\cdot,\cdot) : \mathcal{H} \times \mathcal{H} \to \mathbb{R}$ as

$$\gamma(\mathbf{y}\,|\,\mathbf{z})(\mathbf{v},\mathbf{w}) = Cov\left[\left\langle \frac{\mathbf{y}-\mathbf{Y}}{\|\mathbf{y}-\mathbf{Y}\|},\mathbf{v}\right\rangle, \left\langle \frac{\mathbf{y}-\mathbf{Y}}{\|\mathbf{y}-\mathbf{Y}\|},\mathbf{w}\right\rangle \,\bigg|\, \mathbf{X} = \mathbf{z}\right].$$

Under the following additional assumptions, we shall obtain the asymptotic normality of the conditional sample spatial distribution $\widehat{S}(\mathbf{y}\,|\,\mathbf{x})$ and the rate of convergence of the conditional sample spatial depth $\widehat{SD}(\mathbf{y}\,|\,\mathbf{x})$.

A-1. The bandwidth $h_n$ satisfies $h_n \to 0$ and $(n\phi(h_n\,|\,\mathbf{x}))^{-1}\log n \to 0$ as $n \to \infty$.

A-2. For $\mathbf{y} \in \mathcal{H}$ and for $d(\mathbf{x},\mathbf{z}) \leq C_1$, we have $(d(\mathbf{x},\mathbf{z}))^{-1}\|S(\mathbf{y}\,|\,\mathbf{z}) - S(\mathbf{y}\,|\,\mathbf{x})\| \leq s_1$, where $C_1$ and $s_1$ are positive constants.

We now discuss assumptions A-1 and A-2.

- We can choose a sequence of bandwidths $\{h_n\}$ satisfying condition A-1 whenever the small ball probability function $\phi(h\,|\,\mathbf{x})$ is a continuous function of $h$ for all sufficiently small $h$, and $\phi(h\,|\,\mathbf{x}) > 0$ for all $h > 0$. It follows from the discussion on assumption C(i) that this requirement is satisfied for many covariate distributions.
- Assumption A-2 is a smoothness condition on the conditional spatial distribution $S(\mathbf{y}\,|\,\mathbf{z})$ for $\mathbf{z}$ lying in a neighborhood of $\mathbf{x}$, which will be required to derive the order of convergence of $\widehat{SD}(\mathbf{y}\,|\,\mathbf{x})$ to $SD(\mathbf{y}\,|\,\mathbf{x})$ in Theorem 4.1. Assumption A-2 holds in many models. As an example, consider again the location-scale model: $\mathbf{Y} = \mathbf{m}(\mathbf{X}) + f(\mathbf{X})\mathbf{G}$, with $\mathbf{X}$ and $\mathbf{G}$ being independent random elements in $(\mathcal{M}, d)$ and $\mathcal{H}$ respectively. Assume that $E[\|\mathbf{G}\|] < \infty$, and $\mathbf{m}(\cdot) : \mathcal{M} \to \mathcal{H}$ and $f(\cdot) : \mathcal{M} \to \mathbb{R}$ are both Lipschitz continuous at $\mathbf{x}$ such that $f(\mathbf{x}) > 0$. Assume also that some trivariate marginal distribution of $(\langle\mathbf{G},\mathbf{e}_1\rangle, \langle\mathbf{G},\mathbf{e}_2\rangle, \cdots)$ has a density that is uniformly bounded on bounded subsets. Clearly, this particular requirement is satisfied for $\mathbf{G}$ being any Gaussian process. Then, one can verify that assumption A-2 holds.

**Theorem 4.1.** *Denote*

$$M_n(\mathbf{y} \mid \mathbf{x}) = \frac{\sum_{i=1}^n [S(\mathbf{y} \mid \mathbf{X}_i) - S(\mathbf{y} \mid \mathbf{x})] K(h_n^{-1} d(\mathbf{x}, \mathbf{X}_i))}{\sum_{i=1}^n K(h_n^{-1} d(\mathbf{x}, \mathbf{X}_i))}.$$

*Under assumption A-1,*

$$\sqrt{n\phi(h_n \mid \mathbf{x})}(\widehat{S}(\mathbf{y} \mid \mathbf{x}) - S(\mathbf{y} \mid \mathbf{x}) - M_n(\mathbf{y} \mid \mathbf{x})) \to \mathbf{W}$$

*in distribution as $n \to \infty$, where $\mathbf{W}$ is a Gaussian random element in $\mathcal{H}$ with mean $\mathbf{0}$ and covariance operator $[(E_{(1)}(\mathbf{x}))^{-2} E_{(2)}(\mathbf{x})]\gamma(\mathbf{y} \mid \mathbf{x})(\cdot, \cdot)$. Also, under assumptions A-1 and A-2,*

$$\left|\widehat{SD}(\mathbf{y} \mid \mathbf{x}) - SD(\mathbf{y} \mid \mathbf{x})\right| = O_P\left(\frac{1}{\sqrt{n\phi(h_n \mid \mathbf{x})}} + h_n\right)$$

*as $n \to \infty$.*

$M_n(\mathbf{y} \mid \mathbf{x})$ can be viewed as the bias in the kernel estimate $\widehat{S}(\mathbf{y} \mid \mathbf{x})$ of $S(\mathbf{y} \mid \mathbf{x})$. From assumptions A-1 and A-2, it follows that $\|M_n(\mathbf{y} \mid \mathbf{x})\| = O(h_n)$ as $n \to \infty$ almost surely. So, from Theorem 4.1, we get

$$\left\|\widehat{S}(\mathbf{y} \mid \mathbf{x}) - S(\mathbf{y} \mid \mathbf{x})\right\| = O_P\left(\frac{1}{\sqrt{n\phi(h_n \mid \mathbf{x})}} + h_n\right)$$

as $n \to \infty$.

If $h_n$ satisfies $\sqrt{n\phi(h_n \mid \mathbf{x})}h_n \to 0$ as $n \to \infty$ (cf. [33, p. 163], [10, p. 1563]), then $\sqrt{n\phi(h_n \mid \mathbf{x})}M_n(\mathbf{y} \mid \mathbf{x}) \to \mathbf{0}$ *in probability* as $n \to \infty$, and from Theorem 4.1, we get $\sqrt{n\phi(h_n \mid \mathbf{x})}(\widehat{S}(\mathbf{y} \mid \mathbf{x}) - S(\mathbf{y} \mid \mathbf{x})) \to \mathbf{W}$ *in distribution* as $n \to \infty$. Note that the order of convergence for both $\widehat{S}(\mathbf{y} \mid \mathbf{x})$ and $\widehat{SD}(\mathbf{y} \mid \mathbf{x})$ is $O_P([\sqrt{n\phi(h_n \mid \mathbf{x})}]^{-1} + h_n)$. The terms $[\sqrt{n\phi(h_n \mid \mathbf{x})}]^{-1}$ and $h_n$ in the order of convergence come from the variance and the bias of the estimate $\widehat{S}(\mathbf{y} \mid \mathbf{x})$, respectively. For a choice of bandwidth $h_n$, which balances the bias and the variance, $\sqrt{n\phi(h_n \mid \mathbf{x})}h_n$ will be bounded and bounded away from 0 as $n \to \infty$.

Let the covariate $\mathbf{X}$ be either finite dimensional with a positive density at $\mathbf{x}$ or a fractal-type process. Then, $\phi(h_n \mid \mathbf{x}) = O(h_n^d)$, where $d > 0$. In that case, the choice of the bandwidth $h_n$, which balances the asymptotic order of the bias, i.e., $O(h_n)$, and that of the variance, i.e., $O_P([\sqrt{n\phi(h_n \mid \mathbf{x})}]^{-1})$, is $c_3 n^{-(d+2)^{-1}}$ for some constant $c_3 > 0$ depending on $\mathbf{x}$. It is easy to see that this choice of $h_n$ satisfies assumption A-1, and for this choice, the optimum rate of convergence of both $\widehat{SD}(\mathbf{y} \mid \mathbf{x})$ and $\widehat{S}(\mathbf{y} \mid \mathbf{x})$ is $O_P(n^{-(d+2)^{-1}})$. This optimum rate of convergence is same as the optimum rate in the usual nonparametric mean regression estimate involving a multivariate response and a multivariate covariate (see, e.g., [39], [40]). Now, suppose the covariate $\mathbf{X}$ is a continuous Gaussian Markov process equipped with the $L_2$-norm. In such a case, $\phi(h_n \mid \mathbf{x}) \sim c_1(\mathbf{x}) \exp[-c_2 h_n^{-2}]$ as $h_n \to 0$, where $c_1(\mathbf{x}) > 0$ and $c_2 > 0$. Here, it is easy to see that if $\sqrt{n\phi(h_n \,|\, \mathbf{x})}h_n$ is bounded above, then $(n\phi(h_n \,|\, \mathbf{x}))^{-1}\log n$ is bounded away from 0 as $n \to \infty$. So, there exists no choice of $h_n$ which simultaneously satisfies assumption A-1 and balances the orders of the bias and the variance. It is easy to verify that assumption A-1 is satisfied if $h_n = c_4([\sqrt{\log n}]^{-1})$ for a constant $c_4 > \sqrt{c_2}$. Then, the rate of convergence of both $\widehat{SD}(\mathbf{y} \,|\, \mathbf{x})$ and $\widehat{S}(\mathbf{y} \,|\, \mathbf{x})$ is $O_P([\sqrt{\log n}]^{-1})$. Using arguments similar to those used in [16], it can be shown that this rate is optimal under appropriate conditions and same as the optimal rate of the kernel-based mean regression estimate. Finally, we notice that the dimension of the response does not affect the convergence rate.

We next turn our attention to the conditional sample spatial quantiles. Denote the conditional probability measures of $\mathbf{Y}$ and $\mathbf{Y}^{(n)}$ given $\mathbf{X} = \mathbf{z}$ as $\mu(\cdot \,|\, \mathbf{z})$ and $\mu^{(n)}(\cdot \,|\, \mathbf{z})$, respectively. The following assumptions are required for the subsequent results.

B-1. The bandwidth $h_n$ satisfies $h_n \to 0$ and $(n\phi(h_n \,|\, \mathbf{x}))^{(1/2)-2\alpha}\log n \to 0$ as $n \to \infty$, where $1/4 < \alpha \le 1/2$ is a constant.

B-2. There exists a constant $C_2 > 0$ and a positive integer $N_1$ such that whenever $d(\mathbf{x}, \mathbf{z}) \le C_2$, $\mu^{(N_1)}(\cdot \,|\, \mathbf{z})$ is non-atomic and its support is not contained in a straight line in $\mathcal{H}$, i.e., there exist no $\mathbf{a}, \mathbf{b} \in \mathcal{H}$ such that

$$\mu^{(N_1)}\left(\{\mathbf{v} \in \mathcal{H} : \mathbf{v} = \mathbf{a} + t\mathbf{b}, t \in (-\infty, \infty)\} \,|\, \mathbf{z}\right) = 1.$$

B-3. There exists a constant $C_3 > 0$ and a positive integer $N_2$ such that whenever $d(\mathbf{x}, \mathbf{z}) \le C_3$ and $n \ge N_2$, we have for each $C > 0$, $E[\|\mathbf{Q} - \mathbf{Y}^{(n)}\|^{-2} \,|\, \mathbf{X} = \mathbf{z}] \le s_2(C)$ for all $\mathbf{Q} \in \mathcal{Z}_n$ with $\|\mathbf{Q}\| \le C$. Here $s_2(C)$ is a positive constant depending on $C$.

Below we discuss assumptions B-1, B-2 and B-3.

- We can choose a sequence of bandwidths $\{h_n\}$ satisfying condition B-1 if $\phi(h \,|\, \mathbf{x})$ is continuous in $h$ for all sufficiently small $h$, and $\phi(h \,|\, \mathbf{x}) > 0$ for all $h > 0$. From the discussion on assumption C(i), one can verify that this requirement is met for many covariate distributions.
- Assumption B-2 implies that the probability measure $\mu^{(N_1)}(\cdot \,|\, \mathbf{z})$, which is equivalent to a multivariate probability measure, induces a continuous multivariate distribution, which is not concentrated on a straight line whenever $\mathbf{z}$ lies in some neighborhood of $\mathbf{x}$. Note that $\mu^{(N_1)}(\cdot \,|\, \mathbf{z})$ is non-atomic implies that both $\mu(\cdot \,|\, \mathbf{z})$ and $\mu^{(n)}(\cdot \,|\, \mathbf{z})$ are non-atomic for all $n \ge N_1$. Also, if the support of $\mu^{(N_1)}(\cdot \,|\, \mathbf{z})$ is not contained in a straight line in $\mathcal{H}$, then the supports of both $\mu(\cdot \,|\, \mathbf{z})$ and $\mu^{(n)}(\cdot \,|\, \mathbf{z})$ are not contained in any straight line in $\mathcal{H}$ for all $n \ge N_1$. When the conditional distribution of $\mathbf{Y}$ given $\mathbf{X}$ is Gaussian with its conditional mean and conditional covariance operator being continuous at $\mathbf{X} = \mathbf{x}$, assumption B-2 is satisfied.
- It can be established that assumption B-3 holds when some conditional trivariate marginal distribution of $(\langle \mathbf{Y}, \mathbf{e}_1 \rangle, \langle \mathbf{Y}, \mathbf{e}_2 \rangle, \cdots)$ given $\mathbf{X} = \mathbf{z}$ has a density that is uniformly bounded on bounded subsets for $\mathbf{z}$ satisfying $d(\mathbf{x}, \mathbf{z}) \le C_3$. This requirement holds in all the cases where $\mathbf{Y}$ is conditionally a Gaussian process given $\mathbf{X}$ and its conditional mean and conditional covariance operator are continuous at $\mathbf{X} = \mathbf{x}$.

Define
$$g_n(\mathbf{Q} \,|\, \mathbf{z}) = E[\|\mathbf{Q} - \mathbf{Y}^{(n)}\| - \|\mathbf{Y}^{(n)}\| \,|\, \mathbf{X} = \mathbf{z}] - \langle \boldsymbol{\tau}^{(n)}, \mathbf{Q} \rangle,$$

and
$$\tilde{g}_n(\mathbf{Q} \,|\, \mathbf{x}) = E[(\|\mathbf{Q} - \mathbf{Y}^{(n)}\| - \|\mathbf{Y}^{(n)}\| - \langle \boldsymbol{\tau}^{(n)}, \mathbf{Q} \rangle) E_n^{-1} K(h_n^{-1} d(\mathbf{x}, \mathbf{X}))].$$

For $\mathbf{Q} \in \mathcal{H}$, define the Fréchet derivatives $g^{(1)}(\mathbf{Q} \,|\, \mathbf{z})$, $g_n^{(1)}(\mathbf{Q} \,|\, \mathbf{z})$, $\tilde{g}_n^{(1)}(\mathbf{Q} \,|\, \mathbf{x})$, $\widehat{g}_n^{(1)}(\mathbf{Q} \,|\, \mathbf{x})$ by

$$g^{(1)}(\mathbf{Q} \,|\, \mathbf{z}) = \frac{\partial}{\partial \mathbf{Q}} g(\mathbf{Q} \,|\, \mathbf{z}) = E[\|\mathbf{Q} - \mathbf{Y}\|^{-1}(\mathbf{Q} - \mathbf{Y}) \,|\, \mathbf{X} = \mathbf{z}] - \boldsymbol{\tau},$$

$$g_n^{(1)}(\mathbf{Q} \,|\, \mathbf{z}) = \frac{\partial}{\partial \mathbf{Q}} g_n(\mathbf{Q} \,|\, \mathbf{z}) = E[\|\mathbf{Q} - \mathbf{Y}^{(n)}\|^{-1}(\mathbf{Q} - \mathbf{Y}^{(n)}) \,|\, \mathbf{X} = \mathbf{z}] - \boldsymbol{\tau}^{(n)},$$

$$\tilde{g}_n^{(1)}(\mathbf{Q} \,|\, \mathbf{x}) = \frac{\partial}{\partial \mathbf{Q}} \tilde{g}_n(\mathbf{Q} \,|\, \mathbf{x}) = E[(\|\mathbf{Q} - \mathbf{Y}^{(n)}\|^{-1}(\mathbf{Q} - \mathbf{Y}^{(n)}) - \boldsymbol{\tau}^{(n)}) E_n^{-1} K(h_n^{-1} d(\mathbf{x}, \mathbf{X}))],$$

$$\widehat{g}_n^{(1)}(\mathbf{Q} \,|\, \mathbf{x}) = \frac{\partial}{\partial \mathbf{Q}} \widehat{g}_n(\mathbf{Q} \,|\, \mathbf{x}) = \frac{\sum_{i=1}^n (\|\mathbf{Q} - \mathbf{Y}_i^{(n)}\|^{-1}(\mathbf{Q} - \mathbf{Y}_i^{(n)})) K(h_n^{-1} d(\mathbf{x}, \mathbf{X}_i))}{\sum_{i=1}^n K(h_n^{-1} d(\mathbf{x}, \mathbf{X}_i))} - \boldsymbol{\tau}^{(n)}.$$

Now, for each $\mathbf{Q} \in \mathcal{H}$, define $g^{(2)}(\mathbf{Q} \,|\, \mathbf{z})(\cdot)$, $g_n^{(2)}(\mathbf{Q} \,|\, \mathbf{z})(\cdot)$, $\tilde{g}_n^{(2)}(\mathbf{Q} \,|\, \mathbf{x})(\cdot) : \mathcal{H} \to \mathcal{H}$ by

$$(g^{(2)}(\mathbf{Q} \,|\, \mathbf{z}))(\mathbf{h}) = \Big(\frac{\partial^2}{\partial Q^2} g(\mathbf{Q} \,|\, \mathbf{z})\Big)(\mathbf{h}) = E\left[\frac{\mathbf{h}}{\|\mathbf{Q} - \mathbf{Y}\|} - \frac{\langle \mathbf{h}, \mathbf{Q} - \mathbf{Y} \rangle (\mathbf{Q} - \mathbf{Y})}{\|\mathbf{Q} - \mathbf{Y}\|^3} \,\Big|\, \mathbf{X} = \mathbf{z}\right],$$

$$(g_n^{(2)}(\mathbf{Q} \,|\, \mathbf{z}))(\mathbf{h}) = \Big(\frac{\partial^2}{\partial Q^2} g_n(\mathbf{Q} \,|\, \mathbf{z})\Big)(\mathbf{h})$$
$$= E\left[\frac{\mathbf{h}}{\|\mathbf{Q} - \mathbf{Y}^{(n)}\|} - \frac{\langle \mathbf{h}, \mathbf{Q} - \mathbf{Y}^{(n)} \rangle (\mathbf{Q} - \mathbf{Y}^{(n)})}{\|\mathbf{Q} - \mathbf{Y}^{(n)}\|^3} \,\Big|\, \mathbf{X} = \mathbf{z}\right],$$

$$(\tilde{g}_n^{(2)}(\mathbf{Q} \,|\, \mathbf{x}))(\mathbf{h}) = \Big(\frac{\partial^2}{\partial Q^2} \tilde{g}_n(\mathbf{Q} \,|\, \mathbf{z})\Big)(\mathbf{h})$$
$$= E\left[\left[\frac{h}{\|\mathbf{Q} - \mathbf{Y}^{(n)}\|} - \frac{\langle \mathbf{h}, \mathbf{Q} - \mathbf{Y}^{(n)} \rangle (\mathbf{Q} - \mathbf{Y}^{(n)})}{\|\mathbf{Q} - \mathbf{Y}^{(n)}\|^3}\right] \frac{K(h_n^{-1} d(\mathbf{x}, \mathbf{X}))}{E_n}\right].$$

From assumptions B-1 and B-2, it follows that for $\mathbf{z}$ lying in a neighborhood of $\mathbf{x}$ and for all sufficiently large $n$, $g_n(\mathbf{Q} \,|\, \mathbf{z})$ and $\tilde{g}_n(\mathbf{Q} \,|\, \mathbf{x})$ have unique minimizers $\mathbf{Q}_n(\boldsymbol{\tau} \,|\, \mathbf{z})$ and $\tilde{\mathbf{Q}}_n(\boldsymbol{\tau} \,|\, \mathbf{x})$ respectively in $\mathcal{Z}_n$, and

$$g_n^{(1)}(\mathbf{Q}_n(\boldsymbol{\tau} \,|\, \mathbf{z}) \,|\, \mathbf{z}) = \tilde{g}_n^{(1)}(\tilde{\mathbf{Q}}_n(\boldsymbol{\tau} \,|\, \mathbf{x}) \,|\, \mathbf{x}) = 0.$$

From assumptions B-1, B-2 and B-3, we get that for all sufficiently large $n$ and for $\mathbf{z}$ lying in a neighborhood of $\mathbf{x}$, $g^{(2)}(\mathbf{Q} \,|\, \mathbf{z})(\cdot)$, $g_n^{(2)}(\mathbf{Q} \,|\, \mathbf{z})(\cdot)$ and $\tilde{g}_n^{(2)}(\mathbf{Q} \,|\, \mathbf{x})(\cdot)$ are continuous linear operators on $\mathcal{H}$ for all $\mathbf{Q}$. We now state a Bahadur type asymptotic linearization of $\widehat{\mathbf{Q}}_n(\boldsymbol{\tau} \,|\, \mathbf{x})$.

**Theorem 4.2.** *Let the assumptions B-1 through B-3 hold, and $(n\phi(h_n\,|\,\mathbf{x}))^{-(1-2\alpha)}d_n \to c_5 > 0$ as $n \to \infty$, where $\alpha$ is as described in assumption B-1. Then,*

$$\widehat{\mathbf{Q}}_n(\boldsymbol{\tau}\,|\,\mathbf{x}) - \tilde{\mathbf{Q}}_n(\boldsymbol{\tau}\,|\,\mathbf{x})$$
$$= -[\tilde{g}_n^{(2)}(\tilde{\mathbf{Q}}_n(\boldsymbol{\tau}\,|\,\mathbf{x})\,|\,\mathbf{x})]^{-1}\left[\frac{n^{-1}\sum_{i=1}^n\left[\frac{\tilde{\mathbf{Q}}_n(\boldsymbol{\tau}\,|\,\mathbf{x})-\mathbf{Y}_i^{(n)}}{\|\tilde{\mathbf{Q}}_n(\boldsymbol{\tau}\,|\,\mathbf{x})-\mathbf{Y}_i^{(n)}\|} - \boldsymbol{\tau}^{(n)}\right]E_n^{-1}K(h_n^{-1}d(\mathbf{x},\mathbf{X}_i))}{n^{-1}\sum_{i=1}^n E_n^{-1}K(h_n^{-1}d(\mathbf{x},\mathbf{X}_i))}\right]$$
$$+ R_n(\mathbf{x}),$$

*where $\|R_n(\mathbf{x})\| = O(\epsilon_n^2)$ as $n \to \infty$ almost surely, and $\epsilon_n = (n\phi(h_n\,|\,\mathbf{x}))^{-\alpha}\sqrt{\log n}$.*

Define $\mathbf{B}_n(\boldsymbol{\tau}\,|\,\mathbf{x}) = \tilde{\mathbf{Q}}_n(\boldsymbol{\tau}\,|\,\mathbf{x}) - \mathbf{Q}_n(\boldsymbol{\tau}\,|\,\mathbf{x})$. We view it as a kind of bias in the estimate $\widehat{\mathbf{Q}}_n(\boldsymbol{\tau}\,|\,\mathbf{x})$. Then, one can show that under the assumptions of Theorem 4.2, $\|\mathbf{B}_n(\boldsymbol{\tau}\,|\,\mathbf{x})\| \to 0$ as $n \to \infty$. In addition, suppose that there exist a constant $C_4 > 0$ and a positive integer $N_3$ such that whenever $d(\mathbf{x},\mathbf{z}) \leq C_4$ and $n \geq N_3$, we have, for each $C > 0$,

$$(d(\mathbf{x},\mathbf{z}))^{-1}\|g_n^{(1)}(\mathbf{Q}\,|\,\mathbf{z}) - g_n^{(1)}(\mathbf{Q}\,|\,\mathbf{x})\| \leq s_3(C)$$

for all $\mathbf{Q} \in \mathcal{Z}_n$ with $\|\mathbf{Q}\| \leq C$, where $s_3(C)$ is a positive constant depending on $C$. Then, it can be shown that $\|\mathbf{B}_n(\boldsymbol{\tau}\,|\,\mathbf{x})\| = O(h_n)$ as $n \to \infty$ (see Proposition B.1). It is easy to show that all these hold when the response $\mathbf{Y}$ satisfies a location-scale model as discussed in connection with assumption A-2 earlier.

We can show that $\|\mathbf{Q}_n(\boldsymbol{\tau}\,|\,\mathbf{x}) - \mathbf{Q}(\boldsymbol{\tau}\,|\,\mathbf{x})\| \to 0$ as $n \to \infty$ (see Lemma 2.4 in [17]). As a consequence, we get $\|\widehat{\mathbf{Q}}_n(\boldsymbol{\tau}\,|\,\mathbf{x}) - \mathbf{Q}(\boldsymbol{\tau}\,|\,\mathbf{x})\| \to 0$ as $n \to \infty$ *almost surely* for each $\boldsymbol{\tau}$ with $\|\boldsymbol{\tau}\| < 1$. Recall that we defined an estimate of the conditional spread as $\widehat{D}_2(\boldsymbol{\tau}\,|\,\mathbf{x}) = \|\widehat{\mathbf{Q}}_n(\boldsymbol{\tau}\,|\,\mathbf{x}) - \widehat{\mathbf{Q}}_n(-\boldsymbol{\tau}\,|\,\mathbf{x})\|$. Hence, we get $\widehat{D}_2(\boldsymbol{\tau}\,|\,\mathbf{x}) \to D_2(\boldsymbol{\tau}\,|\,\mathbf{x})$ as $n \to \infty$ *almost surely*.

We next state a result on the asymptotic normality of the conditional sample spatial quantile. Recall $\gamma(\mathbf{y}\,|\,\mathbf{z})(\cdot,\cdot)$ defined before Theorem 4.1. Let $\gamma_0(\mathbf{Q}\,|\,\mathbf{x})(\cdot) : \mathcal{H} \to \mathcal{H}$ be the continuous linear operator obtained from $\gamma(\mathbf{Q}\,|\,\mathbf{x})(\cdot,\cdot)$, i.e., $\langle \gamma_0(\mathbf{Q}\,|\,\mathbf{x})(\mathbf{v}),\mathbf{w}\rangle = \gamma(\mathbf{Q}\,|\,\mathbf{x})(\mathbf{v},\mathbf{w})$ for all $\mathbf{v},\mathbf{w} \in \mathcal{H}$.

**Theorem 4.3.** *Suppose that $\mathbf{Y}$ has the conditional Karhunen-Loeve expansion given by $\mathbf{Y} = \mathbf{m}(\mathbf{x}) + \sum_{k=1}^{\infty}\sqrt{\lambda_k}Z_k\boldsymbol{\psi}_k$, where the $Z_k$'s are conditionally uncorrelated random variables with conditional mean 0 and conditional variance 1 given $\mathbf{X} = \mathbf{x}$, and the $\lambda_k$'s and the $\boldsymbol{\psi}_k$'s are the eigenvalues and the corresponding eigenfunctions of the conditional covariance operator of $\mathbf{Y}$ given $\mathbf{X} = \mathbf{x}$. Also, let $\sqrt{n\phi(h_n\,|\,\mathbf{x})}\|\boldsymbol{\tau} - \boldsymbol{\tau}^{(n)}\| \to 0$, $\sqrt{n\phi(h_n\,|\,\mathbf{x})}\|\mathbf{m}(\mathbf{x}) - (\mathbf{m}(\mathbf{x}))^{(n)}\| \to 0$ and $(n\phi(h_n\,|\,\mathbf{x})\sum_{k>d_n}\lambda_k) \to 0$ as $n \to \infty$. Suppose the assumptions B-1 through B-3 hold, and $(n\phi(h_n\,|\,\mathbf{x}))^{-(1-2\alpha)}d_n \to c_5 > 0$ as $n \to \infty$, where $\alpha$ is as in assumption B-1. Then,*

$$\sqrt{n\phi(h_n\,|\,\mathbf{x})}\bigl[\widehat{\mathbf{Q}}_n(\boldsymbol{\tau}\,|\,\mathbf{x}) - \mathbf{Q}(\boldsymbol{\tau}\,|\,\mathbf{x}) - \mathbf{B}_n(\boldsymbol{\tau}\,|\,\mathbf{x})\bigr] \to \mathbf{W}$$

in distribution as $n \to \infty$, where $\mathbf{W}$ is a Gaussian random element in $\mathcal{H}$ with mean $\mathbf{0}$ and covariance operator

$$\mathbf{\Sigma}(\mathbf{x}) = \frac{E_{(2)}(\mathbf{x})}{(E_{(1)}(\mathbf{x}))^2} \left[ g^{(2)}(\mathbf{Q}(\boldsymbol{\tau} \,|\, \mathbf{x}) \,|\, \mathbf{x}) \right]^{-1} \gamma_0(\mathbf{Q}(\boldsymbol{\tau} \,|\, \mathbf{x}) \,|\, \mathbf{x}) \left[ g^{(2)}(\mathbf{Q}(\boldsymbol{\tau} \,|\, \mathbf{x}) \,|\, \mathbf{x}) \right]^{-1}.$$

The assumption concerning the conditional Karhunen-Loeve expansion of $\mathbf{Y}$ and the conditions $\sqrt{n\phi(h_n \,|\, \mathbf{x})} \|\boldsymbol{\tau} - \boldsymbol{\tau}^{(n)}\| \to 0$, $\sqrt{n\phi(h_n \,|\, \mathbf{x})} \|\mathbf{m}(\mathbf{x}) - (\mathbf{m}(\mathbf{x}))^{(n)}\| \to 0$ and $(n\phi(h_n \,|\, \mathbf{x}) \sum_{k>d_n} \lambda_k) \to 0$ as $n \to \infty$ are required to ensure $\sqrt{n\phi(h_n \,|\, \mathbf{x})} \|\mathbf{Q}_n(\boldsymbol{\tau} \,|\, \mathbf{x}) - \mathbf{Q}(\boldsymbol{\tau} \,|\, \mathbf{x})\| \to 0$ as $n \to \infty$. Note that $\mathbf{Q}_n(\boldsymbol{\tau} \,|\, \mathbf{x}) \in \mathcal{Z}_n$ can be viewed as a finite dimensional approximation of $\mathbf{Q}(\boldsymbol{\tau} \,|\, \mathbf{x})$, and these conditions are necessary to control the asymptotic bias arising from such an approximation. For further insights into these conditions, readers are referred to [8], where the authors used similar assumptions in their Theorem 3.4 to derive asymptotic normality of unconditional spatial quantiles in Hilbert spaces. Note that when the response space $\mathcal{H}$ is finite dimensional, we can take $\alpha = 1/2$, $\mathcal{Z}_n = \mathcal{H}$ and $d_n = dimension(\mathcal{H})$ for all $n$. Then, there is no such bias, and our theorems yield the Bahadur representation and the asymptotic normality for conditional spatial quantiles of a finite dimensional response as a special case.

Suppose that the covariate $\mathbf{X}$ is either finite dimensional with a positive density at $\mathbf{x}$ or a fractal-type process. Then, like in the cases of the conditional spatial distribution and depth, an appropriate choice of $h_n$, which balances the asymptotic order of the bias term $\mathbf{B}_n(\boldsymbol{\tau} \,|\, \mathbf{x})$ and the asymptotic order of the variance of the estimate $\widehat{\mathbf{Q}}_n(\boldsymbol{\tau} \,|\, \mathbf{x})$, is $c_3 n^{-(d+2)^{-1}}$ for some $d > 0$. The optimum rate of convergence of $\widehat{\mathbf{Q}}_n(\boldsymbol{\tau} \,|\, \mathbf{x})$ is $O_P\bigl(n^{-(d+2)^{-1}}\bigr)$ in that case. If $\mathbf{X}$ is a continuous Gaussian Markov process on an interval equipped with the $L_2$-norm, there is no choice of $h_n$, which simultaneously satisfies B-1 and $\sqrt{n\phi(h_n \,|\, \mathbf{x})} h_n = O(1)$ as $n \to \infty$ as observed in the discussion following Theorem 4.1. However, if one chooses $h_n = c_4([\sqrt{\log n}]^{-1})$ with an appropriate constant $c_4 > 0$ in a similar way as done before, then assumption B-1 is satisfied, and the rate of convergence of $\widehat{\mathbf{Q}}_n(\boldsymbol{\tau} \,|\, \mathbf{x})$ becomes $O_P([\sqrt{\log n}]^{-1})$ as $n \to \infty$. The optimality of this convergence rate can be established under appropriate conditions, using arguments similar to those used in [16], where the optimum rates for the kernel-based mean regression estimate are derived. And it can be verified that this optimal rate is same as the optimal rate of convergence of the kernel mean regression estimate in the same setup. Once again, we see that the convergence rates do not depend on the dimension of the response.

Theorem 4.3 can be utilized to construct confidence sets for the conditional quantile $\mathbf{Q}(\boldsymbol{\tau} \,|\, \mathbf{x})$ when the sequence of bandwidths $\{h_n\}$ satisfies $\sqrt{n\phi(h_n \,|\, \mathbf{x})} h_n \to 0$ as $n \to \infty$ in addition to the conditions in Theorem 4.3. Then, we get that $\sqrt{n\phi(h_n \,|\, \mathbf{x})}\bigl[\widehat{\mathbf{Q}}_n(\boldsymbol{\tau} \,|\, \mathbf{x}) - \mathbf{Q}(\boldsymbol{\tau} \,|\, \mathbf{x})\bigr] \to \mathbf{W}$ *in distribution* as $n \to \infty$. Let $0 < r < 1$, and let $K_r$ be a set in $\mathcal{H}$ such that $P[\mathbf{W} \in K_r] = 1 - r$. Then, $P[\mathbf{Q}(\boldsymbol{\tau} \,|\, \mathbf{x}) \in \widehat{\mathbf{Q}}_n(\boldsymbol{\tau} \,|\, \mathbf{x}) + K_r] \to 1 - r$ as $n \to \infty$ so that $\widehat{\mathbf{Q}}_n(\boldsymbol{\tau} \,|\, \mathbf{x}) + K_r$ is asymptotically a $100(1-r)\%$ confidence set for $\mathbf{Q}(\boldsymbol{\tau} \,|\, \mathbf{x})$. We next discuss two ways of constructing $K_r$, which will be closed and convex.

In the first construction, we take $K_r$ to be a closed ball $C(\mathbf{0}, c_r)$ in $\mathcal{H}$ centered at $\mathbf{0} \in \mathcal{H}$ with radius $c_r$ such that $P[\|\mathbf{W}\|^2 \leq c_r^2] = 1 - r$. Using the Kerhunen-Loeve expansion, it is easy to see that $\|\mathbf{W}\|^2 = \sum_{k=1}^{\infty} \zeta_k \chi_k$, where $\{\zeta_k\}$ is the decreasing sequence of eigenvalues of $\mathbf{\Sigma}(\mathbf{x})$, and $\{\chi_k\}$ is a sequence of independent random variables identically distributed as a Chi-square distribution with degree of freedom 1. In practice, we can estimate the sequence $\{\zeta_k\}$ by the decreasing sequence of eigenvalues $\{\widehat{\zeta}_{k,n}\}$ of an estimate of the covariance operator $\mathbf{\Sigma}(\mathbf{x})$. $c_r^2$ can be estimated by the $(1-r)$-quantile of the real valued random variable $\sum_{k=1}^{\infty} \widehat{\zeta}_{k,n} \chi_k$ using the Monte-Carlo method. However, the drawback of this confidence set is that the ball $C(\mathbf{0}, c_r)$ can hardly be visualized in a plot when the response values are random functions in an $L_2$ space.

We present a second construction, where the confidence sets can be conveniently plotted. Again from the Kerhunen-Loeve expansion, we get that $\mathbf{W} = \sum_{k=1}^{\infty} \sqrt{\zeta_k} Z_k \boldsymbol{\xi}_k$, where $\{\boldsymbol{\xi}_k\}$ is the sequence of eigenvectors of $\mathbf{\Sigma}(\mathbf{x})$ corresponding to the decreasing sequence of eigenvalues $\{\zeta_k\}$, and $\{Z_k\}$ is a sequence of independent standard normal random variables. Let $z(\beta)$ denote the $\beta$-quantile of the standard normal distribution, where $0 < \beta < 1$. Let $l_k = z\big((1-(1-r)^{2^{-k}})/2\big)$ and $u_k = z\big(1-(1-(1-r)^{2^{-k}})/2\big)$ for $k = 1, 2, \cdots$. So, $P[l_k \leq Z_k \leq u_k] = (1-r)^{2^{-k}}$ for all $k$, and hence

$$P[l_k \leq Z_k \leq u_k \text{ for all } k] = \prod_{k=1}^{\infty} (1-r)^{2^{-k}} = 1 - r.$$

The set $K_r$ can be constructed as

$$K_r = \left\{ \mathbf{v} = \sum_{k=1}^{\infty} v_k \boldsymbol{\xi}_k \in \mathcal{H} \,\middle|\, l_k \sqrt{\zeta_k} \leq v_k \leq u_k \sqrt{\zeta_k} \text{ for all } k \right\}.$$

Once again, we can substitute $\zeta_k$'s by their estimates obtained from an appropriate estimate of the covariance operator $\mathbf{\Sigma}(\mathbf{x})$. We plot this confidence set for the estimated conditional spatial median for simulated and real data examples in [17].

Both the constructions of the confidence sets for $\mathbf{Q}(\boldsymbol{\tau} \,|\, \mathbf{x})$ require an estimate $\widehat{\mathbf{\Sigma}}_n(\mathbf{x})$ of the covariance operator $\mathbf{\Sigma}(\mathbf{x})$ of $\mathbf{W}$. We can estimate $\mathbf{\Sigma}(\mathbf{x})$ by its sample counter-part

$$\widehat{\mathbf{\Sigma}}_n(\mathbf{x}) = \frac{E_{(2),n}(\mathbf{x})}{(E_{(1),n}(\mathbf{x}))^2} \left[\widehat{g}_n^{(2)}(\widehat{\mathbf{Q}}_n(\boldsymbol{\tau} \,|\, \mathbf{x}) \,|\, \mathbf{x})\right]^{-1} \widehat{\gamma}_n(\widehat{\mathbf{Q}}_n(\boldsymbol{\tau} \,|\, \mathbf{x}) \,|\, \mathbf{x}) \left[\widehat{g}_n^{(2)}(\widehat{\mathbf{Q}}_n(\boldsymbol{\tau} \,|\, \mathbf{x}) \,|\, \mathbf{x})\right]^{-1},$$

where

$$E_{(i),n}(\mathbf{x}) = \frac{\sum_{i=1}^n K^j(h_n^{-1} d(\mathbf{x}, \mathbf{X}_i))}{\sum_{i=1}^n I(d(\mathbf{x}, \mathbf{X}_i) \leq h_n)} \text{ for } i = 1, 2,$$

$\widehat{g}_n^{(2)}(\mathbf{Q} \,|\, \mathbf{x})(\cdot) : \mathcal{H} \to \mathcal{H}$ is defined by

$$\widehat{g}_n^{(2)}(\mathbf{Q} \,|\, \mathbf{x})(\mathbf{v}) = \frac{\sum_{i=1}^n \left[\frac{\mathbf{v}}{\|\mathbf{Q}-\mathbf{Y}_i^{(n)}\|} - \frac{\langle \mathbf{v}, \mathbf{Q}-\mathbf{Y}_i^{(n)}\rangle(\mathbf{Q}-\mathbf{Y}_i^{(n)})}{\|\mathbf{Q}-\mathbf{Y}_i^{(n)}\|^3}\right] K(h_n^{-1} d(\mathbf{x}, \mathbf{X}_i))}{\sum_{i=1}^n K(h_n^{-1} d(\mathbf{x}, \mathbf{X}_i))},$$

and $\widehat{\gamma}_n(\mathbf{Q} \,|\, \mathbf{x})(\cdot) : \mathcal{H} \to \mathcal{H}$ is the sample covariance operator defined by

$$
\begin{aligned}
&\widehat{\gamma}_n(\mathbf{Q} \,|\, \mathbf{x})(\mathbf{v}) \\
&= \frac{\sum_{i=1}^n \left\langle \frac{\mathbf{Q}-\mathbf{Y}_i^{(n)}}{\|\mathbf{Q}-\mathbf{Y}_i^{(n)}\|}, \mathbf{v} \right\rangle \frac{\mathbf{Q}-\mathbf{Y}_i^{(n)}}{\|\mathbf{Q}-\mathbf{Y}_i^{(n)}\|} K(h_n^{-1} d(\mathbf{x}, \mathbf{X}_i))}{\sum_{i=1}^n K(h_n^{-1} d(\mathbf{x}, \mathbf{X}_i))} \\
&\quad - \left[ \frac{\sum_{i=1}^n \left\langle \frac{\mathbf{Q}-\mathbf{Y}_i^{(n)}}{\|\mathbf{Q}-\mathbf{Y}_i^{(n)}\|}, \mathbf{v} \right\rangle K(h_n^{-1} d(\mathbf{x}, \mathbf{X}_i))}{\sum_{i=1}^n K(h_n^{-1} d(\mathbf{x}, \mathbf{X}_i))} \cdot \frac{\sum_{i=1}^n \frac{\mathbf{Q}-\mathbf{Y}_i^{(n)}}{\|\mathbf{Q}-\mathbf{Y}_i^{(n)}\|} K(h_n^{-1} d(\mathbf{x}, \mathbf{X}_i))}{\sum_{i=1}^n K(h_n^{-1} d(\mathbf{x}, \mathbf{X}_i))} \right].
\end{aligned}
$$

As stated in the next theorem, $\widehat{\boldsymbol{\Sigma}}_n(\mathbf{x})$ is a consistent estimate of $\boldsymbol{\Sigma}(\mathbf{x})$.

**Theorem 4.4.** *Under the conditions in Theorem 4.3, $\widehat{\boldsymbol{\Sigma}}_n(\mathbf{x}) \to \boldsymbol{\Sigma}(\mathbf{x})$ in probability as $n \to \infty$.*

## 5. Data Analysis

In this section, quantile regression and conditional maximal depth sets are demonstrated using simulated and real data. We consider three datasets here. The first one is a simulated data generated from a heteroscedastic model. The second dataset is a real data concerning per capita GDP and saving rate in 125 countries over a time period of 26 years. The third dataset is about cigarette sales and net disposable income in 46 states in the USA over a 30-year period.

In all our analysis, we consider the functional response and the functional covariate as random elements in appropriate $L_2$ spaces. For the sake of simplicity, we choose the indicator function on $[0,1]$ as the kernel $K(\cdot)$, and the bandwidth $h$ is chosen by leave-one-out cross validation in the following way. Let $m_n^{(-i)}(\mathbf{z}, h)$ denote the conditional spatial median estimated at $\mathbf{X} = \mathbf{z}$ using the bandwidth $h$ and leaving out the $i$-th sample observation $(\mathbf{X}_i, \mathbf{Y}_i)$. We choose the bandwidth $h_{opt}$ such that $h_{opt} = \arg\min_h n^{-1} \sum_{i=1}^n \|m_n^{(-i)}(\mathbf{X}_i, h) - \mathbf{Y}_i\|$. For a given bandwidth $h$, let $C_n(\mathbf{x}, h) = \{\mathbf{Y}_i \,|\, d(\mathbf{x}, \mathbf{X}_i) \leq h\}$ ($d(\cdot, \cdot)$ is the $L_2$ metric here), and $\#[C_n(\mathbf{x}, h)]$ be the cardinality of $C_n(\mathbf{x}, h)$. While computing $\widehat{\mathbf{Q}}_n(\boldsymbol{\tau} \,|\, \mathbf{x})$ taking bandwidth $h$, we take

$$d_n = \left\lfloor \min\left\{ \sqrt{\#[C_n(\mathbf{x}, h)]},\ 2[\#[C_n(\mathbf{x}, h)]]^{(1/3)} \right\} \right\rfloor,$$

where $\lfloor r \rfloor$ denotes the largest integer less than or equal to $r$. Then, $\alpha = (1/3)$ and $c_5 = 2$, where $\alpha$ and $c_5$ are as in Theorem 4.2. To fix the basis $\{\mathbf{e}_1, \mathbf{e}_2, \cdots, \mathbf{e}_{d_n}\}$, we first estimate the conditional covariance operator of $\mathbf{Y}$ given $\mathbf{X} = \mathbf{x}$ using the same kernel function $K(\cdot)$ and the bandwidth $h$. Then, the eigenfunction corresponding to the $k$-th largest eigenvalue of the estimated conditional covariance operator is taken as $\mathbf{e}_k$ for $k = 1, \cdots, d_n$. The cross-validated bandwidth $h_{opt}$ is used for all further computations of quantiles, maximal depth sets and spread measures.

## 5.1. Simulated Data

We consider a regression model, where the covariate $\mathbf{X}(t) = U exp(t)$ for $t \in [0, 1]$ with $U \sim Uniform[0, 1]$, and the response $\mathbf{Y}(t) = \|\mathbf{X}\|\mathbf{B}(t)$, where $\mathbf{B}(t)$ is the standard Brownian Motion on $[0, 1]$, and $\|\cdot\|$ is the $L_2$ norm. We simulate 100 observations from this model, and construct the quantiles regression estimates and conditional maximal depth sets. The sample size is 100.

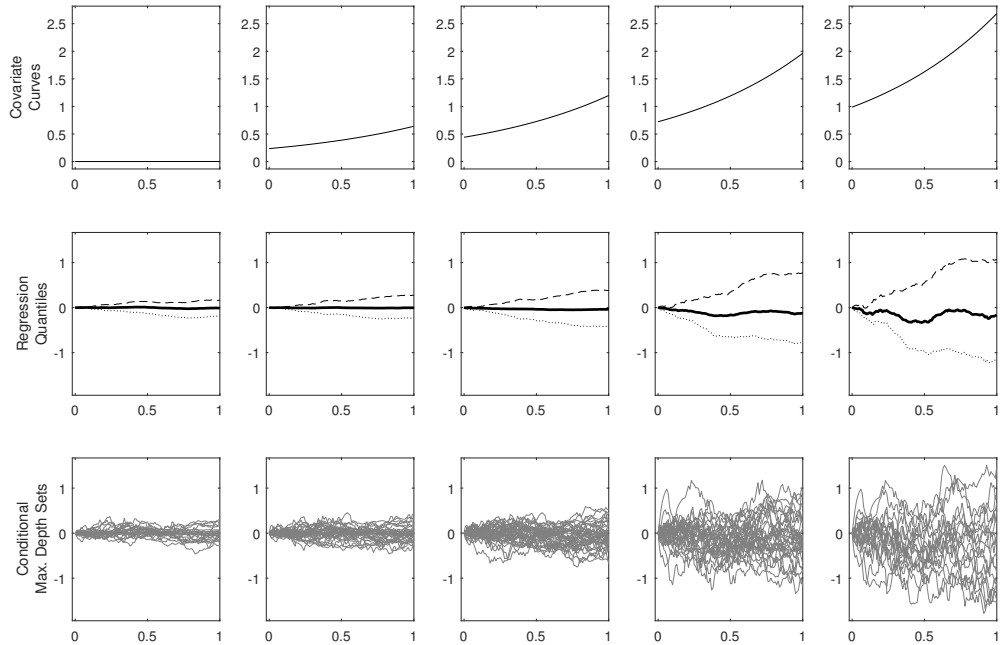

**Figure 2**. Plots of the selected covariate curves (1$^{\text{st}}$ row), the corresponding conditional spatial quantiles (2$^{\text{nd}}$ row) and conditional maximal depth sets (3$^{\text{rd}}$ row) for the simulated data. The dotted, the dashed and the solid curves in the 2$^{\text{nd}}$ row are $\widehat{\mathbf{Q}}_n(-0.5\mathbf{u} \,|\, \mathbf{x})$, $\widehat{\mathbf{Q}}_n(0.5\mathbf{u} \,|\, \mathbf{x})$ and $\widehat{\mathbf{Q}}_n(\mathbf{0} \,|\, \mathbf{x})$, respectively.

The value of the bandwidth $h$ obtained through cross validation is 0.68. We compute the 50% conditional maximal depth sets, the conditional spatial median and conditional spatial quantiles corresponding to $\boldsymbol{\tau} = \pm 0.5\mathbf{u}$, where $\mathbf{u}$ is the first principal component of the estimated conditional covariance operator of the response. To demonstrate the conditional quantile curves and depth sets, we order the covariate curves by their $L_2$ norms, and choose 5 covariate curves whose ranks are equidistant in this ordering. The conditional spatial quantiles and the conditional maximal depth sets for these 5 covariate curves are plotted in Figure 2. We see that the spatial quantiles and the maximal depth sets clearly capture the change of the conditional distributions of the response given the selected covariate curves and the heteroscedasticity present in the data. The conditional spread measures $\widehat{D}_1(p\,|\,\mathbf{x})$ and $\widehat{D}_2(\boldsymbol{\tau}\,|\,\mathbf{x})$ for all the covariate curves are plotted in Figure 3 against the ranks of all the covariate curves in the ordering by their $L_2$ norms. The heteroscedasticity of the model is also evident from these plots.

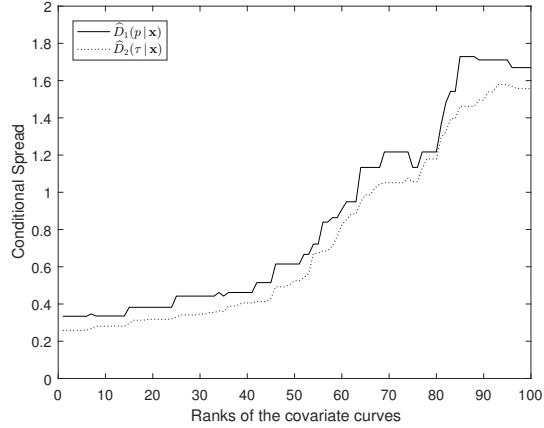

**Figure 3**. Plots of $\widehat{D}_1(p\,|\,\mathbf{x})$ and $\widehat{D}_2(\boldsymbol{\tau}\,|\,\mathbf{x})$ against the ranks of the covariate curves in the ordering by their $L_2$ norms in the heteroscedastic model.

## 5.2. The Penn Table Data

We consider now a real data, which is heteroscedastic in nature. The Penn Table dataset is a panel of 125 observations for the period 1960–1985. It includes real GDP per capita (in 1985 dollars) and saving rate (in percent) of 125 countries for those 26 years. This dataset is available in the R package 'Ecdat' (named as 'SumHes'). We take the saving rate curve as the response and the curve of per capita GDP as covariate and investigate the effect of the covariate on the response. Both the response and the covariate, which are functions of time, are considered as random elements in $L_2[1960, 1985]$.

The curve of per capita GDP indicates the productivity of an average citizen of the nation. The bandwidth $h$ obtained through cross validation is 9565.71. The resulting conditional quantiles and maximal depth sets are plotted in Figure 4 for 5 selected covariate curves.

We note that saving rate increases gradually with the rise in per capita GDP. We notice that there is a decreasing trend in the saving rate after 1980 in all the plots. This trend is more noticeable in middle and higher GDP levels as indicated by the plots in the $4^{\text{th}}$ and the $5^{\text{th}}$ columns of Figure 4. The plots in the $5^{\text{th}}$ column correspond to the countries with very high levels of per capita GDP, and the saving rates start decreasing around 1970 in those countries. This indicates an increase in consumption only after 1980 in all the countries except those with very high levels of per capita GDP. The consumption in countries with very high per capita GDP started increasing even earlier, around 1970.

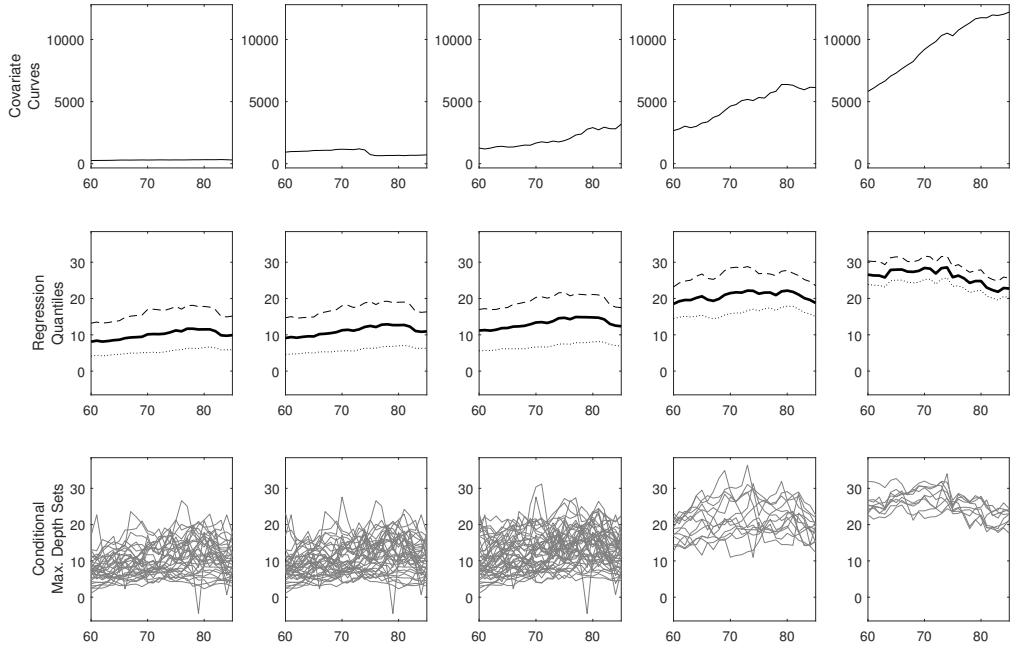

**Figure 4**. Plots of the selected covariate curves (1$^{\text{st}}$ row), the corresponding conditional spatial quantiles (2$^{\text{nd}}$ row) and conditional maximal depth sets (3$^{\text{rd}}$ row) for the Penn Table data. The dotted, the dashed and the solid curves in the 2$^{\text{nd}}$ row are $\widehat{\mathbf{Q}}_n(-0.5\mathbf{u}\,|\,\mathbf{x})$, $\widehat{\mathbf{Q}}_n(0.5\mathbf{u}\,|\,\mathbf{x})$ and $\widehat{\mathbf{Q}}_n(\mathbf{0}\,|\,\mathbf{x})$, respectively.

From the plots of $\widehat{D}_1(p\,|\,\mathbf{x})$ and $\widehat{D}_2(\boldsymbol{\tau}\,|\,\mathbf{x})$ in Figure 5, with the covariate curves arranged by their $L_2$ norms, we notice that the data is heteroscedastic in nature. This observation is also supported by the shrinking difference between the two chosen spatial quantiles and the upper and lower boundaries of the maximal depth sets corresponding to the 4$^{\text{th}}$ and the 5$^{\text{th}}$ chosen covariate curves in Figure 4.

## 5.3. The Cigar Data

Our third example deals with the Cigar Data, which is a panel of 46 observations containing information about the sale of cigarettes in 46 states of the USA for the period from 1963 to 1992. This dataset is available in the 'plm' and the 'Ecdat' packages in R and was analyzed earlier by [1]. In addition to information on sales of cigarettes, it also includes per capita net disposable income (NDI) for those 46 states over the 30 year period. We consider the curve of per capita NDI over time as the covariate and the curve of cigarette sales in packs per capita over time as the response. Both the response and the covariate curves are functions of time over a 30 year period from 1963 to 1992, and we view them as random elements in $L_2[1963, 1992]$. Choosing the covariate and the response this way, we can investigate the effect of income over consumption of cigarettes in different states. The cross validated choice of the bandwidth is 10061.27. The covariate curves are arranged by their $L_2$ norms, and the corresponding conditional quantile curves and the maximal depth sets are plotted in Figure 6 for 5 selected covariate curves corresponding to 5 different states.

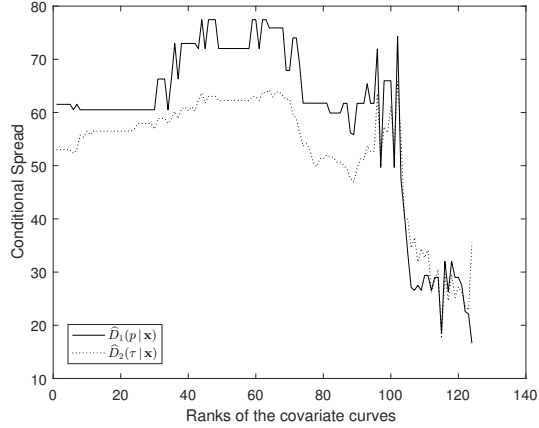

**Figure 5**. Plots of $\widehat{D}_1(p\,|\,\mathbf{x})$ and $\widehat{D}_2(\boldsymbol{\tau}\,|\,\mathbf{x})$ against the ranks of the covariate curves in the ordering by their $L_2$ norms for the Penn Table data.

The plots in Figure 6 illustrate several features of cigarette consumption, both over income and over time, in the states. The difference of the temporal trends in cigarette consumption over different levels of income is noticeable. We can observe from the conditional quantile curves that the sale of cigarettes has a peak around 1980 for each of the selected covariate curves. This peak sale of cigarettes around 1980 is the highest over the time period considered for all the selected covariate curves except the one in the $5^{\text{th}}$ column, which corresponds to the state with the highest income level among the selected covariate curves. In the $5^{\text{th}}$ column of the plots in Figure 6, the sale of cigarettes around 1980 is slightly lower than the sale around 1963, the beginning of the time period considered here. A small dip is noticeable in the spatial quantile curves and the upper and the lower boundaries of the maximal depth sets around 1970 in all the plots. This means a decrease in cigarette consumption around 1970 in the states. The conditional quantile curves start rising again after 1970 and peak around 1980. After that time, those curves are consistently decreasing. The difference of the two conditional quantiles curves is significantly lower after 1980 in the $5^{\text{th}}$ column in Figure 6. This indicates that cigarette consumption decreased more homogeneously in states with very high income levels. In lower income states corresponding to the $1^{\text{st}}$ and the $2^{\text{nd}}$ columns in Figure 6, the high difference between the two conditional quantile curves after 1980 indicates high variation in the prevalence of smoking in those states. This is further supported by the plots of the corresponding conditional maximal depth sets in the 1$^{st}$ and the 2$^{nd}$ columns in Figure 6.

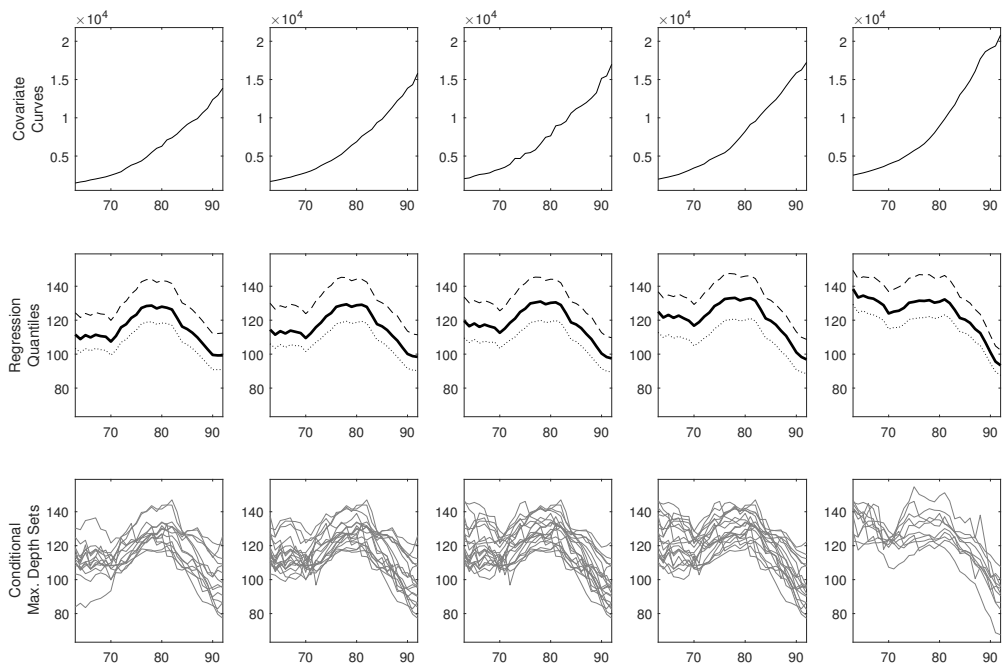

**Figure 6**. Plots of the selected covariate curves (1$^{st}$ row), the corresponding conditional spatial quantiles (2$^{nd}$ row) and conditional maximal depth sets (3$^{rd}$ row) for the Cigar Data. The dotted, the dashed and the solid curves in the 2$^{nd}$ row are $\widehat{\mathbf{Q}}_n(-0.5\mathbf{u}\,|\,\mathbf{x})$, $\widehat{\mathbf{Q}}_n(0.5\mathbf{u}\,|\,\mathbf{x})$ and $\widehat{\mathbf{Q}}_n(\mathbf{0}\,|\,\mathbf{x})$, respectively.

Our preceding observations coincide with several important events in the history of smoking in the US in the previous century (see [4, p. 16]), and those offer additional insights into the effects of these events. It was observed in [4, p. 15] that the per capita cigarette consumption in the US was the highest in 1963. However, we saw in the preceding analysis that the cigarette sales in the low and middle income states did not reach their highest levels until around 1980, while the sales of cigarettes in some very high income states were at their highest levels in 1963. So, that peak of 1963 reported by [4] was probably due to high consumption in some very high income states. In 1964, the US Surgeon General's report asserted that cigarette consumption is a leading cause of cancer, and counter-advertising on television against smoking was run in the period 1967–1970. These are likely reasons for the small drop in cigarette sales around 1970. We saw that sales of cigarette again started to rise in all the states after 1970. This may be due to the renewed effort to increase sales by the tobacco industry, like introducing special cigarette brands targeted at women, and the end of free time to anti-smoking advertisement in television broadcasting. Nonsmokers' rights movement began after 1970, and gained force by the beginning of the '80s. In 1983, Federal tax on cigarette was increased, and the Surgeon General's report in 1986 linked environmental smoking to lung cancer. The decrease in cigarette sales throughout the '80s and the beginning of the '90s may be explained as the combined effect of these events.

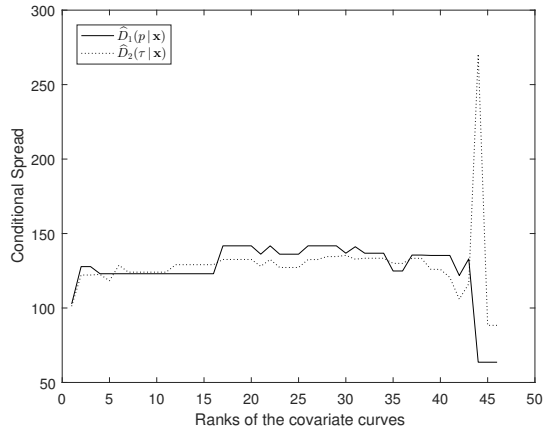

**Figure 7**. Plots of $\widehat{D}_1(p\,|\,\mathbf{x})$ and $\widehat{D}_2(\boldsymbol{\tau}\,|\,\mathbf{x})$ against the ranks of the covariate curves in the ordering by their $L_2$ norms for the Cigar Data.

We plot the conditional spread measures $\widehat{D}_1(p\,|\,\mathbf{x})$ and $\widehat{D}_2(\boldsymbol{\tau}\,|\,\mathbf{x})$ in Figure 7, with the covariate curves arranged by their $L_2$ norms. From these plots, we notice that the data is fairly homoscedastic over the covariate curves except for some extreme covariate curves. The variations in $\widehat{D}_1(p\,|\,\mathbf{x})$ and $\widehat{D}_2(\boldsymbol{\tau}\,|\,\mathbf{x})$ for those extreme covariate curves appear due to the fact that those covariate curves have very few observations in their neighborhoods, and among those observations there seem to be some outliers.

# Appendix A: Computation of $\widehat{\mathbf{Q}}_n(\boldsymbol{\tau}\,|\,\mathbf{x})$

We describe here the computational procedure of $\widehat{\mathbf{Q}}_n(\boldsymbol{\tau}\,|\,\mathbf{x})$ in detail, when the response space $\mathcal{H}$ is infinite dimensional. Observe that

$$\widehat{g}_n(\mathbf{Q}\,|\,\mathbf{x}) = \frac{\sum_{i=1}^{n}(\|\mathbf{Q}-\mathbf{Y}_i^{(n)}\| - \langle \boldsymbol{\tau}^{(n)}, \mathbf{Q}\rangle)w_i}{\sum_{i=1}^{n} w_i},$$

where $w_i = K(h_n^{-1} d(\mathbf{x}, \mathbf{X}_i))$ for $1 \leq i \leq n$. Since $\widehat{g}_n(\mathbf{Q} \,|\, \mathbf{x})$ is convex, $\widehat{\mathbf{Q}}_n(\boldsymbol{\tau} \,|\, \mathbf{x})$ minimizes $\widehat{g}_n(\mathbf{Q} \,|\, \mathbf{x})$ in $\mathcal{Z}_n$ if and only if the Gâteaux derivative

$$\lim_{t \to 0+} t^{-1} \left[ \widehat{g}_n(\widehat{\mathbf{Q}}_n(\boldsymbol{\tau} \,|\, \mathbf{x}) + t\mathbf{h} \,|\, \mathbf{x}) - \widehat{g}_n(\widehat{\mathbf{Q}}_n(\boldsymbol{\tau} \,|\, \mathbf{x}) \,|\, \mathbf{x}) \right] \geq 0$$

for all $\mathbf{h} \in \mathcal{Z}_n$. Denote $I_n = \{i \,|\, \mathbf{Y}_i^{(n)} = \widehat{\mathbf{Q}}_n(\boldsymbol{\tau} \,|\, \mathbf{x})\}$. So, for all $\mathbf{h} \in \mathcal{Z}_n$, we need to have that the Gâteaux derivative, which equals

$$\sum_{i \notin I_n} w_i \left[ \left\langle \frac{\widehat{\mathbf{Q}}_n(\boldsymbol{\tau} \,|\, \mathbf{x}) - \mathbf{Y}_i^{(n)}}{\|\widehat{\mathbf{Q}}_n(\boldsymbol{\tau} \,|\, \mathbf{x}) - \mathbf{Y}_i^{(n)}\|}, \mathbf{h} \right\rangle - \left\langle \boldsymbol{\tau}^{(n)}, \mathbf{h} \right\rangle \right] + \sum_{i \in I_n} w_i \left[ \|\mathbf{h}\| - \left\langle \boldsymbol{\tau}^{(n)}, \mathbf{h} \right\rangle \right] \geq 0.$$

Replacing $\mathbf{h}$ by $-\mathbf{h}$, we get another version of the above inequality. Since $|\|\mathbf{h}\| \pm \langle \boldsymbol{\tau}^{(n)}, \mathbf{h} \rangle| \leq (1 + \|\boldsymbol{\tau}^{(n)}\|)\|\mathbf{h}\|$, we get from the two inequalities that

$$\left\| \sum_{i \notin I_n} w_i \left[ \frac{\widehat{\mathbf{Q}}_n(\boldsymbol{\tau} \,|\, \mathbf{x}) - \mathbf{Y}_i^{(n)}}{\|\widehat{\mathbf{Q}}_n(\boldsymbol{\tau} \,|\, \mathbf{x}) - \mathbf{Y}_i^{(n)}\|} - \boldsymbol{\tau}^{(n)} \right] \right\| \leq \left( 1 + \|\boldsymbol{\tau}^{(n)}\| \right) \sum_{i \in I_n} w_i \tag{A.1}$$

if the set $I_n$ is non-empty. On the other hand, if $I_n$ is an empty set,

$$\sum_{i=1}^{n} w_i \left[ \frac{\widehat{\mathbf{Q}}_n(\boldsymbol{\tau} \,|\, \mathbf{x}) - \mathbf{Y}_i^{(n)}}{\|\widehat{\mathbf{Q}}_n(\boldsymbol{\tau} \,|\, \mathbf{x}) - \mathbf{Y}_i^{(n)}\|} - \boldsymbol{\tau}^{(n)} \right] = 0. \tag{A.2}$$

Now, we state the algorithm for computing $\widehat{\mathbf{Q}}_n(\boldsymbol{\tau} \,|\, \mathbf{x})$, when $\mathbf{Y}_1^{(n)}, \cdots, \mathbf{Y}_n^{(n)}$ do not all lie on a straight line in $\mathcal{Z}_n$. For each $i$, denote the set $J_i = \{j \,|\, \mathbf{Y}_j^{(n)} = \mathbf{Y}_i^{(n)}\}$. In the first step of our computation, we check whether the condition

$$\left\| \sum_{j \notin J_i} w_j \left[ \frac{\mathbf{Y}_i^{(n)} - \mathbf{Y}_j^{(n)}}{\|\mathbf{Y}_i^{(n)} - \mathbf{Y}_j^{(n)}\|} - \boldsymbol{\tau}^{(n)} \right] \right\| \leq \left( 1 + \|\boldsymbol{\tau}^{(n)}\| \right) \sum_{j \in J_i} w_j$$

is satisfied for any $1 \leq i \leq n$. If it is satisfied for some $i$, we take $\widehat{\mathbf{Q}}_n(\boldsymbol{\tau} \,|\, \mathbf{x}) = \mathbf{Y}_i^{(n)}$. Otherwise, we move to the next step and try to solve equation (A.2).

To solve equation (A.2), we take an initial approximation $\mathbf{Q}_1$ of $\widehat{\mathbf{Q}}_n(\boldsymbol{\tau} \,|\, \mathbf{x})$ to start the iteration. $\mathbf{Q}_1$ may be taken as the estimated pointwise conditional median of $\mathbf{Y}_1^{(n)}, \cdots, \mathbf{Y}_n^{(n)}$ given $\mathbf{X} = \mathbf{x}$ if the response is a random function. Let $\mathbf{Q}_1, \cdots, \mathbf{Q}_m$ be the $m$ successive approximations of $\widehat{\mathbf{Q}}_n(\boldsymbol{\tau} \,|\, \mathbf{x})$ obtained in the first $m$ consecutive iterations. Denote $g_m = \min\{\widehat{g}_n(\mathbf{Q}_j \,|\, \mathbf{x}) \,|\, 1 \leq j \leq m\}$. To compute $\mathbf{Q_{m+1}}$, we first compute

$$\mathbf{V} = \sum_{i=1}^{n} w_i \left[ \frac{\mathbf{Q}_m - \mathbf{Y}_i^{(n)}}{\|\mathbf{Q}_m - \mathbf{Y}_i^{(n)}\|} - \boldsymbol{\tau}^{(n)} \right]$$

$$\text{and } \mathbf{A} = \sum_{i=1}^{n} w_i \left[ \frac{Id(\mathcal{Z}_n)}{\|\mathbf{Q}_m - \mathbf{Y}_i^{(n)}\|} - \frac{\left( \cdot, \mathbf{Q}_m - \mathbf{Y}_i^{(n)} \right) \left\langle \mathbf{Q}_m - \mathbf{Y}_i^{(n)}, \cdot \right)}{\|\mathbf{Q}_m - \mathbf{Y}_i^{(n)}\|^3} \right],$$

where $Id(\mathcal{Z}_n)$ is the identity map on $\mathcal{Z}_n$, and $(\cdot, \mathbf{y})\langle \mathbf{z}, \cdot \rangle$ is the outer product of $\mathbf{y}$ and $\mathbf{z}$ in $\mathcal{Z}_n$. Since $\mathcal{Z}_n$ is finite dimensional, $\mathbf{A}$ is a positive definite operator on $\mathcal{Z}_n$ because $\mathbf{Y}_1^{(n)}, \cdots, \mathbf{Y}_n^{(n)}$ do not all lie on a straight line. We set $\mathbf{Q}' = \mathbf{Q}_m - \mathbf{A}^{-1}\mathbf{V}$. If $\widehat{g}_n(\mathbf{Q}' \,|\, \mathbf{x}) \leq g_m$, we take $\mathbf{Q}_{m+1} = \mathbf{Q}'$. Else, we set $\mathbf{Q}_{m+1} = f_m \mathbf{Q}_m + (1 - f_m)\mathbf{Q}'$, where $f_m = \widehat{g}_n(\mathbf{Q}' \,|\, \mathbf{x})/(\widehat{g}_n(\mathbf{Q}' \,|\, \mathbf{x}) + g_m)$. We stop iterating when two successive approximations of $\widehat{\mathbf{Q}}_n(\boldsymbol{\tau} \,|\, \mathbf{x})$ are sufficiently close.

## Appendix B: Mathematical details and proofs

***Proof of Theorem 4.1.*** Let $\widehat{S}(\mathbf{y} \,|\, \mathbf{x}) - S(\mathbf{y} \,|\, \mathbf{x}) - M_n(\mathbf{y} \,|\, \mathbf{x}) = A_n/B_n$, where

$$A_n = n^{-1} \sum_{i=1}^{n} \left[ \frac{\mathbf{y} - \mathbf{Y}_i}{\|\mathbf{y} - \mathbf{Y}_i\|} - S(\mathbf{y} \,|\, \mathbf{X}_i) \right] \frac{K(h_n^{-1} d(\mathbf{x}, \mathbf{X}_i))}{E_n}$$

and $B_n = n^{-1} \sum_{i=1}^{n} E_n^{-1} K(h_n^{-1} d(\mathbf{x}, \mathbf{X}_i))$. Using assumptions C(ii) and A-1, we get $B_n \to 1$ as $n \to \infty$ *almost surely*. From assumption C(iii), we get $\|\gamma(\mathbf{y} \,|\, \mathbf{z})(\cdot, \cdot) - \gamma(\mathbf{y} \,|\, \mathbf{x})(\cdot, \cdot)\| \to 0$ as $d(\mathbf{x}, \mathbf{z}) \to 0$ for any $\mathbf{y} \in \mathcal{H}$. So, using assumption A-1 and Theorem 1.1 in [29], it follows that $\sqrt{n\phi(h_n \,|\, \mathbf{x})} A_n \to \mathbf{W}$ *in distribution* as $n \to \infty$, where $\mathbf{W}$ is as described in Theorem 4.1. Hence,

$$\sqrt{n\phi(h_n \,|\, \mathbf{x})} \left( \widehat{S}(\mathbf{y} \,|\, \mathbf{x}) - S(\mathbf{y} \,|\, \mathbf{x}) - M_n(\mathbf{y} \,|\, \mathbf{x}) \right) \to \mathbf{W}$$

*in distribution* as $n \to \infty$.

Using Taylor expansion of the norm function at $S(\mathbf{y} \,|\, \mathbf{x}) + M_n(\mathbf{y} \,|\, \mathbf{x})$, we get

$$\left[ \widehat{SD}(\mathbf{y} \,|\, \mathbf{x}) - (1 - \|S(\mathbf{y} \,|\, \mathbf{x}) + M_n(\mathbf{y} \,|\, \mathbf{x})\|) \right]$$
$$= -\|S(\mathbf{y} \,|\, \mathbf{x}) + M_n(\mathbf{y} \,|\, \mathbf{x})\|^{-1} \left\langle (S(\mathbf{y} \,|\, \mathbf{x}) + M_n(\mathbf{y} \,|\, \mathbf{x})), (\widehat{S}(\mathbf{y} \,|\, \mathbf{x}) - S(\mathbf{y} \,|\, \mathbf{x}) - M_n(\mathbf{y} \,|\, \mathbf{x})) \right\rangle$$
$$+ o\left( \left\| \widehat{S}(\mathbf{y} \,|\, \mathbf{x}) - S(\mathbf{y} \,|\, \mathbf{x}) - M_n(\mathbf{y} \,|\, \mathbf{x}) \right\| \right).$$

So,

$$\left| \widehat{SD}(\mathbf{y} \,|\, \mathbf{x}) - SD(\mathbf{y} \,|\, \mathbf{x}) \right|$$
$$\leq \left| \widehat{SD}(\mathbf{y} \,|\, \mathbf{x}) - (1 - \|S(\mathbf{y} \,|\, \mathbf{x}) + M_n(\mathbf{y} \,|\, \mathbf{x})\|) \right| + \|M_n(\mathbf{y} \,|\, \mathbf{x})\|$$
$$\leq \left\| \widehat{S}(\mathbf{y} \,|\, \mathbf{x}) - S(\mathbf{y} \,|\, \mathbf{x}) - M_n(\mathbf{y} \,|\, \mathbf{x}) \right\| + o\left( \left\| \widehat{S}(\mathbf{y} \,|\, \mathbf{x}) - S(\mathbf{y} \,|\, \mathbf{x}) - M_n(\mathbf{y} \,|\, \mathbf{x}) \right\| \right) + \|M_n(\mathbf{y} \,|\, \mathbf{x})\|.$$

By the asymptotic normality of $\sqrt{n\phi(h_n \,|\, \mathbf{x})}(\widehat{S}(\mathbf{y} \,|\, \mathbf{x}) - S(\mathbf{y} \,|\, \mathbf{x}) - M_n(\mathbf{y} \,|\, \mathbf{x}))$, we get

$$\left\| \widehat{S}(\mathbf{y} \,|\, \mathbf{x}) - S(\mathbf{y} \,|\, \mathbf{x}) - M_n(\mathbf{y} \,|\, \mathbf{x}) \right\| = O_P\left( \frac{1}{\sqrt{n\phi(h_n \,|\, \mathbf{x})}} \right)$$

as $n \to \infty$. Under assumptions A-1 and A-2, $\|M_n(\mathbf{y}\,|\,\mathbf{x})\| = O(h_n)$ as $n \to \infty$ *almost surely*. Hence,

$$\left|\widehat{SD}(\mathbf{y}\,|\,\mathbf{x}) - SD(\mathbf{y}\,|\,\mathbf{x})\right| = O_P\left(\frac{1}{\sqrt{n\phi(h_n\,|\,\mathbf{x})}} + h_n\right)$$

as $n \to \infty$. □

The following result yields the convergence rate of $\widehat{\mathbf{Q}}_n(\boldsymbol{\tau}\,|\,\mathbf{x})$, and is required for the proof of Theorem 4.2. Some other intermediate results are kept in the supplement [17].

***Proof of Theorem 4.2.*** See the definition of $\bar{\mathbf{Q}}_n$ in (2.1) in the proof of Proposition 2.7 in [17] and the definition of $G_n(\mathbf{Q}\,|\,\mathbf{x})$ in Lemma 2.8 in [17]. Note that

$$\begin{aligned}
&\widehat{g}_n^{(1)}(\bar{\mathbf{Q}}_n\,|\,\mathbf{x}) - \widehat{g}_n^{(1)}(\tilde{\mathbf{Q}}_n(\boldsymbol{\tau}\,|\,\mathbf{x})\,|\,\mathbf{x}) \\
&= \left[\widehat{g}_n^{(1)}(\bar{\mathbf{Q}}_n\,|\,\mathbf{x}) - \widehat{g}_n^{(1)}(\tilde{\mathbf{Q}}_n(\boldsymbol{\tau}\,|\,\mathbf{x})\,|\,\mathbf{x}) - G_n(\bar{\mathbf{Q}}_n\,|\,\mathbf{x})\right] \\
&\quad + \left[G_n(\bar{\mathbf{Q}}_n\,|\,\mathbf{x}) - \tilde{g}_n^{(2)}(\tilde{\mathbf{Q}}_n(\boldsymbol{\tau}\,|\,\mathbf{x})\,|\,\mathbf{x})(\bar{\mathbf{Q}}_n - \tilde{\mathbf{Q}}_n(\boldsymbol{\tau}\,|\,\mathbf{x}))\right] \\
&\quad + \tilde{g}_n^{(2)}(\tilde{\mathbf{Q}}_n(\boldsymbol{\tau}\,|\,\mathbf{x})\,|\,\mathbf{x})(\bar{\mathbf{Q}}_n - \tilde{\mathbf{Q}}_n(\boldsymbol{\tau}\,|\,\mathbf{x})).
\end{aligned}$$

Using Lemma 2.8 and Lemma 2.10 in [17], we get

$$\left[\widehat{g}_n^{(1)}(\bar{\mathbf{Q}}_n\,|\,\mathbf{x}) - \widehat{g}_n^{(1)}(\tilde{\mathbf{Q}}_n(\boldsymbol{\tau}\,|\,\mathbf{x})\,|\,\mathbf{x}) - G_n(\bar{\mathbf{Q}}_n\,|\,\mathbf{x})\right] = O(\epsilon_n^2)$$

and

$$\left[G_n(\bar{\mathbf{Q}}_n\,|\,\mathbf{x}) - \tilde{g}_n^{(2)}(\tilde{\mathbf{Q}}_n(\boldsymbol{\tau}\,|\,\mathbf{x})\,|\,\mathbf{x})(\bar{\mathbf{Q}}_n - \tilde{\mathbf{Q}}_n(\boldsymbol{\tau}\,|\,\mathbf{x}))\right] = O(\epsilon_n^2)$$

as $n \to \infty$ *almost surely*. From inequality (2.3) in the proof of Proposition 2.7 in [17], we have $\widehat{g}_n^{(1)}(\bar{\mathbf{Q}}_n\,|\,\mathbf{x}) = O(\epsilon_n^2)$ for all sufficiently large $n$ *almost surely*. From Lemma 2.5 and Lemma 2.6 in [17], we get that for all sufficiently large $n$, $\tilde{g}_n^{(2)}(\tilde{\mathbf{Q}}_n(\boldsymbol{\tau}\,|\,\mathbf{x})\,|\,\mathbf{x})(\cdot)$ is invertible and $\left\|[\tilde{g}_n^{(2)}(\tilde{\mathbf{Q}}_n(\boldsymbol{\tau}\,|\,\mathbf{x})\,|\,\mathbf{x})]^{-1}\right\| \leq b_M^{-1}$, where $\|\tilde{\mathbf{Q}}_n(\boldsymbol{\tau}\,|\,\mathbf{x})\| \leq M$. Therefore, we get

$$\bar{\mathbf{Q}}_n - \tilde{\mathbf{Q}}_n(\boldsymbol{\tau}\,|\,\mathbf{x}) = -[\tilde{g}_n^{(2)}(\tilde{\mathbf{Q}}_n(\boldsymbol{\tau}\,|\,\mathbf{x})\,|\,\mathbf{x})]^{-1}\left(\widehat{g}_n^{(1)}(\tilde{\mathbf{Q}}_n(\boldsymbol{\tau}\,|\,\mathbf{x})\,|\,\mathbf{x})\right) + r_n(\mathbf{x}),$$

where $\|r_n(\mathbf{x})\| = O(\epsilon_n^2)$ as $n \to \infty$ *almost surely*. Since from (2.2) in the proof of Proposition 2.7 in [17] it follows that $\|\widehat{\mathbf{Q}}_n(\boldsymbol{\tau}\,|\,\mathbf{x}) - \bar{\mathbf{Q}}_n\| = O(\epsilon_n^2)$ *almost surely* for all sufficiently large $n$, we finally have

$$\widehat{\mathbf{Q}}_n(\boldsymbol{\tau}\,|\,\mathbf{x}) - \tilde{\mathbf{Q}}_n(\boldsymbol{\tau}\,|\,\mathbf{x}) = -[\tilde{g}_n^{(2)}(\tilde{\mathbf{Q}}_n(\boldsymbol{\tau}\,|\,\mathbf{x})\,|\,\mathbf{x})]^{-1}(\widehat{g}_n^{(1)}(\tilde{\mathbf{Q}}_n(\boldsymbol{\tau}\,|\,\mathbf{x})\,|\,\mathbf{x})) + R_n(\mathbf{x}),$$

where $\|R_n(\mathbf{x})\| = O(\epsilon_n^2)$ as $n \to \infty$ *almost surely*. □

**Proposition B.1.**  *Under assumptions B-1, B-2 and B-3, $\|\tilde{\mathbf{Q}}_n(\boldsymbol{\tau} \,|\, \mathbf{x}) - \mathbf{Q}_n(\boldsymbol{\tau} \,|\, \mathbf{x})\| \to 0$ as $n \to \infty$. In addition, suppose that there exist a constant $C_4 > 0$ and an integer $N_3 > 0$ such that whenever $d(\mathbf{x}, \mathbf{z}) \leq C_4$ and $n \geq N_3$, we have, for each $C > 0$,*

$$(d(\mathbf{x}, \mathbf{z}))^{-1} \|g_n^{(1)}(\mathbf{Q} \,|\, \mathbf{z}) - g_n^{(1)}(\mathbf{Q} \,|\, \mathbf{x})\| \leq s_3(C)$$

*for all $\mathbf{Q} \in \mathcal{Z}_n$ with $\|\mathbf{Q}\| \leq C$, where $s_3(C)$ is a positive constant depending on $C$. Then, $\|\tilde{\mathbf{Q}}_n(\boldsymbol{\tau} \,|\, \mathbf{x}) - \mathbf{Q}_n(\boldsymbol{\tau} \,|\, \mathbf{x})\| = O(h_n)$ as $n \to \infty$.*

***Proof.*** Define

$$S_n(\mathbf{Q} \,|\, \mathbf{x}) = E[\|\mathbf{Q} - \mathbf{Y}^{(n)}\|^{-1}(\mathbf{Q} - \mathbf{Y}^{(n)}) \,|\, \mathbf{X} = \mathbf{x}]$$

and

$$\tilde{S}_n(\mathbf{Q} \,|\, \mathbf{x}) = E[\|\mathbf{Q} - \mathbf{Y}^{(n)}\|^{-1}(\mathbf{Q} - \mathbf{Y}^{(n)}) E_n^{-1} K(h_n^{-1} d(\mathbf{x}, \mathbf{X}))].$$

By assumptions B-1, B-2 and using arguments similar to those in the proof of Theorem 3.1 in [8], we get that $S_n(\mathbf{Q} \,|\, \mathbf{x})$ and $\tilde{S}_n(\mathbf{Q} \,|\, \mathbf{x})$ are continuous invertible maps with the entire open unit ball in $\mathcal{Z}_n$ as their range, and

$$S_n\left(\mathbf{Q}_n(\boldsymbol{\tau} \,|\, \mathbf{x}) \,|\, \mathbf{x}\right) = \tilde{S}_n\left(\tilde{\mathbf{Q}}_n(\boldsymbol{\tau} \,|\, \mathbf{x}) \,\Big|\, \mathbf{x}\right) = \boldsymbol{\tau}^{(n)}.$$

Under assumption C(iii), we get

$$\left\|S_n\left(\tilde{\mathbf{Q}}_n(\boldsymbol{\tau} \,|\, \mathbf{x}) \,\Big|\, \mathbf{x}\right) - \boldsymbol{\tau}^{(n)}\right\| = \left\|S_n\left(\tilde{\mathbf{Q}}_n(\boldsymbol{\tau} \,|\, \mathbf{x}) \,\Big|\, \mathbf{x}\right) - \tilde{S}_n\left(\tilde{\mathbf{Q}}_n(\boldsymbol{\tau} \,|\, \mathbf{x}) \,\Big|\, \mathbf{x}\right)\right\| \to 0$$

as $n \to \infty$. Using Lemma 2.5 in [17], it follows that the Fréchet derivative of $S_n^{-1}(\cdot \,|\, \mathbf{x})$ exists, and for some $c_6 > 0$,

$$\left\|\tilde{\mathbf{Q}}_n(\boldsymbol{\tau} \,|\, \mathbf{x}) - \mathbf{Q}_n(\boldsymbol{\tau} \,|\, \mathbf{x})\right\| = \left\|S_n^{-1}\left(S_n\left(\tilde{\mathbf{Q}}_n(\boldsymbol{\tau} \,|\, \mathbf{x}) \,\Big|\, \mathbf{x}\right) \,\Big|\, \mathbf{x}\right) - S_n^{-1}\left(\boldsymbol{\tau}^{(n)} \,\Big|\, \mathbf{x}\right)\right\|$$
$$\leq c_6 \left\|S_n\left(\tilde{\mathbf{Q}}_n(\boldsymbol{\tau} \,|\, \mathbf{x}) \,\Big|\, \mathbf{x}\right) - \boldsymbol{\tau}^{(n)}\right\|$$

for all sufficiently large $n$. Therefore, $\|\tilde{\mathbf{Q}}_n(\boldsymbol{\tau} \,|\, \mathbf{x}) - \mathbf{Q}_n(\boldsymbol{\tau} \,|\, \mathbf{x})\| \to 0$ as $n \to \infty$. If the additional condition in Proposition B.1 holds, we have

$$\left\|S_n\left(\tilde{\mathbf{Q}}_n(\boldsymbol{\tau} \,|\, \mathbf{x}) \,\Big|\, \mathbf{x}\right) - \tilde{S}_n\left(\tilde{\mathbf{Q}}_n(\boldsymbol{\tau} \,|\, \mathbf{x}) \,\Big|\, \mathbf{x}\right)\right\| = O(h_n)$$

as $n \to \infty$. So, in that case, $\|\tilde{\mathbf{Q}}_n(\boldsymbol{\tau} \,|\, \mathbf{x}) - \mathbf{Q}_n(\boldsymbol{\tau} \,|\, \mathbf{x})\| = O(h_n)$ as $n \to \infty$. □

***Proof of Theorem 4.3.*** Define

$$T_n = n^{-1} \sum_{i=1}^{n} \left[\frac{\tilde{\mathbf{Q}}_n(\boldsymbol{\tau} \,|\, \mathbf{x}) - \mathbf{Y}_i^{(n)}}{\left\|\tilde{\mathbf{Q}}_n(\boldsymbol{\tau} \,|\, \mathbf{x}) - \mathbf{Y}_i^{(n)}\right\|} - \boldsymbol{\tau}^{(n)}\right] \frac{K(h_n^{-1} d(\mathbf{x}, \mathbf{X}_i))}{E_n}.$$

Clearly, $T_n$ is an average of $n$ *iid* zero mean random elements in $\mathcal{Z}_n$. Define the bilinear operator $\tilde{\gamma}_{(n)}(\mathbf{Q}\,|\,\mathbf{x})(\cdot,\cdot) : \mathcal{H} \times \mathcal{H} \to \mathbb{R}$ as

$$\tilde{\gamma}_{(n)}(\mathbf{Q}\,|\,\mathbf{x})(\mathbf{v},\mathbf{w}) = Cov[\langle \mathbf{V}_n(\mathbf{Q}), \mathbf{v}\rangle, \langle \mathbf{V}_n(\mathbf{Q}), \mathbf{w}\rangle],$$

where

$$\mathbf{V}_n(\mathbf{Q}) = \left[\frac{\mathbf{Q} - \mathbf{Y}^{(n)}}{\|\mathbf{Q} - \mathbf{Y}^{(n)}\|} - \boldsymbol{\tau}^{(n)}\right] \frac{K(h_n^{-1}d(\mathbf{x},\mathbf{X}))}{E_n}.$$

Using assumptions C(iii) and B-1 we get

$$\|\tilde{\gamma}_{(n)}(\tilde{\mathbf{Q}}_n(\boldsymbol{\tau}\,|\,\mathbf{x})\,|\,\mathbf{x})(\cdot,\cdot) - \gamma(\mathbf{Q}(\boldsymbol{\tau}\,|\,\mathbf{x})\,|\,\mathbf{x})(\cdot,\cdot)\| \to 0 \qquad (B.1)$$

as $n \to \infty$. One can show that $g^{(2)}(\mathbf{Q}(\boldsymbol{\tau}\,|\,\mathbf{x})\,|\,\mathbf{x})(\cdot)$ is invertible using similar arguments as in the proofs of Lemma 2.5 and Lemma 2.6 in [17]. Also, using arguments similar to those in the proof of Theorem 3.4 in [8] and assumptions C(iii) and B-3, one can show that

$$\sqrt{n\phi(h_n\,|\,\mathbf{x})}\|[\tilde{g}_n^{(2)}(\tilde{\mathbf{Q}}_n(\boldsymbol{\tau}\,|\,\mathbf{x})\,|\,\mathbf{x})]^{-1}T_n - [g^{(2)}(\mathbf{Q}(\boldsymbol{\tau}\,|\,\mathbf{x})\,|\,\mathbf{x})]^{-1}T_n\| \to 0 \qquad (B.2)$$

*in probability* as $n \to \infty$. Applying Theorem 1.1 in [29] and using equation (B.1), it follows that

$$\sqrt{n\phi(h_n\,|\,\mathbf{x})}[g^{(2)}(\mathbf{Q}(\boldsymbol{\tau}\,|\,\mathbf{x})\,|\,\mathbf{x})]^{-1}T_n \to \mathbf{W} \qquad (B.3)$$

*in distribution* as $n \to \infty$, where $\mathbf{W}$ is the Gaussian random element described in Theorem 4.3. Therefore, from equations (B.2) and (B.3) we get

$$\sqrt{n\phi(h_n\,|\,\mathbf{x})}[\tilde{g}_n^{(2)}(\tilde{\mathbf{Q}}_n(\boldsymbol{\tau}\,|\,\mathbf{x})\,|\,\mathbf{x})]^{-1}T_n \to \mathbf{W}$$

*in distribution* as $n \to \infty$. Since $n^{-1}\sum_{i=1}^n E_n^{-1}K(h_n^{-1}d(\mathbf{x},\mathbf{X}_i)) \to 1$ as $n \to \infty$ *almost surely*, we have

$$\frac{\sqrt{n\phi(h_n\,|\,\mathbf{x})}[\tilde{g}_n^{(2)}(\tilde{\mathbf{Q}}_n(\boldsymbol{\tau}\,|\,\mathbf{x})\,|\,\mathbf{x})]^{-1}T_n}{n^{-1}\sum_{i=1}^n E_n^{-1}K(h_n^{-1}d(\mathbf{x},\mathbf{X}_i))} \to \mathbf{W}$$

*in distribution* as $n \to \infty$. Using the assumptions stated in Theorem 4.3 and the discussion before Theorem 3.4 in [8], one can show that $\sqrt{n\phi(h_n\,|\,\mathbf{x})}[\mathbf{Q}_n(\boldsymbol{\tau}\,|\,\mathbf{x}) - \mathbf{Q}(\boldsymbol{\tau}\,|\,\mathbf{x})] \to \mathbf{0}$ *in probability* as $n \to \infty$. From assumption B-1 and Theorem 4.2, it follows that $\sqrt{n\phi(h_n\,|\,\mathbf{x})}R_n(\mathbf{x}) \to \mathbf{0}$ as $n \to \infty$ *almost surely*. The proof is complete using the representation of $\widehat{\mathbf{Q}}_n(\boldsymbol{\tau}\,|\,\mathbf{x})$ in Theorem 4.2. □

**Proof of Theorem 4.4.** Under the conditions in Theorem 4.3, it can be established using the Bernstein inequality, the Borel-Cantelli Lemma and arguments similar to those in the proof of Lemma 2 in [19] that

$$(E_{(1),n}(\mathbf{x}))^{-2}E_{(2),n}(\mathbf{x}) \to (E_{(1)}(\mathbf{x}))^{-2}E_{(2)}(\mathbf{x})$$

as $n \to \infty$ *almost surely.* Also, using Chebyshev-type arguments, one can show that

$$\widehat{g}_n^{(2)}(\widehat{\mathbf{Q}}_n(\boldsymbol{\tau} \,|\, \mathbf{x}) \,|\, \mathbf{x})(\cdot) \to g^{(2)}(\mathbf{Q}(\boldsymbol{\tau} \,|\, \mathbf{x}) \,|\, \mathbf{x})(\cdot)$$

*in probability* as $n \to \infty$. So, using arguments similar to those in Lemma 2.5 in [17], it follows that

$$[\widehat{g}_n^{(2)}(\widehat{\mathbf{Q}}_n(\boldsymbol{\tau} \,|\, \mathbf{x}) \,|\, \mathbf{x})]^{-1}(\cdot) \to [g^{(2)}(\mathbf{Q}(\boldsymbol{\tau} \,|\, \mathbf{x}) \,|\, \mathbf{x})]^{-1}(\cdot)$$

*in probability* as $n \to \infty$. Similarly, one can establish that

$$\widehat{\gamma}_n(\widehat{\mathbf{Q}}_n(\boldsymbol{\tau} \,|\, \mathbf{x}) \,|\, \mathbf{x}) \to \gamma_0(\mathbf{Q}(\boldsymbol{\tau} \,|\, \mathbf{x}) \,|\, \mathbf{x})$$

*in probability* as $n \to \infty$. Therefore, the estimate $\widehat{\boldsymbol{\Sigma}}_n(\mathbf{x}) \to \boldsymbol{\Sigma}(\mathbf{x})$ *in probability* as $n \to \infty$. □

**Supplementary Material**

**Supplement to "Nonparametric Depth and Quantile Regression for Functional Data"**
(; .pdf). We present here the plots of confidence sets for conditional spatial medians in the examples considered in the paper. In addition, some mathematical details required for the proofs of the theorems are provided.

# Acknowledgements

We thank the Associate Editor and two anonymous referees for their careful reading and valuable comments and suggestions that led to significant improvement of the paper.

# Supplement to "Nonparametric Depth and Quantile Regression for Functional Data"

JOYDEEP CHOWDHURY[1,*] PROBAL CHAUDHURI[1,**]

[1]*Statistics and Mathematics Unit, Indian Statistical Institute, Kolkata, India.*
*E-mail:* [*]joydeepchowdhury01@gmail.com; [**]probal@isical.ac.in

## 1. Confidence Sets for the Conditional Spatial Median

We present the confidence sets for the conditional spatial medians in the datasets considered in section 5 in the main paper. The confidence sets are constructed according to the method described at the end of section 4. In Figure 1, we plot the 95% (i.e., $r = 0.05$) confidence sets for the conditional spatial medians in the simulated data analysed in subsection 5.1. The covariate curves chosen are the same as those in Figure 2 in subsection 5.1. Next, we plot the 95% confidence sets for the conditional spatial medians in the Penn Table Data considered in subsection 5.2 in Figure 2. Again, the chosen covariate curves are the same as those in Figure 4 in subsection 5.2. Finally, in Figure 3, we demonstrate the 95% confidence sets for the conditional spatial medians in the Cigar Data investigated in subsection 5.3 taking the covariate curves same as those in Figure 6 in that subsection.

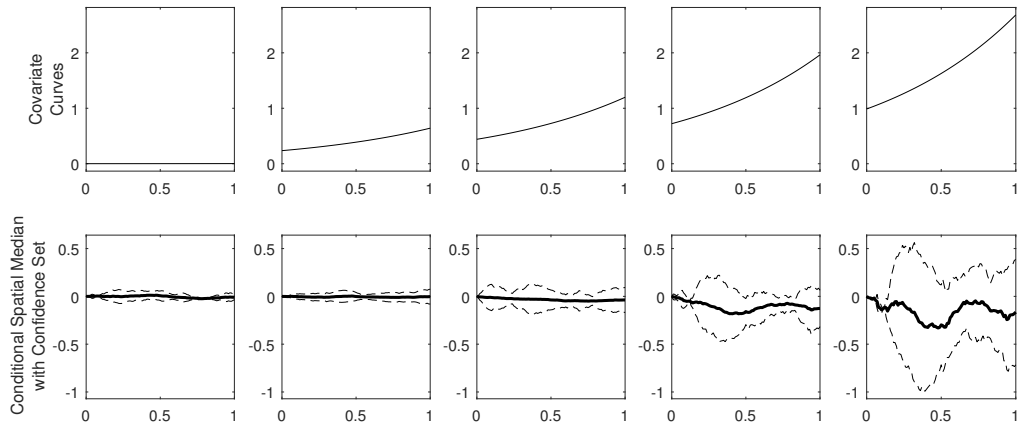

**Figure 1**. Plots of the selected covariate curves (1st row) and the corresponding conditional spatial medians along with the 95% confidence sets (2nd row) for the simulated data in subsection 5.1. The solid curves in the 2nd row are $\widehat{\mathbf{Q}}_n(\mathbf{0} \,|\, \mathbf{x})$, and the two dashed curves above and below $\widehat{\mathbf{Q}}_n(\mathbf{0} \,|\, \mathbf{x})$ constitute the boundaries of the corresponding confidence sets.

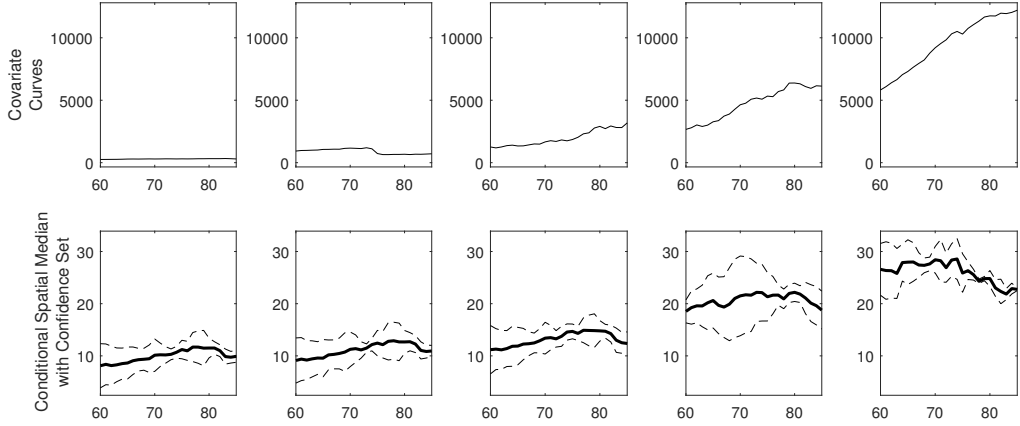

**Figure 2**. Plots of the selected covariate curves (1st row) and the corresponding conditional spatial medians along with the 95% confidence sets (2nd row) for the Penn Table Data in subsection 5.2. The solid curves in the 2nd row are $\widehat{\mathbf{Q}}_n(\mathbf{0}\,|\,\mathbf{x})$, and the two dashed curves above and below $\widehat{\mathbf{Q}}_n(\mathbf{0}\,|\,\mathbf{x})$ constitute the boundaries of the corresponding confidence sets.

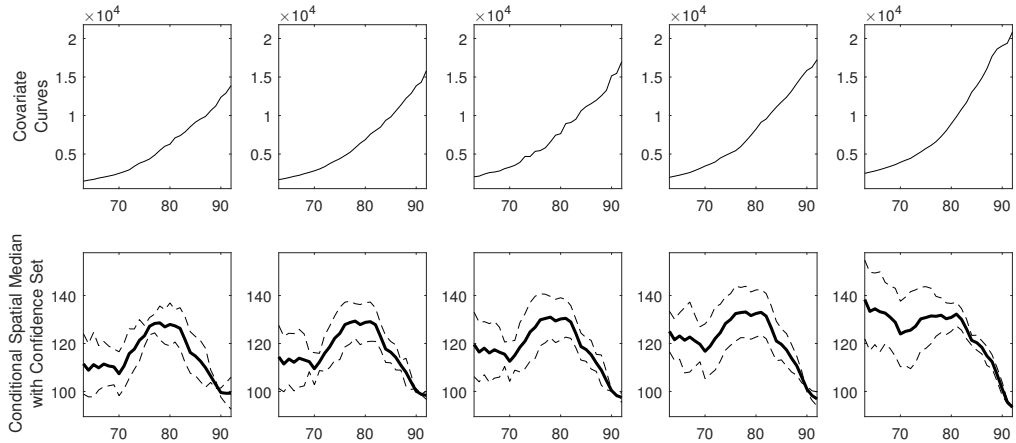

**Figure 3**. Plots of the selected covariate curves (1st row) and the corresponding conditional spatial medians along with the 95% confidence sets (2nd row) for the Cigar Data in subsection 5.3. The solid curves in the 2nd row are $\widehat{\mathbf{Q}}_n(\mathbf{0}\,|\,\mathbf{x})$, and the two dashed curves above and below $\widehat{\mathbf{Q}}_n(\mathbf{0}\,|\,\mathbf{x})$ constitute the boundaries of the corresponding confidence sets.

## 2. Additional Mathematical Details

The following additional results are required for the proofs in Appendix B in the main paper.

**Lemma 2.1.** *If $P[\|\mathbf{Y}\| = w \mid \mathbf{X} = \mathbf{x}] = 0$, then,*

$$\left|P[\|\mathbf{Y}^{(n)}\| \leq w \mid \mathbf{X} = \mathbf{x}] - P[\|\mathbf{Y}^{(n)}\| \leq w \mid \mathbf{X} = \mathbf{z}]\right| \to 0$$

*as $n \to \infty$ and $d(\mathbf{x}, \mathbf{z}) \to 0$.*

**Proof.** Define $(r)^+ = rI[r \geq 0]$ for $r \in \mathbb{R}$. Note that

$$1 - (1 - (\delta^{-1}(w - \|\mathbf{y}^{(n)}\|))^+)^+ \leq I[\|\mathbf{y}^{(n)}\| \leq w] \leq (1 - (\delta^{-1}(\|\mathbf{y}^{(n)}\| - w))^+)^+$$

for any positive $\delta$. So, we get

$$\left|P[\|\mathbf{Y}^{(n)}\| \leq w \mid \mathbf{X} = \mathbf{x}] - P[\|\mathbf{Y}^{(n)}\| \leq w \mid \mathbf{X} = \mathbf{z}]\right|$$
$$\leq P[w - \delta < \|\mathbf{Y}^{(n)}\| \leq w + \delta \mid \mathbf{X} = \mathbf{x}]$$
$$+ \left|\int (1 - (\delta^{-1}(\|\mathbf{y}^{(n)}\| - w))^+)^+ (\mu(d\mathbf{y} \mid \mathbf{x}) - \mu(d\mathbf{y} \mid \mathbf{z}))\right|.$$

Note that the function $f(\mathbf{y}) = (1 - (\delta^{-1}(\|\mathbf{y}^{(n)}\| - w))^+)^+$ satisfies

$$|f(\mathbf{y}_1) - f(\mathbf{y}_2)| \leq \delta^{-1}\|\mathbf{y}_1 - \mathbf{y}_2\|$$

and $0 \leq f(\mathbf{y}) \leq 1$ for all $\mathbf{y}$. Therefore,

$$\left|\int (1 - (\delta^{-1}(\|\mathbf{y}^{(n)}\| - w))^+)^+ (\mu(d\mathbf{y} \mid \mathbf{x}) - \mu(d\mathbf{y} \mid \mathbf{z}))\right| \leq \delta^{-1} d_{BL}(\mu(\cdot \mid \mathbf{x}), \mu(\cdot \mid \mathbf{z}))$$

if $\delta \leq 1$, where $d_{BL}(\cdot, \cdot)$ is the Bounded Lipschitz metric (see, e.g., [4, p. 74]). Given $\epsilon > 0$, choose $\delta > 0$ such that $\delta \leq 1$,

$$P[\|\mathbf{Y}\| = w - \delta \mid \mathbf{X} = \mathbf{x}] = P[\|\mathbf{Y}\| = w + \delta \mid \mathbf{X} = \mathbf{x}] = 0$$
$$\text{and } P[w - \delta < \|\mathbf{Y}\| \leq w + \delta \mid \mathbf{X} = \mathbf{x}] \leq \epsilon/6.$$

Choose $N$ such that for all $n \geq N$,

$$\left|P[\|\mathbf{Y}^{(n)}\| \leq w - \delta \mid \mathbf{X} = \mathbf{x}] - P[\|\mathbf{Y}\| \leq w - \delta \mid \mathbf{X} = \mathbf{x}]\right| \leq \epsilon/6$$
$$\text{and } \left|P[\|\mathbf{Y}^{(n)}\| \leq w + \delta \mid \mathbf{X} = \mathbf{x}] - P[\|\mathbf{Y}\| \leq w + \delta \mid \mathbf{X} = \mathbf{x}]\right| \leq \epsilon/6.$$

Therefore, for all $n \geq N$,

$$P[w - \delta < \|\mathbf{Y}^{(n)}\| \leq w + \delta \mid \mathbf{X} = \mathbf{x}] \leq \epsilon/2.$$

Since $\mathcal{H}$ is a separable Hilbert space, under assumption C(iii) we get that $d_{BL}(\mu(\cdot\,|\,\mathbf{x}), \mu(\cdot\,|\,\mathbf{z})) \to 0$ as $d(\mathbf{x}, \mathbf{z}) \to 0$. Choose $\delta_1 > 0$ such that $d_{BL}(\mu(\cdot\,|\,\mathbf{x}), \mu(\cdot\,|\,\mathbf{z})) < (\epsilon/2)\delta$ if $d(\mathbf{x}, \mathbf{z}) < \delta_1$. Therefore,

$$\left| P[\|\mathbf{Y}^{(n)}\| \leq w \,|\, \mathbf{X} = \mathbf{x}] - P[\|\mathbf{Y}^{(n)}\| \leq w \,|\, \mathbf{X} = \mathbf{z}] \right| < \epsilon$$

if $n \geq N$ and $d(\mathbf{x}, \mathbf{z}) < \delta_1$. $\square$

**Lemma 2.2.** *Under assumption B-1, there exists a constant $M > 0$ such that $\|\widehat{\mathbf{Q}}_n(\boldsymbol{\tau}\,|\,\mathbf{x})\| \leq M$ for all sufficiently large $n$ almost surely.*

**Proof.** For a constant $M > 0$, we shall show that if $\|\mathbf{Q}\| > M$, $\widehat{g}_n(\mathbf{Q}\,|\,\mathbf{x}) > \widehat{g}_n(\mathbf{0}\,|\,\mathbf{x})$ for all sufficiently large $n$ *almost surely*. Then, since $\widehat{\mathbf{Q}}_n(\boldsymbol{\tau}\,|\,\mathbf{x})$ minimizes $\widehat{g}_n(\mathbf{Q}\,|\,\mathbf{x})$, we must have $\|\widehat{\mathbf{Q}}_n(\boldsymbol{\tau}\,|\,\mathbf{x})\| \leq M$ for all sufficiently large $n$ *almost surely*. Note that

$$\widehat{g}_n(\mathbf{Q}\,|\,\mathbf{x}) - \widehat{g}_n(\mathbf{0}\,|\,\mathbf{x})$$
$$= \frac{n^{-1} \sum_{i=1}^n [\|\mathbf{Q} - \mathbf{Y}_i^{(n)}\| - \|\mathbf{Y}_i^{(n)}\| - \langle \boldsymbol{\tau}^{(n)}, \mathbf{Q} \rangle] E_n^{-1} K(h_n^{-1} d(\mathbf{x}, \mathbf{X}_i))}{n^{-1} \sum_{i=1}^n E_n^{-1} K(h_n^{-1} d(\mathbf{x}, \mathbf{X}_i))}.$$

By assumption B-1, $n^{-1} \sum_{i=1}^n E_n^{-1} K(h_n^{-1} d(\mathbf{x}, \mathbf{X}_i)) \to 1$ as $n \to \infty$ *almost surely*. Choose an integer $m > \sqrt{2}[\sqrt{2} - \sqrt{1 + \|\boldsymbol{\tau}\|}]^{-1}$. Choose $M > 0$ such that

$$P[\|\mathbf{Y}\| > m^{-1} M \,|\, \mathbf{X} = \mathbf{x}] < (1/3) m^{-1} \text{ and } P[\|\mathbf{Y}\| = m^{-1} M \,|\, \mathbf{X} = \mathbf{x}] = 0.$$

We have

$$n^{-1} \sum_{i=1}^n [\|\mathbf{Q} - \mathbf{Y}_i^{(n)}\| - \|\mathbf{Y}_i^{(n)}\| - \langle \boldsymbol{\tau}^{(n)}, \mathbf{Q} \rangle] E_n^{-1} K(h_n^{-1} d(\mathbf{x}, \mathbf{X}_i))$$
$$= n^{-1} \sum_{i=1}^n [\|\mathbf{Q} - \mathbf{Y}_i^{(n)}\| - \|\mathbf{Y}_i^{(n)}\| - \langle \boldsymbol{\tau}^{(n)}, \mathbf{Q} \rangle] E_n^{-1} K(h_n^{-1} d(\mathbf{x}, \mathbf{X}_i)) I[\|\mathbf{Y}_i^{(n)}\| \leq m^{-1} M]$$
$$+ n^{-1} \sum_{i=1}^n [\|\mathbf{Q} - \mathbf{Y}_i^{(n)}\| - \|\mathbf{Y}_i^{(n)}\| - \langle \boldsymbol{\tau}^{(n)}, \mathbf{Q} \rangle] E_n^{-1} K(h_n^{-1} d(\mathbf{x}, \mathbf{X}_i)) I[\|\mathbf{Y}_i^{(n)}\| > m^{-1} M].$$

Let

$$A_n = n^{-1} \sum_{i=1}^n E_n^{-1} K(h_n^{-1} d(\mathbf{x}, \mathbf{X}_i)) I[\|\mathbf{Y}_i^{(n)}\| > m^{-1} M].$$

Using assumption B-1, we get $0 \leq A_n \leq E[A_n] + (1/3) m^{-1}$ for all sufficiently large $n$ *almost surely*. Denote $p_n = P[\|\mathbf{Y}^{(n)}\| > m^{-1} M \,|\, \mathbf{X} = \mathbf{x}]$. From Lemma 2.1, it follows that $E[A_n] - p_n \to 0$ as $n \to \infty$. Since $\|\mathbf{Y}^{(n)}\| \to \|\mathbf{Y}\|$ as $n \to \infty$ *almost surely*, we have

$$p_n \to P[\|\mathbf{Y}\| > m^{-1} M \,|\, \mathbf{X} = \mathbf{x}] < (1/3) m^{-1}$$

as $n \to \infty$. Therefore, $0 \leq A_n < (2/3)m^{-1}$ for all sufficiently large $n$ *almost surely*. So,

$$\left| n^{-1} \sum_{i=1}^{n} [\|\mathbf{Q} - \mathbf{Y}_i^{(n)}\| - \|\mathbf{Y}_i^{(n)}\| - \langle \boldsymbol{\tau}^{(n)}, \mathbf{Q}\rangle] E_n^{-1} K(h_n^{-1} d(\mathbf{x}, \mathbf{X}_i)) I[\|\mathbf{Y}_i^{(n)}\| > m^{-1} M] \right|$$
$$\leq \|\mathbf{Q}\|(1 + \|\boldsymbol{\tau}\|) A_n < \|\mathbf{Q}\|(1 + \|\boldsymbol{\tau}\|) m^{-1}$$

for all sufficiently large $n$ *almost surely*. Since

$$n^{-1} \sum_{i=1}^{n} E_n^{-1} K(h_n^{-1} d(\mathbf{x}, \mathbf{X}_i)) > 1 - (1/3)m^{-1}$$

for all sufficiently large $n$ *almost surely*, we have, if $\|\mathbf{Q}\| > M$,

$$n^{-1} \sum_{i=1}^{n} [\|\mathbf{Q} - \mathbf{Y}_i^{(n)}\| - \|\mathbf{Y}_i^{(n)}\| - \langle \boldsymbol{\tau}^{(n)}, \mathbf{Q}\rangle] E_n^{-1} K(h_n^{-1} d(\mathbf{x}, \mathbf{X}_i)) I[\|\mathbf{Y}_i^{(n)}\| \leq m^{-1} M]$$
$$\geq \|\mathbf{Q}\|[1 - 2m^{-1} - \|\boldsymbol{\tau}\|] \left[ n^{-1} \sum_{i=1}^{n} E_n^{-1} K(h_n^{-1} d(\mathbf{x}, \mathbf{X}_i)) - A_n \right]$$
$$> \|\mathbf{Q}\|[1 - 2m^{-1} - \|\boldsymbol{\tau}\|](1 - m^{-1})$$

for all sufficiently large $n$ *almost surely*. Therefore, when $\|\mathbf{Q}\| > M$, we get

$$n^{-1} \sum_{i=1}^{n} [\|\mathbf{Q} - \mathbf{Y}_i^{(n)}\| - \|\mathbf{Y}_i^{(n)}\| - \langle \boldsymbol{\tau}^{(n)}, \mathbf{Q}\rangle] E_n^{-1} K(h_n^{-1} d(\mathbf{x}, \mathbf{X}_i))$$
$$> \|\mathbf{Q}\|[(1 - 2m^{-1} - \|\boldsymbol{\tau}\|)(1 - m^{-1}) - (1 + \|\boldsymbol{\tau}\|)m^{-1}]$$
$$= \|\mathbf{Q}\|[2(1 - m^{-1})^2 - (1 + \|\boldsymbol{\tau}\|)] > 0$$

for all sufficiently large $n$ *almost surely*, by the choice of $m$. Hence, $\widehat{g}_n(\mathbf{Q} \,|\, \mathbf{x}) > \widehat{g}_n(\mathbf{0} \,|\, \mathbf{x})$ for all sufficiently large $n$ *almost surely* if $\|\mathbf{Q}\| > M$. □

**Lemma 2.3.** *under assumption B-1, $\|\tilde{\mathbf{Q}}_n(\boldsymbol{\tau} \,|\, \mathbf{x})\| \leq M$ for all sufficiently large $n$, where $M$ is the constant defined in Lemma 2.2.*

**Proof.** Consider the integer $m$ and the constant $M$ defined in the proof of Lemma 2.2. So, by Lemma 2.1, there exists $\delta_1 > 0$ and an integer $n_1 \geq 1$ such that whenever $d(\mathbf{x}, \mathbf{z}) < \delta_1$ and $n \geq n_1$,

$$P[\|\mathbf{Y}^{(n)}\| > m^{-1} M \,|\, \mathbf{X} = \mathbf{z}] < (2/3)m^{-1}.$$

Since $h_n \to 0$ by assumption B-1, there exists an integer $n_2 \geq n_1$ such that whenever $n \geq n_2$,

$$P[\|\mathbf{Y}^{(n)}\| > m^{-1} M \,|\, \mathbf{X}] I(d(\mathbf{x}, \mathbf{X}) \leq h_n) \leq (2/3)m^{-1} I(d(\mathbf{x}, \mathbf{X}) \leq h_n).$$

Note that

$$\tilde{g}_n(\mathbf{Q}\,|\,\mathbf{x}) - \tilde{g}_n(\mathbf{0}\,|\,\mathbf{x})$$
$$= E[(\|\mathbf{Q} - \mathbf{Y}^{(n)}\| - \|\mathbf{Y}^{(n)}\| - \langle \boldsymbol{\tau}^{(n)}, \mathbf{Q}\rangle)I(\|\mathbf{Y}^{(n)}\| \leq m^{-1}M)E_n^{-1}K(h_n^{-1}d(\mathbf{x}, \mathbf{X}))]$$
$$+ E[(\|\mathbf{Q} - \mathbf{Y}^{(n)}\| - \|\mathbf{Y}^{(n)}\| - \langle \boldsymbol{\tau}^{(n)}, \mathbf{Q}\rangle)I(\|\mathbf{Y}^{(n)}\| > m^{-1}M)E_n^{-1}K(h_n^{-1}d(\mathbf{x}, \mathbf{X}))].$$

Now, if $n \geq n_2$,

$$\left| E[(\|\mathbf{Q} - \mathbf{Y}^{(n)}\| - \|\mathbf{Y}^{(n)}\| - \langle \boldsymbol{\tau}^{(n)}, \mathbf{Q}\rangle)I(\|\mathbf{Y}^{(n)}\| > m^{-1}M)E_n^{-1}K(h_n^{-1}d(\mathbf{x}, \mathbf{X}))] \right|$$
$$\leq \|\mathbf{Q}\|(1 + \|\boldsymbol{\tau}\|)E[P(\|\mathbf{Y}^{(n)}\| > m^{-1}M\,|\,\mathbf{X})E_n^{-1}K(h_n^{-1}d(\mathbf{x}, \mathbf{X}))]$$
$$< \|\mathbf{Q}\|(1 + \|\boldsymbol{\tau}\|)m^{-1}.$$

Also, if $\|\mathbf{Q}\| > M$ and $n \geq n_2$,

$$E[(\|\mathbf{Q} - \mathbf{Y}^{(n)}\| - \|\mathbf{Y}^{(n)}\| - \langle \boldsymbol{\tau}^{(n)}, \mathbf{Q}\rangle)I(\|\mathbf{Y}^{(n)}\| \leq m^{-1}M)E_n^{-1}K(h_n^{-1}d(\mathbf{x}, \mathbf{X}))]$$
$$\geq E[(\|\mathbf{Q}\| - 2\|\mathbf{Y}^{(n)}\| - \|\boldsymbol{\tau}^{(n)}\|\|\mathbf{Q}\|)I(\|\mathbf{Y}^{(n)}\| \leq m^{-1}M)E_n^{-1}K(h_n^{-1}d(\mathbf{x}, \mathbf{X}))]$$
$$> \|\mathbf{Q}\|(1 - 2m^{-1} - \|\boldsymbol{\tau}\|)(1 - m^{-1}).$$

So, if $\|\mathbf{Q}\| > M$ and $n \geq n_2$,

$$\tilde{g}_n(\mathbf{Q}\,|\,\mathbf{x}) - \tilde{g}_n(\mathbf{0}\,|\,\mathbf{x}) > \|\mathbf{Q}\|(1 - 2m^{-1} - \|\boldsymbol{\tau}\|)(1 - m^{-1}) - \|\mathbf{Q}\|(1 + \|\boldsymbol{\tau}\|)m^{-1} > 0$$

by the choice of $m$. Since $\tilde{\mathbf{Q}}_n(\boldsymbol{\tau}\,|\,\mathbf{x})$ minimizes $\tilde{g}_n(\mathbf{Q}\,|\,\mathbf{x})$, we must have $\|\tilde{\mathbf{Q}}_n(\boldsymbol{\tau}\,|\,\mathbf{x})\| \leq M$ for all $n \geq n_2$. □

**Lemma 2.4.** *Under assumption B-2, $\|\mathbf{Q}_n(\boldsymbol{\tau}\,|\,\mathbf{x}) - \mathbf{Q}(\boldsymbol{\tau}\,|\,\mathbf{x})\| \to 0$ as $n \to \infty$.*

***Proof.*** First, we shall show that $\|\mathbf{Q}_n(\boldsymbol{\tau}\,|\,\mathbf{x})\| \leq M$ for all $n$, where $M > 0$ is the constant defined in the proof of Lemma 2.2. Consider the integer $m$ defined in the proof of Lemma 2.2. Note that $P[\|\mathbf{Y}\| > m^{-1}M\,|\,\mathbf{X} = \mathbf{x}] < m^{-1}$. Since $\|\mathbf{Y}^{(n)}\| \leq \|\mathbf{Y}\|$, $P[\|\mathbf{Y}^{(n)}\| > m^{-1}M\,|\,\mathbf{X} = \mathbf{x}] < m^{-1}$ for all $n$. Note that

$$g_n(\mathbf{Q}\,|\,\mathbf{x}) - g_n(\mathbf{0}\,|\,\mathbf{x})$$
$$= g_n(\mathbf{Q}\,|\,\mathbf{x})$$
$$= E[(\|\mathbf{Q} - \mathbf{Y}^{(n)}\| - \|\mathbf{Y}^{(n)}\| - \langle \boldsymbol{\tau}^{(n)}, \mathbf{Q}\rangle)I(\|\mathbf{Y}^{(n)}\| \leq m^{-1}M)\,|\,\mathbf{X} = \mathbf{x}]$$
$$+ E[(\|\mathbf{Q} - \mathbf{Y}^{(n)}\| - \|\mathbf{Y}^{(n)}\| - \langle \boldsymbol{\tau}^{(n)}, \mathbf{Q}\rangle)I(\|\mathbf{Y}^{(n)}\| > m^{-1}M)\,|\,\mathbf{X} = \mathbf{x}].$$

Also,

$$\left| E\left[(\|\mathbf{Q} - \mathbf{Y}^{(n)}\| - \|\mathbf{Y}^{(n)}\| - \langle \boldsymbol{\tau}^{(n)}, \mathbf{Q}\rangle)I(\|\mathbf{Y}^{(n)}\| > m^{-1}M)\,\middle|\,\mathbf{X} = \mathbf{x}\right] \right|$$
$$< \|\mathbf{Q}\|(1 + \|\boldsymbol{\tau}\|)m^{-1}.$$

If $\|\mathbf{Q}\| > M$, then

$$E\left[(\|\mathbf{Q}-\mathbf{Y}^{(n)}\|-\|\mathbf{Y}^{(n)}\|-\langle\boldsymbol{\tau}^{(n)},\mathbf{Q}\rangle)I(\|\mathbf{Y}^{(n)}\|\leq m^{-1}M)\;\Big|\;\mathbf{X}=\mathbf{x}\right]$$
$$> \|\mathbf{Q}\|(1-2m^{-1}-\|\boldsymbol{\tau}\|)(1-m^{-1}).$$

So, if $\|\mathbf{Q}\| > M$,

$$g_n(\mathbf{Q}\,|\,\mathbf{x}) > \|\mathbf{Q}\|[(1-2m^{-1}-\|\boldsymbol{\tau}\|)(1-m^{-1})-(1+\|\boldsymbol{\tau}\|)m^{-1}]>0$$

by the choice of $m$. Since $\mathbf{Q}_n(\boldsymbol{\tau}\,|\,\mathbf{x})$ minimizes $g_n(\mathbf{Q}\,|\,\mathbf{x})$, we must have $\|\mathbf{Q}_n(\boldsymbol{\tau}\,|\,\mathbf{x})\|\leq M$ for all $n$.

Note that $\mathbf{Q}(\boldsymbol{\tau}\,|\,\mathbf{x})$ exists and is unique under assumption B-2. Using Theorem 1 and Theorem 3 in [1], it is enough to show $g(\mathbf{Q}_n(\boldsymbol{\tau}\,|\,\mathbf{x})\,|\,\mathbf{x}) \to g(\mathbf{Q}(\boldsymbol{\tau}\,|\,\mathbf{x})\,|\,\mathbf{x})$ as $n\to\infty$ to complete the proof. Since $\mathbf{Q}(\boldsymbol{\tau}\,|\,\mathbf{x})$ minimizes $g(\mathbf{Q}\,|\,\mathbf{x})$, we have

$$0 \leq g(\mathbf{Q}_n(\boldsymbol{\tau}\,|\,\mathbf{x})\,|\,\mathbf{x}) - g(\mathbf{Q}(\boldsymbol{\tau}\,|\,\mathbf{x})\,|\,\mathbf{x})$$
$$= [g(\mathbf{Q}_n(\boldsymbol{\tau}\,|\,\mathbf{x})\,|\,\mathbf{x}) - g_n(\mathbf{Q}_n(\boldsymbol{\tau}\,|\,\mathbf{x})\,|\,\mathbf{x})] + [g_n(\mathbf{Q}(\boldsymbol{\tau}\,|\,\mathbf{x})\,|\,\mathbf{x}) - g(\mathbf{Q}(\boldsymbol{\tau}\,|\,\mathbf{x})\,|\,\mathbf{x})]$$
$$+ [g_n(\mathbf{Q}_n(\boldsymbol{\tau}\,|\,\mathbf{x})\,|\,\mathbf{x}) - g_n(\mathbf{Q}^{(n)}(\boldsymbol{\tau}\,|\,\mathbf{x})|\,\mathbf{x})] + [g_n(\mathbf{Q}^{(n)}(\boldsymbol{\tau}\,|\,\mathbf{x})|\,\mathbf{x}) - g_n(\mathbf{Q}(\boldsymbol{\tau}\,|\,\mathbf{x})\,|\,\mathbf{x})],$$

where $\mathbf{Q}^{(n)}(\boldsymbol{\tau}\,|\,\mathbf{x})$ is the projection of $\mathbf{Q}(\boldsymbol{\tau}\,|\,\mathbf{x})$ in $\mathcal{Z}_n$. Now,

$$|g(\mathbf{Q}\,|\,\mathbf{x}) - g_n(\mathbf{Q}\,|\,\mathbf{x})| \leq 2E[\|\mathbf{Y}-\mathbf{Y}^{(n)}\|\,|\,\mathbf{X}=\mathbf{x}] + \|\boldsymbol{\tau}^{(n)}-\boldsymbol{\tau}\|\|\mathbf{Q}\|.$$

So, for every $C > 0$, $\sup_{\|\mathbf{Q}\|\leq C}|g(\mathbf{Q}\,|\,\mathbf{x}) - g_n(\mathbf{Q}\,|\,\mathbf{x})| \to 0$ as $n \to \infty$. Since $\|\mathbf{Q}_n(\boldsymbol{\tau}\,|\,\mathbf{x})\|\leq M$ for all $n$, we get

$$[g(\mathbf{Q}_n(\boldsymbol{\tau}\,|\,\mathbf{x})\,|\,\mathbf{x}) - g_n(\mathbf{Q}_n(\boldsymbol{\tau}\,|\,\mathbf{x})\,|\,\mathbf{x})] \to 0 \text{ and } [g_n(\mathbf{Q}(\boldsymbol{\tau}\,|\,\mathbf{x})\,|\,\mathbf{x}) - g(\mathbf{Q}(\boldsymbol{\tau}\,|\,\mathbf{x})\,|\,\mathbf{x})] \to 0$$

as $n \to \infty$. Since $\mathbf{Q}_n(\boldsymbol{\tau}\,|\,\mathbf{x})$ minimizes $g_n(\mathbf{Q}\,|\,\mathbf{x})$ in $\mathcal{Z}_n$ and $\mathbf{Q}^{(n)}(\boldsymbol{\tau}\,|\,\mathbf{x}) \in \mathcal{Z}_n$, we have

$$[g_n(\mathbf{Q}_n(\boldsymbol{\tau}\,|\,\mathbf{x})\,|\,\mathbf{x}) - g_n(\mathbf{Q}^{(n)}(\boldsymbol{\tau}\,|\,\mathbf{x})|\,\mathbf{x})] \leq 0$$

for all $n$. Also,

$$|g_n(\mathbf{Q}^{(n)}(\boldsymbol{\tau}\,|\,\mathbf{x})|\,\mathbf{x}) - g_n(\mathbf{Q}(\boldsymbol{\tau}\,|\,\mathbf{x})\,|\,\mathbf{x})| \leq 2\|\mathbf{Q}^{(n)}(\boldsymbol{\tau}\,|\,\mathbf{x}) - \mathbf{Q}(\boldsymbol{\tau}\,|\,\mathbf{x})\| \to 0$$

as $n \to \infty$. Therefore, $g(\mathbf{Q}_n(\boldsymbol{\tau}\,|\,\mathbf{x})\,|\,\mathbf{x}) \to g(\mathbf{Q}(\boldsymbol{\tau}\,|\,\mathbf{x})\,|\,\mathbf{x})$ as $n\to\infty$. $\square$

**Lemma 2.5.** *Under assumptions B-2 and B-3, there exists $C_5 > 0$ and an integer $N_4 > 0$ such that whenever $d(\mathbf{x},\mathbf{z}) \leq C_5$ and $n \geq N_4$, we have, for every $C > 0$, $b_C\|\mathbf{h}\| \leq \|g_n^{(2)}(\mathbf{Q}\,|\,\mathbf{z})(\mathbf{h})\| \leq B_C\|\mathbf{h}\|$ for any $\mathbf{Q},\mathbf{h} \in \mathcal{Z}_n$ with $\|\mathbf{Q}\| \leq C$. Here $0 < b_C < B_C < \infty$ are constants depending on $C$.*

**Proof.** Assumption B-3 ensures that $g_n^{(2)}(\mathbf{Q}\,|\,\mathbf{z})(\cdot)$ is well-defined for $n \geq N_2$ and $d(\mathbf{x},\mathbf{z}) \leq C_3$. First, we shall show that for every $C > 0$, we have $b_C\|\mathbf{h}\|^2 \leq \langle g_n^{(2)}(\mathbf{Q}\,|\,\mathbf{z})(\mathbf{h}),\mathbf{h}\rangle \leq$ $B_C \|\mathbf{h}\|^2$ for any $\mathbf{Q}, \mathbf{h} \in \mathcal{Z}_n$ with $\|\mathbf{Q}\| \leq C$, whenever $n \geq N_4$ and $d(\mathbf{x}, \mathbf{z}) \leq C_5$. Here $C_5 > 0$ is a constant and $N_4 > 0$ is an integer. Without loss of generality, consider any $\mathbf{h}$ with $\|\mathbf{h}\| = 1$. Define $P_\mathbf{h}(\cdot)$ to be the projection operator on the orthogonal complement of $\mathbf{h}$. Then,

$$\langle (g_n^{(2)}(\mathbf{Q} \,|\, \mathbf{z}))(\mathbf{h}), \mathbf{h} \rangle$$
$$= E\left[\|\mathbf{Q} - \mathbf{Y}^{(n)}\|^{-1}\left[1 - \|\mathbf{Q} - \mathbf{Y}^{(n)}\|^{-2}\langle \mathbf{h}, \mathbf{Q} - \mathbf{Y}^{(n)}\rangle^2\right] \,\Big|\, \mathbf{X} = \mathbf{z}\right]$$
$$= E\left[\|\mathbf{Q} - \mathbf{Y}^{(n)}\|^{-3}\left\|P_\mathbf{h}\left(\mathbf{Q} - \mathbf{Y}^{(n)}\right)\right\|^2 \,\Big|\, \mathbf{X} = \mathbf{z}\right].$$

So, from assumption B-3, it follows that whenever $n \geq N_2$ and $d(\mathbf{x}, \mathbf{z}) \leq C_3$, we have

$$\langle (g_n^{(2)}(\mathbf{Q} \,|\, \mathbf{z}))(\mathbf{h}), \mathbf{h} \rangle \leq E\left[\|\mathbf{Q} - \mathbf{Y}^{(n)}\|^{-1} \,\Big|\, \mathbf{X} = \mathbf{z}\right] \leq \sqrt{s_2(C)}$$

for any $\mathbf{Q} \in \mathcal{Z}_n$ with $\|\mathbf{Q}\| \leq C$. Set $B_C = \sqrt{s_2(C)}$ and $N_4 = \max\{N_1, N_2\}$, where $N_1$ is the constant from assumption B-2. For the other inequality, using assumption B-2 and following the arguments in the proof of Proposition 2.1 in [2], we get that there exists a subspace $\mathcal{S} \subset \mathcal{Z}_{N_1}$ of dimension 2 such that $\langle \mathbf{u}, \mathbf{Y}^{(N_1)} \rangle$ is non-degenerate for every $\mathbf{u} \in \mathcal{S}$. This implies that $\langle \mathbf{u}, \mathbf{Y} \rangle$ is non-degenerate for every $\mathbf{u} \in \mathcal{S}$. So, either $E[|\langle \mathbf{u}, \mathbf{Y} \rangle| \,|\, \mathbf{X} = \mathbf{x}] = \infty$ for every $\mathbf{u} \in \mathcal{S}$, or $\inf\left\{Var[\langle \mathbf{u}, \mathbf{Y} \rangle \,|\, \mathbf{X} = \mathbf{x}] \,\big|\, \mathbf{u} \in \mathcal{S}, \|\mathbf{u}\| = 1 \text{ with } E[|\langle \mathbf{u}, \mathbf{Y} \rangle| \,|\, \mathbf{X} = \mathbf{x}] < \infty\right\} = b > 0.$

Since $\mathcal{Z}_n \subset \mathcal{Z}_{n+1}$ for all $n$, we have $\mathcal{S} \subset \mathcal{Z}_n$ for all $n \geq N_4$. Also, there is $\mathbf{v} \in \mathcal{S}$ such that $\|\mathbf{v}\| = 1$ and $\langle \mathbf{h}, \mathbf{v} \rangle = 0$. Then, for any $\mathbf{y}$, $\|P_\mathbf{h}(\mathbf{y})\|^2 \geq \langle \mathbf{y}, \mathbf{v} \rangle^2$. So, for $n \geq N_4$,

$$\|P_\mathbf{h}(\mathbf{Q} - \mathbf{Y}^{(n)})\|^2 \geq \langle \mathbf{Q} - \mathbf{Y}^{(n)}, \mathbf{v} \rangle^2 = \langle \mathbf{Q} - \mathbf{Y}, \mathbf{v} \rangle^2$$

as $\mathbf{Q}, \mathbf{Y}^{(n)}, \mathbf{v} \in \mathcal{Z}_n$. Also, for $\mathbf{Q} \in \mathcal{Z}_n$, $\|\mathbf{Q} - \mathbf{Y}^{(n)}\| \leq \|\mathbf{Q} - \mathbf{Y}\|$ as $(\mathbf{Q} - \mathbf{Y}^{(n)}) \in \mathcal{Z}_n$. Note that for any $M_1 > 0$,

$$\langle \mathbf{Q} - \mathbf{Y}, \mathbf{v} \rangle^2 I(\|\mathbf{Y}\| \leq M_1) = \langle \mathbf{Q} - \mathbf{Y} I(\|\mathbf{Y}\| \leq M_1), \mathbf{v} \rangle^2 - \langle \mathbf{Q}, \mathbf{v} \rangle^2 I(\|\mathbf{Y}\| > M_1).$$

Therefore, if $\mathbf{Q} \in \mathcal{Z}_n$ with $\|\mathbf{Q}\| \leq C$ and $n \geq N_4$, we have

$$E\left[\left\|\mathbf{Q} - \mathbf{Y}^{(n)}\right\|^{-3}\left\|P_\mathbf{h}\left(\mathbf{Q} - \mathbf{Y}^{(n)}\right)\right\|^2 \,\Big|\, \mathbf{X} = \mathbf{x}\right]$$
$$\geq (C + M_1)^{-3}\left[E\left[\langle \mathbf{Q} - \mathbf{Y} I(\|\mathbf{Y}\| \leq M_1), \mathbf{v}\rangle^2 \,\big|\, \mathbf{X} = \mathbf{x}\right] - C^2 P\left[\|\mathbf{Y}\| > M_1 \,|\, \mathbf{X} = \mathbf{x}\right]\right].$$

We shall now consider the two cases, viz., $E[|\langle \mathbf{Y}, \mathbf{v} \rangle| \,|\, \mathbf{X} = \mathbf{x}] < \infty$ and $E[|\langle \mathbf{Y}, \mathbf{v} \rangle| \,|\, \mathbf{X} = \mathbf{x}] = \infty$, separately.

First, suppose $E[|\langle \mathbf{Y}, \mathbf{v} \rangle| \,|\, \mathbf{X} = \mathbf{x}] < \infty$. We have

$$E\left[\langle \mathbf{Q} - \mathbf{Y} I(\|\mathbf{Y}\| \leq M_1), \mathbf{v}\rangle^2 \,\big|\, \mathbf{X} = \mathbf{x}\right] \geq Var\left[\langle \mathbf{Y} I(\|\mathbf{Y}\| \leq M_1), \mathbf{v}\rangle \,|\, \mathbf{X} = \mathbf{x}\right].$$

Note that
$$Var[\langle \mathbf{Y}I(\|\mathbf{Y}\| \leq M_1), \mathbf{v}\rangle \,|\, \mathbf{X} = \mathbf{x}] \to Var[\langle \mathbf{Y}, \mathbf{v}\rangle \,|\, \mathbf{X} = \mathbf{x}] \geq b > 0$$

as $M_1 \to \infty$. Choose $M_1 > 0$ such that
$$(C + M_1)^{-3}\big[Var[\langle \mathbf{Y}I(\|\mathbf{Y}\| \leq M_1), \mathbf{v}\rangle \,|\, \mathbf{X} = \mathbf{x}] - C^2 P[\|\mathbf{Y}\| > M_1 \,|\, \mathbf{X} = \mathbf{x}]\big] > 0$$
and $P[\|\mathbf{Y}\| = M_1 \,|\, \mathbf{X} = \mathbf{x}] = 0$.

Using assumption C(iii), we get
$$\big[Var[\langle \mathbf{Y}I(\|\mathbf{Y}\| \leq M_1), \mathbf{v}\rangle \,|\, \mathbf{X} = \mathbf{z}] - C^2 P[\|\mathbf{Y}\| > M_1 \,|\, \mathbf{X} = \mathbf{z}]\big]$$
$$\to \big[Var[\langle \mathbf{Y}I(\|\mathbf{Y}\| \leq M_1), \mathbf{v}\rangle \,|\, \mathbf{X} = \mathbf{x}] - C^2 P[\|\mathbf{Y}\| > M_1 \,|\, \mathbf{X} = \mathbf{x}]\big]$$

as $d(\mathbf{x}, \mathbf{z}) \to 0$.

Next, suppose $E[|\langle \mathbf{Y}, \mathbf{v}\rangle| \,|\, \mathbf{X} = \mathbf{x}] = \infty$. Then, it is easy to show that
$$E[\langle \mathbf{Q} - \mathbf{Y}I(\|\mathbf{Y}\| \leq M_1), \mathbf{v}\rangle^2 \,|\, \mathbf{X} = \mathbf{x}]$$
$$> -C[5C + 2] + E\big[|\langle \mathbf{Y}, \mathbf{v}\rangle| I(\|\mathbf{Y}\| \leq M_1, |\langle \mathbf{Y}, \mathbf{v}\rangle| > \max(2C, 1)) \,|\, \mathbf{X} = \mathbf{x}\big] \to \infty$$

as $M_1 \to \infty$. So, we can find $M_1$ such that
$$(C + M_1)^{-3} \left\{ -C(5C + 2) + E\big[|\langle \mathbf{Y}, \mathbf{v}\rangle| I(\|\mathbf{Y}\| \leq M_1, |\langle \mathbf{Y}, \mathbf{v}\rangle| > \max(2C, 1)) \,|\, \mathbf{X} = \mathbf{x}\big] \right\} > 0$$
and $P[\|\mathbf{Y}\| = M_1 \,|\, \mathbf{X} = \mathbf{x}] = P[|\langle \mathbf{Y}, \mathbf{v}\rangle| = \max(2C, 1) \,|\, \mathbf{X} = \mathbf{x}] = 0$.

Again, using assumption C(iii), we get
$$E\big[|\langle \mathbf{Y}, \mathbf{v}\rangle| I(\|\mathbf{Y}\| \leq M_1, |\langle \mathbf{Y}, \mathbf{v}\rangle| > \max(2C, 1)) \,|\, \mathbf{X} = \mathbf{z}\big]$$
$$\to E\big[|\langle \mathbf{Y}, \mathbf{v}\rangle| I(\|\mathbf{Y}\| \leq M_1, |\langle \mathbf{Y}, \mathbf{v}\rangle| > \max(2C, 1)) \,|\, \mathbf{X} = \mathbf{x}\big]$$

as $d(\mathbf{x}, \mathbf{z}) \to 0$.

Therefore, there exist $C_5 > 0$ such that for every $C$, we can find $b_C > 0$ satisfying $\langle (g_n^{(2)}(\mathbf{Q} \,|\, \mathbf{z}))(\mathbf{h}), \mathbf{h}\rangle \geq b_C$ for all $\mathbf{Q} \in \mathcal{Z}_n$ with $\|\mathbf{Q}\| \leq C$ whenever $n \geq N_4$ and $d(\mathbf{x}, \mathbf{z}) \leq C_5$.

Finally, from the arguments in the proofs of Theorem 4.3.13 and Theorem 4.3.16 in [3], it follows that whenever $n \geq N_4$ and $d(\mathbf{x}, \mathbf{z}) \leq C_5$, for every $C > 0$, we have $b_C \|\mathbf{h}\| \leq \|g_n^{(2)}(\mathbf{Q} \,|\, \mathbf{z})(\mathbf{h})\| \leq B_C \|\mathbf{h}\|$ for any $\mathbf{Q}, \mathbf{h} \in \mathcal{Z}_n$ with $\|\mathbf{Q}\| \leq C$. □

**Lemma 2.6.** *Under assumptions B-2 and B-3, $g_n^{(2)}(\mathbf{Q}_n(\boldsymbol{\tau} \,|\, \mathbf{x}) \,|\, \mathbf{x})(\cdot)$ is invertible for all sufficiently large $n$. In addition, if assumption B-1 holds, then $\tilde{g}_n^{(2)}(\tilde{\mathbf{Q}}_n(\boldsymbol{\tau} \,|\, \mathbf{x}) \,|\, \mathbf{x})(\cdot)$ is invertible for all sufficiently large $n$.*

**Proof.** We have $\|\tilde{\mathbf{Q}}_n(\boldsymbol{\tau} \,|\, \mathbf{x})\| \leq M$ and $\|\mathbf{Q}_n(\boldsymbol{\tau} \,|\, \mathbf{x})\| \leq M$ for all sufficiently large $n$ from Lemma 2.3 and Lemma 2.4, respectively. It follows from Lemma 2.5 that $b_M \|\mathbf{h}\| \leq \|g_n^{(2)}(\mathbf{Q}_n(\boldsymbol{\tau} \,|\, \mathbf{x}) \,|\, \mathbf{x})(\mathbf{h})\|$ and $b_M \|\mathbf{h}\| \leq \|\tilde{g}_n^{(2)}(\tilde{\mathbf{Q}}_n(\boldsymbol{\tau} \,|\, \mathbf{x}) \,|\, \mathbf{x})(\mathbf{h})\|$ for all $\mathbf{h} \in \mathcal{Z}_n$ and all sufficiently large $n$. Therefore, by the Inverse Mapping Theorem, $g_n^{(2)}(\mathbf{Q}_n(\boldsymbol{\tau} \,|\, \mathbf{x}) \,|\, \mathbf{x})(\cdot)$ and $\tilde{g}_n^{(2)}(\tilde{\mathbf{Q}}_n(\boldsymbol{\tau} \,|\, \mathbf{x}) \,|\, \mathbf{x})(\cdot)$ are invertible for all sufficiently large $n$. □

**Proposition 2.7.** *Under assumptions B-1, B-2 and B-3, $\|\widehat{\mathbf{Q}}_n(\boldsymbol{\tau} \,|\, \mathbf{x}) - \tilde{\mathbf{Q}}_n(\boldsymbol{\tau} \,|\, \mathbf{x})\| = O(\epsilon_n)$ as $n \to \infty$ almost surely, where $\epsilon_n = (n\phi(h_n \,|\, \mathbf{x}))^{-\alpha}\sqrt{\log n}$.*

***Proof.*** From Lemma 2.2 and Lemma 2.3, $\|\widehat{\mathbf{Q}}_n(\boldsymbol{\tau} \,|\, \mathbf{x}) - \tilde{\mathbf{Q}}_n(\boldsymbol{\tau} \,|\, \mathbf{x})\| \le 2M = M_2$ (say) for all sufficiently large $n$ *almost surely*. Let $\mathbb{Z}$ denote the set of all integers. Define

$$G_n = \left\{ \tilde{\mathbf{Q}}_n(\boldsymbol{\tau} \,|\, \mathbf{x}) + \sum_{j=1}^{d_n} \beta_j \mathbf{e}_j \,\bigg|\, \beta_j \in [-M_2, M_2],\, n^4 \beta_j \in \mathbb{Z} \text{ and } \bigg\|\sum_{j=1}^{d_n} \beta_j \mathbf{e}_j\bigg\| \le M_2 \right\}.$$

Denote a point in $G_n$ that is nearest to $\widehat{\mathbf{Q}}_n(\boldsymbol{\tau} \,|\, \mathbf{x})$ as $\bar{\mathbf{Q}}_n$, i.e.,

$$\bar{\mathbf{Q}}_n = \arg\min\left\{ \left\|\mathbf{Q} - \widehat{\mathbf{Q}}_n(\boldsymbol{\tau} \,|\, \mathbf{x})\right\| \,\bigg|\, \mathbf{Q} \in G_n \right\}. \tag{2.1}$$

So, by the choice of $d_n$ in Theorem 4.2 in the main paper,

$$\|\widehat{\mathbf{Q}}_n(\boldsymbol{\tau} \,|\, \mathbf{x}) - \bar{\mathbf{Q}}_n\| \le d_n n^{-4} \tag{2.2}$$

for all sufficiently large $n$ *almost surely*. We now have

$$\left\| n^{-1} \sum_{i=1}^{n} \left[ \frac{\bar{\mathbf{Q}}_n - \mathbf{Y}_i^{(n)}}{\|\bar{\mathbf{Q}}_n - \mathbf{Y}_i^{(n)}\|} - \boldsymbol{\tau}^{(n)} \right] \frac{K(h_n^{-1} d(\mathbf{x}, \mathbf{X}_i))}{E_n} \right\|$$
$$\le \left\| n^{-1} \sum_{i=1}^{n} \left[ \frac{\widehat{\mathbf{Q}}_n(\boldsymbol{\tau} \,|\, \mathbf{x}) - \mathbf{Y}_i^{(n)}}{\|\widehat{\mathbf{Q}}_n(\boldsymbol{\tau} \,|\, \mathbf{x}) - \mathbf{Y}_i^{(n)}\|} - \boldsymbol{\tau}^{(n)} \right] \frac{K(h_n^{-1} d(\mathbf{x}, \mathbf{X}_i))}{E_n} \right\|$$
$$+ n^{-1} \sum_{i=1}^{n} \left\| \frac{\bar{\mathbf{Q}}_n - \mathbf{Y}_i^{(n)}}{\|\bar{\mathbf{Q}}_n - \mathbf{Y}_i^{(n)}\|} - \frac{\widehat{\mathbf{Q}}_n(\boldsymbol{\tau} \,|\, \mathbf{x}) - \mathbf{Y}_i^{(n)}}{\|\widehat{\mathbf{Q}}_n(\boldsymbol{\tau} \,|\, \mathbf{x}) - \mathbf{Y}_i^{(n)}\|} \right\| \frac{K(h_n^{-1} d(\mathbf{x}, \mathbf{X}_i))}{E_n}.$$

Note that when $\widehat{\mathbf{Q}}_n(\boldsymbol{\tau} \,|\, \mathbf{x}) \ne \mathbf{Y}_i^{(n)}$ and $\bar{\mathbf{Q}}_n \ne \mathbf{Y}_i^{(n)}$, we have

$$\left\| \frac{\widehat{\mathbf{Q}}_n(\boldsymbol{\tau} \,|\, \mathbf{x}) - \mathbf{Y}_i^{(n)}}{\|\widehat{\mathbf{Q}}_n(\boldsymbol{\tau} \,|\, \mathbf{x}) - \mathbf{Y}_i^{(n)}\|} - \frac{\bar{\mathbf{Q}}_n - \mathbf{Y}_i^{(n)}}{\|\bar{\mathbf{Q}}_n - \mathbf{Y}_i^{(n)}\|} \right\|$$
$$\le \left\| \frac{\widehat{\mathbf{Q}}_n(\boldsymbol{\tau} \,|\, \mathbf{x}) - \mathbf{Y}_i^{(n)}}{\|\widehat{\mathbf{Q}}_n(\boldsymbol{\tau} \,|\, \mathbf{x}) - \mathbf{Y}_i^{(n)}\|} - \frac{\widehat{\mathbf{Q}}_n(\boldsymbol{\tau} \,|\, \mathbf{x}) - \mathbf{Y}_i^{(n)}}{\|\bar{\mathbf{Q}}_n - \mathbf{Y}_i^{(n)}\|} \right\| + \left\| \frac{\widehat{\mathbf{Q}}_n(\boldsymbol{\tau} \,|\, \mathbf{x}) - \mathbf{Y}_i^{(n)}}{\|\bar{\mathbf{Q}}_n - \mathbf{Y}_i^{(n)}\|} - \frac{\bar{\mathbf{Q}}_n - \mathbf{Y}_i^{(n)}}{\|\bar{\mathbf{Q}}_n - \mathbf{Y}_i^{(n)}\|} \right\|$$
$$\le 2 \frac{\|\widehat{\mathbf{Q}}_n(\boldsymbol{\tau} \,|\, \mathbf{x}) - \bar{\mathbf{Q}}_n\|}{\|\bar{\mathbf{Q}}_n - \mathbf{Y}_i^{(n)}\|}.$$

Also, when $\|\bar{\mathbf{Q}}_n - \mathbf{Y}_i^{(n)}\| > n^{-2}$, we have $\widehat{\mathbf{Q}}_n(\boldsymbol{\tau} \,|\, \mathbf{x}) \ne \mathbf{Y}_i^{(n)}$. So,

$$n^{-1} \sum_{i=1}^{n} \left\| \frac{\bar{\mathbf{Q}}_n - \mathbf{Y}_i^{(n)}}{\|\bar{\mathbf{Q}}_n - \mathbf{Y}_i^{(n)}\|} - \frac{\widehat{\mathbf{Q}}_n(\boldsymbol{\tau} \,|\, \mathbf{x}) - \mathbf{Y}_i^{(n)}}{\|\widehat{\mathbf{Q}}_n(\boldsymbol{\tau} \,|\, \mathbf{x}) - \mathbf{Y}_i^{(n)}\|} \right\| \frac{K(h_n^{-1} d(\mathbf{x}, \mathbf{X}_i))}{E_n}$$

$$\leq \frac{2}{n} \sum_{i=1}^{n} \frac{\|\widehat{\mathbf{Q}}_n(\boldsymbol{\tau} \,|\, \mathbf{x}) - \bar{\mathbf{Q}}_n\|}{\|\bar{\mathbf{Q}}_n - \mathbf{Y}_i^{(n)}\|} I\left(\|\bar{\mathbf{Q}}_n - \mathbf{Y}_i^{(n)}\| > \frac{1}{n^2}\right) \frac{K(h_n^{-1} d(\mathbf{x}, \mathbf{X}_i))}{E_n}$$

$$+ \frac{2}{n} \sum_{i=1}^{n} I\left(\|\bar{\mathbf{Q}}_n - \mathbf{Y}_i^{(n)}\| \leq \frac{1}{n^2}\right) \frac{K(h_n^{-1} d(\mathbf{x}, \mathbf{X}_i))}{E_n}$$

$$\leq \frac{2d_n}{n^2} n^{-1} \sum_{i=1}^{n} \frac{K(h_n^{-1} d(\mathbf{x}, \mathbf{X}_i))}{E_n} + 2n^{-1} \sum_{i=1}^{n} I(\|\bar{\mathbf{Q}}_n - \mathbf{Y}_i^{(n)}\| \leq n^{-2}) \frac{K(h_n^{-1} d(\mathbf{x}, \mathbf{X}_i))}{E_n}.$$

Denote

$$p_n(\mathbf{Q}) = E[P(\|\mathbf{Q} - \mathbf{Y}^{(n)}\| \leq n^{-2} \,|\, \mathbf{X}) E_n^{-1} K(h_n^{-1} d(\mathbf{x}, \mathbf{X}))].$$

Note that for $\mathbf{Q} \in G_n$, $\|\mathbf{Q}\| \leq 2M_2$ for all $n$. Since $G_n \subset \mathcal{Z}_n$, using Markov inequality and assumption B-3, we get that for all sufficiently large $n$, $\max_{\mathbf{Q} \in G_n} p_n(\mathbf{Q}) \leq n^{-4} s_2(2M_2)$. Therefore, using assumption C(ii) and the Bernstein inequality, we have

$$P\left[\max_{\mathbf{Q} \in G_n} n^{-1} \sum_{i=1}^{n} I(\|\mathbf{Q} - \mathbf{Y}_i^{(n)}\| \leq n^{-2}) E_n^{-1} K(h_n^{-1} d(\mathbf{x}, \mathbf{X}_i)) > b_1 \epsilon_n^2\right]$$

$$\leq \sum_{\mathbf{Q} \in G_n} P\left[\sum_{i=1}^{n} n^{-1} I(\|\mathbf{Q} - \mathbf{Y}_i^{(n)}\| \leq n^{-2}) E_n^{-1} K(h_n^{-1} d(\mathbf{x}, \mathbf{X}_i)) - p_n(\mathbf{Q}) > 2^{-1} b_1 \epsilon_n^2\right]$$

$$\leq (3M_2 n^4)^{d_n} \exp[-b_2 n \phi(h_n \,|\, \mathbf{x}) \epsilon_n^2]$$

for all sufficiently large $n$ and any $b_1 > 0$, where $b_2 = [8L]^{-1} l b_1$. Using assumption B-1, the Borel-Cantelli Lemma, the choice of $d_n$ and choosing $b_1$ appropriately, we get

$$\max_{\mathbf{Q} \in G_n} n^{-1} \sum_{i=1}^{n} I(\|\mathbf{Q} - \mathbf{Y}_i^{(n)}\| \leq n^{-2}) E_n^{-1} K(h_n^{-1} d(\mathbf{x}, \mathbf{X}_i)) \leq b_1 \epsilon_n^2$$

for all sufficiently large $n$ *almost surely*. Also, note that $n^{-1} \sum_{i=1}^{n} E_n^{-1} K(h_n^{-1} d(\mathbf{x}, \mathbf{X}_i)) < 2$ for all sufficiently large $n$ *almost surely*. Hence, we get

$$n^{-1} \sum_{i=1}^{n} \left\| \frac{\bar{\mathbf{Q}}_n - \mathbf{Y}_i^{(n)}}{\|\bar{\mathbf{Q}}_n - \mathbf{Y}_i^{(n)}\|} - \frac{\widehat{\mathbf{Q}}_n(\boldsymbol{\tau} \,|\, \mathbf{x}) - \mathbf{Y}_i^{(n)}}{\|\widehat{\mathbf{Q}}_n(\boldsymbol{\tau} \,|\, \mathbf{x}) - \mathbf{Y}_i^{(n)}\|} \right\| \frac{K(h_n^{-1} d(\mathbf{x}, \mathbf{X}_i))}{E_n} \leq \frac{4d_n}{n^2} + 2b_1 \epsilon_n^2 < 3b_1 \epsilon_n^2$$

for all sufficiently large $n$ *almost surely*. Also, assumptions B-1 and B-2 imply that the $\mathbf{Y}_i^{(n)}$s for which $d(\mathbf{x}, \mathbf{X}_i) \leq h_n$, are distinct *almost surely* for all sufficiently large $n$. So, putting $w_i = n^{-1} E_n^{-1} K(h_n^{-1} d(\mathbf{x}, \mathbf{X}_i))$ in inequality (A.1), it follows that

$$\left\| n^{-1} \sum_{i=1}^{n} \left[ \frac{\widehat{\mathbf{Q}}_n(\boldsymbol{\tau} \,|\, \mathbf{x}) - \mathbf{Y}_i^{(n)}}{\|\mathbf{Y}_i^{(n)} - \widehat{\mathbf{Q}}_n(\boldsymbol{\tau} \,|\, \mathbf{x})\|} - \boldsymbol{\tau}^{(n)} \right] \frac{K(h_n^{-1} d(\mathbf{x}, \mathbf{X}_i))}{E_n} \right\| \leq (ln\phi(h_n \,|\, \mathbf{x}))^{-1} 3L < b_1 \epsilon_n^2$$

*almost surely* for all sufficiently large $n$. Hence, we get

$$\left\| n^{-1} \sum_{i=1}^{n} \left[ \frac{\bar{\mathbf{Q}}_n - \mathbf{Y}_i^{(n)}}{\|\bar{\mathbf{Q}}_n - \mathbf{Y}_i^{(n)}\|} - \boldsymbol{\tau}^{(n)} \right] \frac{K(h_n^{-1} d(\mathbf{x}, \mathbf{X}_i))}{E_n} \right\| < 4 b_1 \epsilon_n^2 \tag{2.3}$$

*almost surely* for all sufficiently large $n$. Next, denote

$$\mathbf{V}_{i,n}(\mathbf{Q})$$
$$= \left[ \frac{\mathbf{Q} - \mathbf{Y}_i^{(n)}}{\|\mathbf{Q} - \mathbf{Y}_i^{(n)}\|} - \boldsymbol{\tau}^{(n)} \right] \frac{K(h_n^{-1} d(\mathbf{x}, \mathbf{X}_i))}{E_n} - E\left[ \left[ \frac{\mathbf{Q} - \mathbf{Y}^{(n)}}{\|\mathbf{Q} - \mathbf{Y}^{(n)}\|} - \boldsymbol{\tau}^{(n)} \right] \frac{K(h_n^{-1} d(\mathbf{x}, \mathbf{X}))}{E_n} \right].$$

In view of assumption C(ii), $\|\mathbf{V}_{i,n}(\mathbf{Q})\| \leq (l \phi(h_n \,|\, \mathbf{x}))^{-1} 4L$ for all $i$. Also, $E[\|\mathbf{V}_{i,n}(\mathbf{Q})\|^2] \leq [l^2 \phi(h_n \,|\, \mathbf{x})]^{-1}(16 L^2)$ for all $i$ and $n$. So,

$$E[\|\mathbf{V}_{i,n}(\mathbf{Q})\|^m] \leq [(l \phi(h_n \,|\, \mathbf{x}))^{-1} 4L]^{m-2} (l^2 \phi(h_n \,|\, \mathbf{x}))^{-1}(16 L^2)$$

for all sufficiently large $n$ and for all $m \geq 2$. Using the Bernstein inequality for random elements in a separable Hilbert space (see [5, Corollary in p. 491]), we get

$$P\left[ \max_{\mathbf{Q} \in G_n} \left\| n^{-1} \sum_{i=1}^{n} \mathbf{V}_{i,n}(\mathbf{Q}) \right\| > b_3 \epsilon_n \right]$$
$$\leq \sum_{\mathbf{Q} \in G_n} P\left[ \left\| \sum_{i=1}^{n} \mathbf{V}_{i,n}(\mathbf{Q}) \right\| > b_3 n \epsilon_n \right]$$
$$\leq 2 (3 M_2 n^4)^{d_n} \exp[-b_4 n \phi(h_n \,|\, \mathbf{x}) \epsilon_n^2]$$

for all sufficiently large $n$ and any $b_3 > 0$, where $b_4 = (20 L^2)^{-1} l^2 b_3^2$. Therefore, choosing an appropriate $b_3$ and using assumption B-1, the choice of $d_n$ and the Borel-Cantelli Lemma, we get

$$\max_{\mathbf{Q} \in G_n} \left\| n^{-1} \sum_{i=1}^{n} \mathbf{V}_{i,n}(\mathbf{Q}) \right\| \leq b_3 \epsilon_n \tag{2.4}$$

for all sufficiently large $n$ *almost surely*. Let $\mathbf{Q} \in G_n$ and $\|\mathbf{Q} - \tilde{\mathbf{Q}}_n(\boldsymbol{\tau} \,|\, \mathbf{x})\| > b_5 \epsilon_n$, where $b_5 > 0$. From a Taylor expansion and Lemma 2.5, we have

$$\left\| E\left[ \left[ \frac{\mathbf{Q} - \mathbf{Y}^{(n)}}{\|\mathbf{Q} - \mathbf{Y}^{(n)}\|} - \boldsymbol{\tau}^{(n)} \right] \frac{K(h_n^{-1} d(\mathbf{x}, \mathbf{X}))}{E_n} \right] \right\| \geq b_{2 M_2} \|\mathbf{Q} - \tilde{\mathbf{Q}}_n(\boldsymbol{\tau} \,|\, \mathbf{x})\| > b_{2 M_2} b_5 \epsilon_n$$

for all sufficiently large $n$. Then, from inequality (2.4), we get

$$\left\| n^{-1} \sum_{i=1}^{n} \left[ \frac{\mathbf{Q} - \mathbf{Y}_i^{(n)}}{\|\mathbf{Q} - \mathbf{Y}_i^{(n)}\|} - \boldsymbol{\tau}^{(n)} \right] \frac{K(h_n^{-1} d(\mathbf{x}, \mathbf{X}_i))}{E_n} \right\| > (b_{2 M_2} b_5 - b_3) \epsilon_n$$

for all sufficiently large $n$ *almost surely*. Choosing $b_5$ such that

$$(b_{2M_2}b_5 - b_3) > 5b_1 \tag{2.5}$$

and from inequality (2.3), we see that we must have

$$\|\bar{\mathbf{Q}}_n - \tilde{\mathbf{Q}}_n(\boldsymbol{\tau} \,|\, \mathbf{x})\| \leq b_5 \epsilon_n \tag{2.6}$$

for all sufficiently large $n$ *almost surely*. Therefore, we have

$$\|\widehat{\mathbf{Q}}_n(\boldsymbol{\tau} \,|\, \mathbf{x}) - \tilde{\mathbf{Q}}_n(\boldsymbol{\tau} \,|\, \mathbf{x})\| \leq \|\widehat{\mathbf{Q}}_n(\boldsymbol{\tau} \,|\, \mathbf{x}) - \bar{\mathbf{Q}}_n\| + \|\bar{\mathbf{Q}}_n - \tilde{\mathbf{Q}}_n(\boldsymbol{\tau} \,|\, \mathbf{x})\| \leq (1 + b_5)\epsilon_n$$

for all sufficiently large $n$ *almost surely*. $\square$

**Lemma 2.8.** *Recall $\bar{\mathbf{Q}}_n$ defined in (2.1) in the proof of Proposition 2.7. Define $G_n(\mathbf{Q} \,|\, \mathbf{x})$ as*

$$G_n(\mathbf{Q} \,|\, \mathbf{x}) = \frac{E\left[\left[\frac{\mathbf{Q} - \mathbf{Y}^{(n)}}{\|\mathbf{Q} - \mathbf{Y}^{(n)}\|} - \frac{\tilde{\mathbf{Q}}_n(\boldsymbol{\tau} \,|\, \mathbf{x}) - \mathbf{Y}^{(n)}}{\|\tilde{\mathbf{Q}}_n(\boldsymbol{\tau} \,|\, \mathbf{x}) - \mathbf{Y}^{(n)}\|}\right] E_n^{-1} K(h_n^{-1} d(\mathbf{x}, \mathbf{X}))\right]}{n^{-1} \sum_{i=1}^n E_n^{-1} K(h_n^{-1} d(\mathbf{x}, \mathbf{X}_i))}.$$

*Under assumptions B-1, B-2 and B-3, we have $\widehat{g}_n^{(1)}(\bar{\mathbf{Q}}_n \,|\, \mathbf{x}) - \widehat{g}_n^{(1)}(\tilde{\mathbf{Q}}_n(\boldsymbol{\tau} \,|\, \mathbf{x}) \,|\, \mathbf{x}) - G_n(\bar{\mathbf{Q}}_n \,|\, \mathbf{x}) = O(\epsilon_n^2)$ as $n \to \infty$ almost surely, where $\epsilon_n$ is as defined in Proposition 2.7.*

**Proof.** Denote $H_n = \{\mathbf{Q} \in G_n \,|\, \|\mathbf{Q} - \tilde{\mathbf{Q}}_n(\boldsymbol{\tau} \,|\, \mathbf{x})\| \leq b_6 \epsilon_n\}$, where $b_6 = (1 + b_5)$ and $b_5$ is the constant defined in (2.5) in the proof of Proposition 2.7. From (2.6) in the proof of Proposition 2.7, we see that $\bar{\mathbf{Q}}_n \in H_n$ for all sufficiently large $n$ *almost surely*. Now,

$$\widehat{g}_n^{(1)}(\mathbf{Q} \,|\, \mathbf{x}) - \widehat{g}_n^{(1)}(\tilde{\mathbf{Q}}_n(\boldsymbol{\tau} \,|\, \mathbf{x}) \,|\, \mathbf{x}) - G_n(\mathbf{Q} \,|\, \mathbf{x}) = \frac{n^{-1} \sum_{i=1}^n \{A_{i,n}(\mathbf{Q}) - E[A_{i,n}(\mathbf{Q})]\}}{n^{-1} \sum_{i=1}^n E_n^{-1} K(h_n^{-1} d(\mathbf{x}, \mathbf{X}_i))},$$

where

$$A_{i,n}(\mathbf{Q}) = \left[\frac{\mathbf{Q} - \mathbf{Y}_i^{(n)}}{\|\mathbf{Q} - \mathbf{Y}_i^{(n)}\|} - \frac{\tilde{\mathbf{Q}}_n(\boldsymbol{\tau} \,|\, \mathbf{x}) - \mathbf{Y}_i^{(n)}}{\|\tilde{\mathbf{Q}}_n(\boldsymbol{\tau} \,|\, \mathbf{x}) - \mathbf{Y}_i^{(n)}\|}\right] E_n^{-1} K(h_n^{-1} d(\mathbf{x}, \mathbf{X}_i)).$$

Note that

$$\|A_{i,n}(\mathbf{Q})\| \leq 2\|\tilde{\mathbf{Q}}_n(\boldsymbol{\tau} \,|\, \mathbf{x}) - \mathbf{Y}_i^{(n)}\|^{-1} \|\mathbf{Q} - \tilde{\mathbf{Q}}_n(\boldsymbol{\tau} \,|\, \mathbf{x})\| E_n^{-1} K(h_n^{-1} d(\mathbf{x}, \mathbf{X}_i))$$

*almost surely* for all $i$, $\mathbf{Q}$ and for all sufficiently large $n$. Let $U_{i,n}(\mathbf{Q}) = A_{i,n}(\mathbf{Q}) - E[A_{i,n}(\mathbf{Q})]$. Note that $\|U_{i,n}(\mathbf{Q})\| \leq (l\phi(h_n \,|\, \mathbf{x}))^{-1} 4L$ for all $i$ and for all sufficiently large $n$. Also,

$$\|U_{i,n}(\mathbf{Q})\|^2 \leq [\|A_{i,n}(\mathbf{Q})\| + \|E[A_{i,n}(\mathbf{Q})]\|]^2 \leq 2[\|A_{i,n}(\mathbf{Q})\|^2 + \|E[A_{i,n}(\mathbf{Q})]\|^2].$$

So,

$$E[\|U_{i,n}(\mathbf{Q})\|^2] \leq 2E[\|A_{i,n}(\mathbf{Q})\|^2 + \|E[A_{i,n}(\mathbf{Q})]\|^2]$$
$$\leq 4E\left[\|A_{i,n}(\mathbf{Q})\|^2\right]$$
$$\leq [l^2\phi(h_n\,|\,\mathbf{x})]^{-1}[16L^2 s_2(M)]\|\mathbf{Q}-\tilde{\mathbf{Q}}_n(\boldsymbol{\tau}\,|\,\mathbf{x})\|^2$$

for all sufficiently large $n$ using assumption B-3 and and Lemma 2.3. Therefore,

$$E[\|U_{i,n}(\mathbf{Q})\|^m] \leq \left[\frac{4L}{l\phi(h_n\,|\,\mathbf{x})}\right]^{m-2}\frac{16L^2 s_2(M)}{l^2\phi(h_n\,|\,\mathbf{x})}\|\mathbf{Q}-\tilde{\mathbf{Q}}_n(\boldsymbol{\tau}\,|\,\mathbf{x})\|^2.$$

Using assumption B-1, the Bernstein inequality in a separable Hilbert space, the choice of $d_n$ in Theorem 4.2 and the Borel-Cantelli Lemma, we get $\max_{\mathbf{Q}\in H_n}\left\|n^{-1}\sum_{i=1}^n U_{i,n}(\mathbf{Q})\right\| \leq b_7\epsilon_n^2$ for all sufficiently large $n$ *almost surely*, for an appropriate positive constant $b_7$. Since $n^{-1}\sum_{i=1}^n E_n^{-1}K(h_n^{-1}d(\mathbf{x},\mathbf{X}_i)) > (1/2)$ for all sufficiently large $n$ *almost surely*, we have

$$\left\|\widehat{g}_n^{(1)}(\bar{\mathbf{Q}}_n\,|\,\mathbf{x}) - \widehat{g}_n^{(1)}(\tilde{\mathbf{Q}}_n(\boldsymbol{\tau}\,|\,\mathbf{x})\,|\,\mathbf{x}) - G_n(\bar{\mathbf{Q}}_n\,|\,\mathbf{x})\right\|$$
$$\leq \frac{\max_{\mathbf{Q}\in H_n}\left\|n^{-1}\sum_{i=1}^n U_{i,n}(\mathbf{Q})\right\|}{n^{-1}\sum_{i=1}^n E_n^{-1}K(h_n^{-1}d(\mathbf{x},\mathbf{X}_i))} \leq 2b_7\epsilon_n^2$$

for all sufficiently large $n$ *almost surely*. $\square$

**Lemma 2.9.** *Under assumptions B-2 and B-3, for every $C > 0$ and and for all sufficiently large $n$, we have $\|(\tilde{g}_n^{(2)}(\mathbf{Q}_1\,|\,\mathbf{x}))(\mathbf{h}) - (\tilde{g}_n^{(2)}(\mathbf{Q}_2\,|\,\mathbf{x}))(\mathbf{h})\| \leq 6s_2(C)\|\mathbf{h}\|\|\mathbf{Q}_1-\mathbf{Q}_2\|$ for any $\mathbf{Q}_1,\mathbf{Q}_2 \in \mathcal{Z}_n$ with $\|\mathbf{Q}_1\|,\|\mathbf{Q}_2\| \leq C$ and $\mathbf{h} \in \mathcal{Z}_n$.*

**Proof.** Assumption B-2 ensures that for any $\mathbf{Q}_1$ and $\mathbf{Q}_2$, $\mathbf{Y}^{(n)} \neq \mathbf{Q}_1$ and $\mathbf{Y}^{(n)} \neq \mathbf{Q}_2$ *almost surely* for all $n \geq N_1$, and

$$(\tilde{g}_n^{(2)}(\mathbf{Q}_1\,|\,\mathbf{x}))(\mathbf{h}) - (\tilde{g}_n^{(2)}(\mathbf{Q}_2\,|\,\mathbf{x}))(\mathbf{h}) = E[[W_n(\mathbf{Q}_1)(\mathbf{h}) - W_n(\mathbf{Q}_2)(\mathbf{h})]E_n^{-1}K(h_n^{-1}d(\mathbf{x},\mathbf{X}))],$$

where

$$W_n(\mathbf{Q})(\mathbf{h}) = \left[\frac{h}{\|\mathbf{Q}-\mathbf{Y}^{(n)}\|} - \frac{1}{\|\mathbf{Q}-\mathbf{Y}^{(n)}\|}\left\langle \mathbf{h}, \frac{\mathbf{Q}-\mathbf{Y}^{(n)}}{\|\mathbf{Q}-\mathbf{Y}^{(n)}\|}\right\rangle \frac{\mathbf{Q}-\mathbf{Y}^{(n)}}{\|\mathbf{Q}-\mathbf{Y}^{(n)}\|}\right].$$

Note that when $\mathbf{Q}_1 \neq \mathbf{Y}^{(n)}$ and $\mathbf{Q}_2 \neq \mathbf{Y}^{(n)}$, we have

$$\left\|\frac{\mathbf{Q}_2-\mathbf{Y}^{(n)}}{\|\mathbf{Q}_2-\mathbf{Y}^{(n)}\|} - \frac{\mathbf{Q}_1-\mathbf{Y}^{(n)}}{\|\mathbf{Q}_1-\mathbf{Y}^{(n)}\|}\right\| \leq 2\frac{\|\mathbf{Q}_2-\mathbf{Q}_1\|}{\|\mathbf{Q}_2-\mathbf{Y}^{(n)}\|}. \tag{2.7}$$

In that case, from inequality (2.7) it follows that

$$\|W_n(\mathbf{Q}_1)(\mathbf{h}) - W_n(\mathbf{Q}_2)(\mathbf{h})\|$$
$$\leq \frac{6\|\mathbf{h}\|\|\mathbf{Q}_2-\mathbf{Q}_1\|}{\|\mathbf{Q}_1-\mathbf{Y}^{(n)}\|\|\mathbf{Q}_2-\mathbf{Y}^{(n)}\|} \tag{2.8}$$
$$\leq 3\|\mathbf{h}\|\|\mathbf{Q}_2-\mathbf{Q}_1\|\left[\frac{1}{\|\mathbf{Q}_1-\mathbf{Y}^{(n)}\|^2} + \frac{1}{\|\mathbf{Q}_2-\mathbf{Y}^{(n)}\|^2}\right]$$

for any $\mathbf{Q}_1, \mathbf{Q}_2, \mathbf{h} \in \mathcal{H}$. Using assumption B-3, we get that whenever $\mathbf{Q}_1, \mathbf{Q}_2 \in \mathcal{Z}_n$ with $\|\mathbf{Q}_1\|, \|\mathbf{Q}_2\| \leq C$,

$$E\left[\frac{1}{\|\mathbf{Q}_i - \mathbf{Y}^{(n)}\|^2}\frac{K(h_n^{-1}d(\mathbf{x}, \mathbf{X}))}{E_n}\right] \leq s_2(C) \tag{2.9}$$

for all sufficiently large $n$, where $i = 1, 2$. Hence, using inequalities (2.8) and (2.9), it follows that for all sufficiently large $n$, and for any $\mathbf{Q}_1, \mathbf{Q}_2 \in \mathcal{Z}_n$ with $\|\mathbf{Q}_1\|, \|\mathbf{Q}_2\| \leq C$, we have

$$\left\|(\tilde{g}_n^{(2)}(\mathbf{Q}_1 \,|\, \mathbf{x}))(\mathbf{h}) - (\tilde{g}_n^{(2)}(\mathbf{Q}_2 \,|\, \mathbf{x}))(\mathbf{h})\right\| \leq 6s_2(C)\|\mathbf{h}\|\|\mathbf{Q}_2 - \mathbf{Q}_1\|.$$

$\square$

**Lemma 2.10.** *Let $G_n(\mathbf{Q} \,|\, \mathbf{x})$ be as defined in Lemma 2.8. Under assumptions B-1, B-2 and B-3, we have*

$$\left[G_n(\bar{\mathbf{Q}}_n \,|\, \mathbf{x}) - \tilde{g}_n^{(2)}(\tilde{\mathbf{Q}}_n(\boldsymbol{\tau} \,|\, \mathbf{x}) \,|\, \mathbf{x})(\bar{\mathbf{Q}}_n - \tilde{\mathbf{Q}}_n(\boldsymbol{\tau} \,|\, \mathbf{x}))\right] = O(\epsilon_n^2)$$

*as $n \to \infty$ almost surely, where $\epsilon_n$ is as defined in Proposition 2.7.*

**Proof.** Note that

$$G_n(\bar{\mathbf{Q}}_n \,|\, \mathbf{x}) - \tilde{g}_n^{(2)}(\tilde{\mathbf{Q}}_n(\boldsymbol{\tau} \,|\, \mathbf{x}) \,|\, \mathbf{x})(\bar{\mathbf{Q}}_n - \tilde{\mathbf{Q}}_n(\boldsymbol{\tau} \,|\, \mathbf{x}))$$
$$= \frac{[\tilde{g}_n^{(1)}(\bar{\mathbf{Q}}_n \,|\, \mathbf{x}) - \tilde{g}_n^{(1)}(\tilde{\mathbf{Q}}_n(\boldsymbol{\tau} \,|\, \mathbf{x}) \,|\, \mathbf{x})] - \tilde{g}_n^{(2)}(\tilde{\mathbf{Q}}_n(\boldsymbol{\tau} \,|\, \mathbf{x}) \,|\, \mathbf{x})(\bar{\mathbf{Q}}_n - \tilde{\mathbf{Q}}_n(\boldsymbol{\tau} \,|\, \mathbf{x}))}{n^{-1}\sum_{i=1}^n E_n^{-1}K(h_n^{-1}d(\mathbf{x}, \mathbf{X}_i))}$$
$$+ \tilde{g}_n^{(2)}(\tilde{\mathbf{Q}}_n(\boldsymbol{\tau} \,|\, \mathbf{x}) \,|\, \mathbf{x})(\bar{\mathbf{Q}}_n - \tilde{\mathbf{Q}}_n(\boldsymbol{\tau} \,|\, \mathbf{x})) \left[\frac{1 - n^{-1}\sum_{i=1}^n E_n^{-1}K(h_n^{-1}d(\mathbf{x}, \mathbf{X}_i))}{n^{-1}\sum_{i=1}^n E_n^{-1}K(h_n^{-1}d(\mathbf{x}, \mathbf{X}_i))}\right].$$

Using a Taylor expansion and Lemma 2.9, we have

$$\left\|[\tilde{g}_n^{(1)}(\bar{\mathbf{Q}}_n \,|\, \mathbf{x}) - \tilde{g}_n^{(1)}(\tilde{\mathbf{Q}}_n(\boldsymbol{\tau} \,|\, \mathbf{x}) \,|\, \mathbf{x})] - \tilde{g}_n^{(2)}(\tilde{\mathbf{Q}}_n(\boldsymbol{\tau} \,|\, \mathbf{x}) \,|\, \mathbf{x})(\bar{\mathbf{Q}}_n - \tilde{\mathbf{Q}}_n(\boldsymbol{\tau} \,|\, \mathbf{x}))\right\|$$
$$\leq 6s_2(2M)\|\bar{\mathbf{Q}}_n - \tilde{\mathbf{Q}}_n(\boldsymbol{\tau} \,|\, \mathbf{x})\|^2$$

for all sufficiently large $n$. From assumption B-1, it follows that

$$\left|1 - n^{-1}\sum_{i=1}^n E_n^{-1}K(h_n^{-1}d(\mathbf{x}, \mathbf{X}_i))\right| = O\left(\sqrt{(n\phi(h_n \,|\, \mathbf{x}))^{-1}\log n}\right)$$

as $n \to \infty$ *almost surely*. Lemma 2.5 implies

$$\left\|\tilde{g}_n^{(2)}(\tilde{\mathbf{Q}}_n(\boldsymbol{\tau} \,|\, \mathbf{x}) \,|\, \mathbf{x})(\bar{\mathbf{Q}}_n - \tilde{\mathbf{Q}}_n(\boldsymbol{\tau} \,|\, \mathbf{x}))\right\| \leq B_M\|\bar{\mathbf{Q}}_n - \tilde{\mathbf{Q}}_n(\boldsymbol{\tau} \,|\, \mathbf{x})\|$$

for all sufficiently large $n$ *almost surely*. From (2.6) in the proof of Proposition 2.7, we get $\|\bar{\mathbf{Q}}_n - \tilde{\mathbf{Q}}_n(\boldsymbol{\tau} \,|\, \mathbf{x})\| = O(\epsilon_n)$ as $n \to \infty$ *almost surely*. Therefore,

$$\left\|G_n(\bar{\mathbf{Q}}_n \,|\, \mathbf{x}) - \tilde{g}_n^{(2)}(\tilde{\mathbf{Q}}_n(\boldsymbol{\tau} \,|\, \mathbf{x}) \,|\, \mathbf{x})(\bar{\mathbf{Q}}_n - \tilde{\mathbf{Q}}_n(\boldsymbol{\tau} \,|\, \mathbf{x}))\right\| = O(\epsilon_n^2)$$

as $n \to \infty$ *almost surely*. $\square$

**Additional References**